\documentclass[11pt]{revtex4}

\usepackage{amssymb,amsmath,amsfonts,amstext,mathtools}

\usepackage[a4paper, textheight=26.1cm, textwidth=16.4cm, tmargin=2cm
]{geometry}

\def\al{\alpha}

\def\de{\delta}
\def\ep{\epsilon}
\def\ga{\gamma}
\def\Ga{\Gamma}
\def\om{\omega}
\def\vph{\varphi}

\def\th{\theta}

\def\bfmg{\mathbf{G}}
\def\bfy{\mathbf{y}}
\def\bfg{\mathbf{g}}
\def\bfmb{\mathbf{B}}
\def\bfma{\mathbf{A}}

\def\bfmx{\mathbf{X}}
\def\bfz{\mathbf{z}}
\def\bfe{\mathbf{e}}
\def\bfp{\mathbf{p}}
\def\bfq{\mathbf{q}}
\def\bfmr{\mathbf{R}}

\def\sfml{\mathsf{L}}
\def\sfmr{\mathsf{R}}
\def\sfmm{\mathsf{M}}
\def\sfmg{\mathsf{G}}
\def\sfb{\mathsf{b}}
\def\sfa{\mathsf{a}}
\def\sfc{\mathsf{c}}
\def\sfe{\mathsf{e}}

\def\call{\mathcal{L}}
\def\bbj{\mathbb{J}}

\def\p{\partial}
\def\na{\nabla}
\def\x{\times}
\DeclarePairedDelimiter\norm{\lVert}{\rVert}
\def\mun{^{-1}}
\def\cdott{\mathord{\cdot}}
\def\apos{ \hspace{-0.15ex}\raisebox{1.5ex}{,}\hspace{-0.25ex}
}

\def\avg{\texttt{avg}}
\def\osc{\texttt{osc}}

\def\bq{\begin{equation}}
\def\eq{\end{equation}}
\def\bqy{\begin{eqnarray}}
\def\eqy{\end{eqnarray}}

\def\ncr{ \nonumber \\ }

\def\cbc{\bar{\sfc}\sfb\apos\sfc}
\def\cba{\bar{\sfc}\sfb\apos\sfa}
\def\abc{\bar{\sfa}\sfb\apos\sfc}
\def\aba{\bar{\sfa}\sfb\apos\sfa}
\def\aca{\bar{\sfa}\sfc\apos\sfa}

\def\cbb{\bar{\sfc}\sfb\apos\sfb}
\def\abb{\bar{\sfa}\sfb\apos\sfb}

\begin{document}

\title[Intrinsic guiding-center to arbitrary order]{Gyro-gauge-independent formulation
\\
of the guiding-center reduction
\\
to arbitrary order in the Larmor radius}

\author
{L.~de~Guillebon}

\email{de-guillebon@cpt.univ-mrs.fr}

\author
{M.~Vittot}

\affiliation{\makebox[0ex][c]{Centre de Physique Th\'eorique, Aix-Marseille Universit\'e, CNRS, UMR 7332, 13288 Marseille, France}\\
\makebox[0ex][c]{Universit\'e de Toulon, CNRS, CPT, UMR 7332, 83957 La Garde, France}}

\begin{abstract}
\vskip3ex
{\bf Abstract:} The guiding-center reduction is studied using gyro-gauge-independent coordinates. The Lagrangian $1$-form of charged particle dynamics is Lie transformed without introducing a gyro-gauge, but using directly the unit vector of the component of the velocity perpendicular to the magnetic field as the coordinate corresponding to Larmor gyration. The reduction is shown to provide a maximal reduction for the Lagrangian and to work to all orders in the Larmor radius, following exactly the same procedure as when working with the standard gauge-dependent coordinate. 

 The gauge-dependence is removed from the coordinate system by using a constrained variable for the gyro-angle. The closed $1$-form $d\th$ is replaced by a more general non-closed $1$-form, which is equal to $d\th$ in the gauge-dependent case. The gauge vector is replaced by a more general connection in the definition of the gradient, which behaves as a covariant derivative, in perfect agreement with the circle-bundle picture. This explains some results of previous works, whose gauge-independent expressions did not correspond to a gauge fixing but indeed correspond to a connection fixing. 

 In addition, some general results are obtained for the guiding-center reduction. The expansion is polynomial in the cotangent of the pitch-angle as an effect of the structure of the Lagrangian, preserved by Lie derivatives. The induction for the reduction is shown to rely on the inversion of a matrix which is the same for all orders higher than three. It is inverted and explicit induction relations are obtained to go to arbitrary order in the perturbation expansion. The Hamiltonian and symplectic representations of the guiding-center reduction are recovered, but conditions for the symplectic representation at each order are emphasized.

\bigskip

\noindent{\it Keywords\/}: guiding center, Lie transform, Lagrangian $1$-form, maximal reduction, explicit induction, gauge-independent coordinates, gyro-gauge, connection, covariant derivative, principal circle bundle.
\end{abstract}

\maketitle
\newpage

	\section{Introduction}
	\label{intro}

 The dynamics of a charged particle in a strong magnetic field shows up a separation of time-scales, with the existence of a fast Larmor rotation, slower drifts, and an adiabatic invariant. This can be used to isolate completely the slow part of the dynamics from the single fast coordinate, the gyro-angle, and to build a true constant of motion, the magnetic moment \cite{NortAdia, CaryBriz}. Thus the guiding-center reduction brings a slow reduced motion and decreases by two the effective dimension of the dynamics. It is the starting point of gyrokinetics, which is a kinetic description of plasma dynamics in a strong magnetic field, and a key model in the study of plasma micro-turbulence \cite{BrizHahm07}. The most efficient results for the guiding-center reduction are obtained by Lie transforming the Lagrangian $1$-form in an expansion in a small parameter related with the magnetic inhomogeneity at the scale of the Larmor radius \cite{CaryBriz, Little79, Little81, Little83}. 
  
 In this paper, we clarify some aspects of guiding-center theory, especially some difficulties about the definition of the gyro-angle, by showing that the reduction can be done at arbitrary order in the Larmor radius using gyro-gauge-independent coordinates. \\

 Indeed, the introduction of a scalar coordinate for the angle measuring the Larmor rotation, namely the \textsl{gyro-angle}, imposes to choose at each point in the configuration space a direction, an axis, in the plane perpendicular to the magnetic field, defined as the zero from which the angle is measured. This corresponds to fixing a gauge in the theory, the so-called gyro-gauge, and the presence of this gauge in the theory raised some non-trivial questions \cite{Little88, Sugi08, Krom09, Sugi09, BurbQin12}, even the bare existence of a global choice of gauge can fail. So, it is interesting to consider the reduction from a more intrinsic point of view. 
 
 In a previous work \cite{GuilGCmin}, we proposed a guiding-center reduction which avoided to introduce a gyro-gauge. The idea was to Lie transform directly the equations of motion instead of Lie transforming the Lagrangian, as is usually done \cite{CaryLie, Little82}. This brought a much simplified reduction.  Especially, it provided the minimal guiding-center reduction which concerned only four coordinates (instead of six): the transformation generator had no gyro-angle component. So, for the gyro-angle, the initial gauge-independent coordinate could be used, and no gauge fixing was needed. This physical coordinate is the unit vector $\sfc$ of the component of the velocity orthogonal to the magnetic field, which defines the direction of the perpendicular velocity. 
 
 A limitation of that approach was that it was not suited to non-minimal guiding-center reductions, for which the method relying on Lie transforming the Lagrangian appeared as necessary, or at least much more efficient. 
 
 For instance, it is interesting to have the slow reduced motion Hamiltonian, but the Hamiltonian structure of the reduced model is hard to deal with when working on the equations of motion. In addition, the magnetic moment is usually taken as one of the reduced coordinates; this can be done by working on the equations of motion, but it is not so straightforward \cite{GuilMagMom}. In a deeper way, the freedom involved in the gyro-averaged part of the coordinate change can be employed for the reduced motion to be as strongly reduced as possible, to make the guiding-center dynamics as simplified as possible.

 Obtaining such a maximal guiding-center reduction by Lie transforming the equations of motion is far from simple, especially because it implies to solve non-trivial secular differential equations. On the contrary, Lie transforming the Lagrangian $1$-form basically relies on algebraic equations, and the requirements for a maximal reduction are not much more difficult to get than the minimal ones. This method also guarantees that the 4-dimensional slow reduced motion is Hamiltonian, by working on a quarter-canonical structure in the Poisson bracket. 
 
 So, the goal is to use the physical gauge-independent coordinate also when Lie transforming the Lagrangian, in order to consider a gauge-independent maximal guiding-center reduction. \\

 The introduction of a vectorial quantity $\sfc$ for the gyro-angle coordinate raises some questions, because the coordinate system becomes constrained: the variable $\sfc$ has to remain normalized and perpendicular to the magnetic field. Changing the spatial position $\bfq$ implies to change the coordinate $\sfc$ at the same time. This induces a connection for a covariant derivative on a space-dependent circle, which is related to the circle-bundle picture underlying in the gyro-angle coordinate \cite{BurbQin12, Krus62}. It was already present in \cite{GuilGCmin} when Lie transforming the equations of motion, but it is more involved to deal with for the full guiding-center reduction, because the coordinate $\sfc$ will be changed as well, and not only derivatives or vector fields are involved, but also differential forms.

 The resulting reduction will naturally provide gauge-independent results, whereas in the usual approach, they were obtained only for a part of the reduced quantities. This fact can shed interesting light on previous guiding-center results, especially those related to gauge invariance. For instance, in the usual approach, the gradient is not gauge-independent, and the reduced Poisson bracket involves a gauge-independent corrected gradient. A comparison with the results of the gauge-independent formulation is a way to get an intrinsic interpretation for this corrected gradient. 
 
 The results of the gauge-independent formulation can also be used to explore other questions about the gauge-dependent approach, for instance related to gauge arbitrariness and anholonomy \cite{Little81, Little88}. To avoid confusion, they will be the topic of another paper \cite{GuilAnholon}. Here, we show how a maximal guiding-center reduction can be derived in a gauge-independent formulation to arbitrary order in the Larmor radius. 

 The proof relies on explicit induction relations to all orders, because the induction can be written as a matrix product, with some coefficients being differential operators. Through inversion of this matrix, a maximal reduction can be studied, towards a more complete reduction and a more general viewpoint on special reductions considered in previous works, such as the so-called Hamiltonian and symplectic representations identified in \cite{BrizTron12}. 
 
 In the derivation, the cotangent of the pitch-angle is used as the coordinate corresponding to the parallel velocity (component of the velocity parallel to the magnetic field), since this coordinate simplified computations for the minimal guiding-center reduction in \cite{GuilGCmin} and made all formulae polynomials. This will clarify why this polynomiality can be observed in the results of the full guiding-center reduction as well.\\

Taking care of the reduction at higher order is interesting, if not needed, for two reasons. On the first hand, it is necessary to validate the gauge-independent approach. Indeed, the gauge issues became more sensitive when addressing the second-order reduction \cite{Sugi08, BurbySquir13}. An acceptable solution to these issues, which is the goal of the present paper, needs to be appropriate for higher orders. 

On the other hand, it is motivated because standard works proceeded only up to part of the second-order reduction, but recent results emphasize the importance of higher-order terms, for instance because they are involved in the conservation of angular momentum \cite{BurbySquir13, ParrCatt10} and are crucial for a proper description of intrinsic rotation of tokamak plasmas, a key phenomenon to stabilize turbulence and increase the energy confinement time, which is the main goal of magnetic fusion. 
\\

 The paper is organized as follows. In sect. 2, a few facts are reminded about the initial dynamics, the choice of coordinates for the gyro-angle, the method of Lie transforming the Lagrangian $1$-form, and the hierarchy of requirements involved in the guiding-center reduction.
 
 For the sake of completeness and clarity, the mechanism of Lie transforming the Lagrangian through an expansion in a small parameter is described in an appendix, with emphasis on the three steps it involves: an initialization for the lowest orders, whose choices are the key to make the reduction work and possibly be optimal; an algorithm which applies for higher orders, is purely mechanical and can be applied to study the reduction to arbitrary order; and an intermediate step in between. 

 In sect. 3, the method is applied to the guiding-center reduction in case the gyro-angle coordinate is chosen as the physical variable $\sfc$. The derivation is written in matrix form, which emphasizes both the lowest-order choices that allow the reduction to work and to be maximal for the Lagrangian, and the algorithmic character of the procedure at higher orders. The full derivation is explained because we are interested in the reduction at arbitrary order in the Larmor radius, which implies to use all the ingredients of the detailed mechanism at work. For orders lower than $3$, the procedure follows the same lines as when working with the gauge-dependent coordinate, but formulae have to be used in their intrinsic version, for instance because the basis of $1$-forms involves non-closed $1$-forms. Finally, for orders higher than 2, explicit formulae are given for the induction relations, allowing to go to arbitrary order, and to give a unified framework where recent results may seem to discord somehow with each other \cite{BrizTron12, ParrCalv11}. 

 In sect. 4, the results are compared with previous works, either Lie transforming the equations of motion, or using a gauge-dependent gyro-angle.
 
 For the sake of simplicity, we consider the special case where there is no electric field, but the generalization for a non-zero electric field is straightforward, as will be shown in subsect. \ref{SectElecField}.

	\section{Coordinates, method and requirements}
 	\label{p1}

 The dynamical system is simply a charged particle with position $\bfq$, momentum $\bfp$, mass $m$ and charge $e$, under the influence of a static inhomogeneous magnetic field $\bfmb$. The motion is given by the Lorentz force
\begin{align}
	\dot \bfq &= \tfrac {\bfp}{m}
	\,,\ncr
	\dot \bfp &= \tfrac {\bfp}{m}\x e\bfmb
	\,.\notag
\end{align}

 When the magnetic field is strong, the motion implies a separation of time-scales. This is best seen by choosing convenient coordinates for the momentum space, for instance
\begin{align}
	p &:= \norm \bfp
	\,,\ncr
	\vph &:= \arccos \big(\tfrac{\bfp\cdott\sfb}{\norm{\bfp}}\big)
	\,,\ncr
	\sfc &:= \tfrac{\bfp_\perp}{\norm{\bfp_\perp}}
	\,,\notag
\end{align}
where $\sfb:= \tfrac{\bfmb}{\norm{\bfmb}}$ is the unit vector of the magnetic field, and $\bfp_\perp:=\bfp-(\bfp\cdott\sfb)\sfb$ is the so-called \textsl{perpendicular momentum}, i.e. the orthogonal projection of the momentum onto the plane $\bfmb^\perp$ perpendicular to the magnetic field. The coordinate $p$ is the norm of the momentum; the coordinate $\vph$ is the so-called \textsl{pitch-angle}, i.e. the angle between the velocity and the magnetic field. The last coordinate $\sfc$ is the unit vector of the perpendicular velocity.

 Then, the equations of motion write
\begin{align}
	\dot \bfq &= \tfrac {\bfp}{m}
	\,,\ncr
	\dot p & = 0
	\,,\ncr
	\dot \vph & = -\tfrac{\bfp}{m}\cdott\na\sfb\cdott\sfc
	\,,\ncr
	\dot \sfc & = - \tfrac{e B}{m}\sfa 
	- \tfrac{\bfp}{m}\cdott\na\sfb\cdott(\sfc\sfb + \sfa\sfa \cot \vph )
	\,,\notag
\end{align}
where $\bfp$ is now a shorthand for $p(\sfb \cos \vph + \sfc \sin \vph)$, the norm of the magnetic field $\norm\bfmb$ is denoted by $B$, and following Littlejohn's notations \cite{Little79, Little81, Little83}, the vector $\sfa:=\sfb\x\sfc$ is the unit vector of the Larmor radius, so that $(\sfa,\sfb,\sfc)$ is a right-handed orthonormal 
frame (rotating with the momentum). 

 In the case of a strong magnetic field, the only fast term is the Larmor frequency $\om_L:=\tfrac{e B}{m}$. Writing the dynamics as $\dot\bfz\cdott\p_\bfz$, all other terms as $\om_L$ appear to be of order $\tfrac{p}{m}\na$, which means that the small parameter of the theory is of order $\tfrac{p}{m\om_L}\na=\tfrac{p}{eB}\na$. A more detailed study (e.g. \cite{NortAdia, CaryBriz, Little81, GuilGCmin}) shows that it is rather 
\begin{equation}
	\ep:=\tfrac{p\sin\varphi}{eB}\na
	\,.
	\label{DefinitEpsilon}
\end{equation}
It is an operator, but the gradient $\na$ has only a meaning for orderings; it acts on the magnetic field and can be given a more precise meaning by relations such as $\na\approx\tfrac{\norm{\na^{n+1}\bfmb}}{\norm{\na^{n}\bfmb}}$. The ordering parameter $\ep$ is related to the magnetic inhomogeneity at the scale of the Larmor radius $r_L:=\tfrac{p\sin\varphi}{eB}$. By abuse of language, it is often considered as just the Larmor radius, or as the inverse charge $e^{-1}$ \cite{NortAdia, CaryBriz, Little81}.

\subsection{Choice for the gyro-angle coordinate}

 The Larmor frequency $\om_L:=\tfrac{e B}{m}$ concerns only one coordinate, $\sfc$, the direction of the vector $\bfp_\perp$ in the $2$-dimensional plane perpendicular to the magnetic field. This corresponds to an angle, the so-called \textsl{gyro-angle}, and measures the Larmor gyration of the particle momentum around the magnetic field. 

 To get a true scalar angle instead of the vector $\sfc$, one chooses at each point $\bfq$ in space a direction which will be considered as the reference axis $\sfe_1(\bfq) \in \bfmb^\perp(\bfq)$. This corresponds to fixing a gauge in the theory, the so-called \textsl{gyro-gauge}. Then, the gyro-angle $\theta$ is defined from the oriented angle between the chosen reference axis $\sfe_1(\bfq)$ and the vector $\sfc$ through the following relation:
\bq
	\sfc = - \sin \theta \sfe_1 - \cos \theta \sfe_2
	\,.
\label{DefThetaPerp}
\eq
The equation of motion for $\theta$ is
$$
	\dot \theta = 
	\tfrac{e B}{m} 
	+ \cot \vph \tfrac{\bfp}{m}\cdott\na\sfb\cdott\sfa 
	+ \tfrac{\bfp}{m}\cdott\na\sfe_1\cdott \sfe_2
	\,,
$$
with $\sfe_2:=\sfb\x\sfe_1$ the unit vector such that $(\sfb,\sfe_1,\sfe_2)$ is a (fixed) right-handed orthonormal frame \cite{Little81}.\\

 The coordinate $\theta$ is not intrinsic, it depends on the chosen gauge $\sfe_1(\bfq)$, which raised some questions about the gauge-invariance of the theory, about the failure of global existence for $\sfe_1$, as well as about the presence of an anholonomic phase in the coordinate system \cite{Little81, Little88, BurbQin12, LiuQin11, BrizGuil12}. All these difficulties originate because $\theta$ is not given by the physics, it is not a purely intrinsic coordinate. For physical results, what is needed is not $\theta$ but only $\sfc$. This is clearly illustrated by all the results of guiding-center theory, e.g. \cite{NortAdia, CaryBriz, Little81, Little83}, where $\theta$ intervenes only through the quantity $\sfc$ everywhere (except in its own definition and in subsequent relations). So, we will avoid this coordinate and keep the corresponding initial variable, $\sfc$, as in \cite{GuilGCmin}. The quantity $\theta$ will be used only with a symbolic meaning for the fast angle, or when making comparison with the gauge-dependent approach.\\

 The use of a unit vector avoids having to fix a gauge for the zero of the angle, and it allows to work with a physical quantity: the unit vector of the perpendicular velocity
\bq
	\sfc
	:=\tfrac{\bfp_\perp}{\norm{\bfp_\perp}}
	=\tfrac{\bfp-\sfb(\sfb \cdott \bfp~)}{p \sin \vph}
	\,,
\label{DefPerp}
\eq
which indeed corresponds to the direction of the perpendicular velocity, and is a coordinate measuring the Larmor gyration. It is an angle, since it is a unit vector in a plane, namely the plane $\bfmb^\perp$, orthogonal to the local magnetic field. But this unit vector is immersed into $\mathbb R ^3$, which means it is in $\mathbb S ^1(\bfq)$. This spatial dependence implies that the coordinate space is constrained: the gyro-angle $\sfc$ is not independent of the spatial position. 
 
 When the coordinate $\bfq$ is changed, the coordinate $\sfc$ cannot be kept unchanged, otherwise it may get out of $\bfmb^\perp$.  Differentiating relation (\ref{DefPerp}) with respect to $\bfq$ gives 
\bq
	\na\sfc 
	= - \na\sfb\cdott(\sfc\sfb+\sfa\sfa\cot\vph)
	\,.
\label{ConnectNatur}
\eq
This formula can be obtained more easily by noticing that in the change of coordinates $(\bfq,\bfp)\longrightarrow(\bfq,p,\vph,\sfc)$, the following relations hold
\begin{align}
	& 
	\hspace{-3ex}
	-\sin\vph\na\vph 
	=\na\cos\vph
	=\na\sfb\cdott\tfrac{\bfp}{p}
	=\na\sfb\cdott\sfc\sin\vph
	\,,
\ncr
	& 
	\Rightarrow \na\vph 
	= 
	-\na\sfb\cdott\sfc
	\,,
\ncr
	& 
	\Rightarrow 0 
	= 
	\na\left(\tfrac{\bfp}{p}\right)
\ncr
	&
	\hspace{5ex} 
	= 
	\na\sfc\sin\vph+\na\sfb\cos\vph
	+\na\vph(-\sfb\sin\vph+\sfc\cos\vph)
\ncr
	&
	\hspace{5ex} 
	= 
	(\na\sfc +\na\sfb\cdott\sfc\sfb)\sin\vph
	+ \na\sfb\cdott(1-\sfc\sfc)\cos\vph
	\,,
\ncr
	& 
	\Rightarrow \na\sfc 
	= 
	- \na\sfb\cdott(\sfc\sfb+\sfa\sfa\cot\vph)
	\,, 
	\label{GradC}
\end{align}
in which $\na$ means differentiation with respect to $\bfq$ while keeping $\bfp$ constant, and we used that $1=\sfa\sfa+\sfb\sfb+\sfc\sfc$ and that $\sfb$ is a unit vector, which implies $\na\sfb\cdott\sfb=\na\left(\tfrac{\sfb^2}{2}\right)=0$.

 Formula (\ref{ConnectNatur}) is not well defined where $\varphi=0 ~(\text{mod } \pi)$, i.e. where $\bfp$ is parallel to $\bfmb$. But this is no trouble, since it fits in with a usual limitation of guiding-center theories. For instance, guiding-center transformations are not defined where $\varphi$ is zero, since they involve many $\sin \varphi$ as denominators \cite{NortAdia, CaryBriz, Little81, Little83}. At the points where $\varphi=0 ~(\text{mod } \pi)$, the vector $\sfc$ itself is not defined, neither is the coordinate $\theta$. It is an implicit assumption in all gyro-kinetics and guiding-center works that those points are excluded from the theory. \\

 If the coordinate $\theta$ was used, formula (\ref{ConnectNatur}) would be replaced by
\bq
	\na\sfc 
	= - \na\sfb\cdott\sfc\sfb +\na\sfe_1\cdott\sfe_2 ~\sfa
	\,,
\label{ConnectGauge}
\eq
where $\na\sfe_1\cdott\sfe_2$ is the so-called \textsl{gauge-vector}, which is usually denoted by $\bfmr$, depends only of the position $\bfq$, and is related to the choice of gauge. 

To fit in with both coordinates, we define a more general function $\bfmr_g(\bfq,\bfp)$:
$$
	\bfmr_g:=\na\sfc\cdott\sfa 
	\,.
$$
Then, in any coordinates, the previous formulae (\ref{ConnectNatur}) and  (\ref{ConnectGauge}) write 
\bq
	\na\sfc 
	= - \na\sfb\cdott\sfc\sfb
	+ \bfmr_g \sfa
	\,.
\label{ConnectBoth}
\eq
The value $\bfmr_g=- \na\sfb\cdott\sfa\sfa\cot\vph$ corresponds to the physical definition of $\sfc$, and the corresponding formula (\ref{ConnectNatur}). The value $\bfmr_g=\na\sfe_1\cdott\sfe_2=\bfmr$ will be linked with the usual case relying on the gauge-dependent coordinate $\theta$, according to formula (\ref{ConnectGauge}).

 The spatial dependence of the vector $\sfc$ through formula (\ref{ConnectBoth}) must be taken into account each time a gradient acts on a function that depends on the fast angle $\sfc$, e.g. in total derivatives, in Lie transforms, and when computing the action of a spatial component of the change of variables.
 
 A more detailed study of the coordinate system \cite{GuilAnholon} shows that $\bfmr_g$ is the connection associated to the gyro-angle circle bundle, which indeed can be any function of the phase space, and that $\na$ is the (spatial part of the) corresponding covariant derivative.

\subsection{Lie transforming the Lagrangian $1$-form}

 The goal is to isolate the slow part of the dynamics from the fast angle $\theta$ (or $\sfc$), by performing a near-identity change of coordinates such that the dynamics of the remaining coordinates does not depend on $\theta$; this averaging procedure is the primary requirement for the guiding-center reduction. We showed in \cite{GuilGCmin} how it can be derived by Lie transforming the equations of motion when keeping the physical coordinate $\sfc$. It appeared to be very straightforward for the minimal guiding-center reduction, but the additional requirements (e.g. the use of the magnetic moment as a coordinate or a further simplification of the reduced dynamics) were not so easy to obtain. They are more efficiently obtained by Lie transforming the Lagrangian $1$-form, which relies on the Hamiltonian structure of this dynamical system. \\

 The corresponding Hamiltonian function is just the particle kinetic energy 
$$
	H:=\tfrac{p^2}{2m}
	\,.
$$

 The Poisson bracket is non-canonical, it contains the gyro-magnetic coupling term
\bq
\{F,G\}=  \p_\bfq  F \cdott \p_\bfp  G -\p_\bfp  F\cdott  \p_\bfq  G -  \p_\bfp  F\cdott e\bfmb \x  \p_\bfp  G	\, .
\label{Crochet}   
\eq

 Together, the Poisson bracket and the Hamiltonian induce the equations of motion through Hamilton's equations
\bq
   \dot \bfz = \{\bfz,H\} 
   \,,
\eq  
where all the coordinates were grouped in a single phase-space vector $\bfz:=(\bfp,\bfq)$.\\

 Instead of working with the Poisson bracket and the Hamiltonian, it is easier to work on the \textsl{Poincar\'e-Cartan $1$-form} $\Ga$ \cite{Little83}. This last is usually called just a Lagrangian $1$-form, or simply a \textsl{Lagrangian}. It concentrates all the information on the Hamiltonian structure into one single quantity, which in addition is much less constrained than a Poisson bracket. It is a $1$-form, defined over a $7$-dimensional space $\bfy:=(\bfp,\bfq,t)$ by \cite{CaryBriz, Little83, Little82}
\bq
	\Ga:=(e\bfma+\bfp)\cdott d\bfq -H dt
	\,,
\label{LagrIni}
\eq
which yields a variational formulation of the dynamics with the action \cite{Little81}:
$$
	\mathcal A:= \int \Ga
	\,.
$$

 The symplectic part of $\Ga$ is $\Ga_s:=(e\bfma+\bfp)\cdott d\bfq$ \cite{CaryBriz,Little83}. It is a $1$-form $\Ga_s=\Ga_s^i dz^i$ in the usual $6$-dimensional phase-space, which gives the Lagrange $2$-form through exterior derivative
\bq
	\om_s:=d\Ga_s=(\p_i\Ga_s^j-\p_j\Ga_s^i)~dz^i \otimes dz^j
	\,,
\label{ExtDeriv}
\eq
where $\otimes$ means tensorial product, and Einstein's convention is used: there is an implicit summation over repeated indices. In turn, $\om_s$ is invertible, and its inverse gives the Poisson bracket
$$
	\bbj:=\om_s^{-1}
	\,,
$$
with $\bbj$ the bivector defined by the relation $\{F,G\}=\p_i F  \bbj^{ij}  \p_j G$.\\

 In the presence of a strong magnetic field, the small parameter $\ep$ allows for a perturbation expansion of all quantities. The Lagrangian $\Ga$ from formula (\ref{LagrIni}) can be written 
$$
	\Ga=\Ga_{-1}+\Ga_0
	\,,
$$
where the index refers to the order in the magnetic field, (or in $e^{-1}$, following Northrop's work \cite{NortAdia})
\bqy
	\Ga_{-1}&:=&e\bfma\cdott d\bfq
	\,,
	\ncr
	\Ga_0&:=&\tfrac{\bfp}{m}\cdott d\bfq - \tfrac{p^2}{2m}dt
	\,,
\label{LagrOrdering}
\eqy
whose ratio is indeed of order $\ep$. The lowest order being $-1$ reminds that the guiding-center is a singular perturbation theory.\\

 The Lagrangian is transformed by the exponential of a Lie transform
 $$
	\Ga\longrightarrow \overline \Ga:=e^{\call_{\bfmx}}\Ga 
	\,,
$$
where $\call_{\bfmx}$ is the Lie derivative along the vector field $\bfmx$, whose inverse is the generator of the near-identity coordinate transformation:
$$ 
	\bfy\longrightarrow \overline \bfy:=e^{-\bfmx}\bfy 
	\,.
$$ 

 The generator can be expanded $\bfmx=\bfmx _1+ \bfmx _2+...$, where $\bfmx_n$ is the term of $\bfmx$ that is of order $n$ in $\ep$. In fact, it is equivalent but simpler to replace the single transformation $e^{\call_{\bfmx_1}+\call_{\bfmx_2}+...}$ with a complicated generator by a series of transformations with a simple generator for each of them 
$$ 
	\bfy\longrightarrow \overline \bfy:=...e^{-\bfmg_2}e^{-\bfmg_1}\bfy 
	\,,
$$ 
where $-\bfmg_n$ is the vector field generating the $n$-th transformation, and is purely of order $\ep^n$. The Lagrangian then transforms as 
$$
	\Ga\longrightarrow \overline \Ga:=...e^{\sfml_2}e^{\sfml_1}\Ga 
	\,,
$$
where $\sfml_n:=\call_{\bfmg_n}$ is the Lie derivative along the vector field $\bfmg_n$. Notice that $\call_{\bfmx}\neq\sfml_1+\sfml_2+...$ because $\bfmg_n\neq\bfmx_n$.\\

 As the Lagrangian is time-independent, it is interesting to use time-independent perturbation theory, by imposing a transformation that does not depend on time and does not affect the time coordinate: $\bfmg$ is constant in time and $\bfmg^t=0$. 
 
 This implies that the Hamiltonian will transform as a scalar function. Indeed, for any time-independent vector field $\bfmg$ and any $1$-form,
\begin{align}
	(\call_\bfmg \Ga)_t
	& =[i_\bfmg\cdott d \Ga]_t+ [d( i_\bfmg\Ga)]_t
	\ncr
	& = 
	\left[
	\bfmg^i\p_{\bfy^i}(\Ga_j) d\bfy^j
	- 
	\bfmg^i\p_{\bfy^j}(\Ga_i)
	d\bfy^j 
	\right]_t 
	+ \p_t(\bfmg\cdott\Ga)
	\ncr
	& 
	= \bfmg\cdott\p_\bfy\Ga_t
	+ \p_t(\bfmg)\cdott\Ga
	= \bfmg\cdott\p_\bfy\Ga_t
	\,,
	\notag
\end{align}
where formula (\ref{LieDeriv}) was used for Lie derivatives, and formula (\ref{ExtDeriv}) for exterior derivatives (here in the $7$-dimensional space).

 Another consequence is that at each order $n$ in $\ep$, there will be seven requirements (see next subsection), one for each component of the reduced Lagrangian $\overline\Ga_n$, and only six freedoms, one for each component of the time-independent transformation generator $\bfmg_n$. One freedom is missing in $\bfmg_n$ and must be looked for elsewhere. Now, the Lagrangian is defined only to within a total derivative, since only its exterior derivative has a physical meaning, and $d (\Ga+dS) = d\Ga+d^2 S =d \Ga$ for any function $S$, which is called a \textsl{gauge function} \cite{CaryBriz}. Be careful, this gauge has nothing to see with the gyro-gauge nor the gyro-angle, it is just an arbitrariness in the definition of the Lagrangian. Rather, it corresponds to the electromagnetic gauge, since it can be absorbed in a redefinition of the potential $\bfma$ (together with $\Phi$ when there is a non-zero electric field, see page \pageref{LagrElecField}): the Lagrangian is expressed in terms of the potential, and does depend on the electromagnetic gauge, but the dynamics, as well as all physical quantities, are electromagnetic-gauge invariant, they depend only on the electromagnetic field. 

 The freedom embodied in $S$ is needed to obtain the maximal reduction, since it gives the expected seventh freedom. Then the reduced Lagrangian is 
:
$$
	\overline \Ga:= \left(... e^{\sfml_2}e^{\sfml_1}\right)~ (\Ga_{-1} + \Ga_0 )  
	+ \big( dS_{-1}+dS_0 + ... \big)
	\,.
$$
It will be determined order by order in $\ep^n$:
\begin{align}
	\overline \Ga_{-1}
	&=
	\Ga_{-1} + dS_{-1}
\label{ReducLagr}	
\\
	\overline \Ga_{0}
	&=
	\sfml_1\Ga_{-1}+\Ga_0 + dS_0
\ncr
	\overline \Ga_{1}
	&=
	\left(\sfml_2+\tfrac{\sfml_1^2}{2}\right)\Ga_{-1}+\sfml_1\Ga_0
	 + dS_1
\ncr
	\overline \Ga_{2}
	&=
	\big(\sfml_3+\sfml_2\sfml_1
	+\tfrac{\sfml_1^3}{6}\big)\Ga_{-1}
	+\left(\sfml_2+\tfrac{\sfml_1^2}{2}\right)\Ga_0
	 + dS_2
\ncr
	\overline \Ga_{3}
	&=
	\Big[
	\sfml_4
	+\sfml_3\sfml_1
	+\sfml_2
		\big(\tfrac{\sfml_2}{2}+\tfrac{\sfml_1^2}{2}\big)
	+\tfrac{\sfml_1^4}{24}
	\Big]\Ga_{-1}
\ncr
	& \hspace{15ex}
	 +\Big[\sfml_3+\sfml_2\sfml_1+\tfrac{\sfml_1^3}{6}\Big]\Ga_0
	 + dS_3
\ncr
	... & 
	\notag
\end{align}

 In principle, these are differential equations for $\bfmg$, because the action of a Lie derivative $\call_\bfmg$ over a $1$-form $\ga$ writes
\bq
	\call_\bfmg \ga 
	= (i_\bfmg d + d  ~i_\bfmg)\ga
	\,,
\label{LieDeriv}
\eq
where the operator $i_\bfmg$ is the interior product, e.g.
\bq
	i_\bfmg \ga = \bfmg\cdott\ga=\ga(\bfmg)
	\,.
\label{InteriorProduct}
\eq
So, the first term in (\ref{LieDeriv}) is algebraic in $\bfmg$, but the second one is differential in $\bfmg$. 

 However, the differential operators can be avoided by the following argument. The last term in (\ref{LieDeriv}) involves an exterior derivative, and can be removed by redefining the gauge function $S$. In addition, formula (\ref{LieDeriv}) together with the property $d^2=0$ imply that the exterior derivative and the Lie derivative commute:
$$
	\call_\bfmg d 
	= d \call_\bfmg
	= d ~i_\bfmg d
	\,.
$$ 
This means that for any vector fields $\bfmx$ and $\bfmg$
\begin{align}
	\call_\bfmg \call_\bfmx \Ga 
	& = (i_\bfmg d + d i_\bfmg) \call_\bfmx \Ga
	= i_\bfmg d (i_\bfmx d + d i_\bfmx) \Ga 
	+ d (i_\bfmg \call_\bfmx \Ga)
	\ncr
	& = i_\bfmg d i_\bfmx d \Ga 
	+ d (i_\bfmg \call_\bfmx \Ga)
	\,.
\notag
\end{align}
In computations for $\overline\Ga$, the last term can again be removed by redefining the gauge function $S$. By induction, it is now easy to see that in formula (\ref{ReducLagr}), exponentials of Lie derivatives can be considered as just exponentials of interior products provided the gauge function $S_n$ is defined in a convenient way at each order $\ep^n$ to absorb all the exterior derivatives involved in equation (\ref{ReducLagr}): 
\begin{align}
	\call_{\bfmg_{n_1}} \call_{\bfmg_{n_2}} ... \call_{\bfmg_{n_k}} 
	& \Ga + d S
	\ncr
	& = (i_{\bfmg_{n_1}}d) (i_{\bfmg_{n_2}}d) ... (i_{\bfmg_{n_k}} d) \Ga+ dS'
	\,.
\notag
\end{align}
For the following, we will redefine $S$ according to this rule, but for simplicity, we drop the prime and write $S$ for $S'$.

 Using this rule and the fact that in the equation for $\overline\Ga_n$, the $\overline\Ga_{i<n}$ are already known, equations (\ref{ReducLagr}) can be written
\begin{align}
	\overline \Ga_{-1}
	&=
	\Ga_{-1} + dS_{-1}
\label{ReducLagrAlg}	
\\
	\overline \Ga_{0}
	&=
	\bfmg_1\cdott\om_{-1}+\Ga_0 + dS_0
\ncr
	\overline \Ga_{1}
	&=
	 \bfmg_2\cdott\om_{-1}
	+ \tfrac{\bfmg_1}{2}\cdott(\om_0+\overline\om_0)
	+ dS_1
\ncr
	\overline \Ga_{2}
	&=
	\bfmg_3\cdott\om_{-1}
	+ \bfmg_2\cdott\overline\om_0
	+ \tfrac{\bfmg_1}{6}\cdott d
	\big[\bfmg_1\cdott(2\om_0+\overline\om_0)\big]
	+ dS_2
\ncr
	\overline \Ga_{3}
	&=
	\bfmg_4\cdott\om_{-1}
	+ \bfmg_3\cdott\overline\om_0
	+ \bfmg_2\cdott\overline\om_1
	- \tfrac{(\bfmg_2\cdott d)^2}{2}\om_{-1}
\ncr
	& \hspace{20ex} + \tfrac{(\bfmg_1\cdott d)^2}{24}
	\bfmg_1\cdott(3\om_0+\overline\om_0)
	+ dS_3
\ncr
	... & 
	\notag
\end{align}
where the notation $\bfmg\cdott:=i_\bfmg$ is used for the interior product, as in formula (\ref{InteriorProduct}). In addition, the $n$-$th$-order Lagrange $2$-form was defined in the natural way:
$$
	\om_n:=d\Ga_n
	\,.
$$
 
In the next subsection, we study the properties that are wished for $\overline\Ga$. In the next section, the unknowns $\bfmg_n$ and $S_n$ will be determined such that $\overline\Ga_n$ has those desired properties.

\subsection{The hierarchy of requirements}

 A) 
 The primary requirement for the guiding-center reduction is to isolate the slow dynamics of the coordinates $(\overline \bfq, \overline \varphi)$ from the fast gyro-angle $\overline\theta$. From the point of view of the Lagrangian $\overline\Ga$, it may be obtained by making $\overline\Ga$ independent of $\overline\theta$. This is actually stronger than the strict minimal requirement, since it implies to average the dynamics of $\overline\theta$ as well. However, in the Lagrangian approach, contrary to when working on the equations of motion, the minimal requirement would be difficult to get, if not impossible, and it is quite easier to average all the reduced dynamics. 
 
 So, the goal is that the reduced Lagrangian does not depend on the reduced gyro-angle, which means that all its non-zero Fourier components (i.e. purely oscillatory terms) are zero: 
\bq
	\osc (\overline\Ga_n) = 0
	\,,
	\label{RequirPrior1}
\eq
where following Littlejohn's notations, $\osc=1-\avg$ is the projector onto gyro-fluctuations, with $\avg$ the complementary projector onto gyro-averages:
$$
	\avg (f) = \tfrac{1}{2 \pi}\int_0^{2\pi}\! d\theta~ f
	\,,
$$
for any function $f$. This average can be computed without introducing any gauge: either using the intrinsic calculus introduced in \cite{GuilMagMom}, either using the matrix calculus introduced in \cite{GuilGCmin}. Also, the coordinate $\theta$ can be used as an intermediate quantity for this computation, which is made at constant $\bfq$, so that the presence of a gauge (only for the intermediate computation) is of no consequence.\\

 B) 
 Averaging the motion or the Lagrangian does not determine the average components of the coordinate change, as is clear in \cite{GuilGCmin}, for instance. This lets some freedom in the procedure and suggests to impose stronger requirements for the reduction. The basic idea is to use the available freedoms to make the reduced dynamics as simplified as possible. 
 
 A natural prospect is to make trivial the reduced dynamics 
$$
 	\dot{\overline\bfz}^j=0
 	\,,
$$
for some components $j$, by including constants of motion in the reduced coordinates. For the remaining coordinates, one can consider putting their reduced dynamics to zero just for orders higher than $2$ or $3$ for instance:
$$
 	\dot{\overline\bfz}^j_n=0
 	\,,
$$
for all higher orders, where the index $n$ refers to the order in $\ep^n$ and the exponent $j$ indicates the component of the vector $\dot{\overline\bfz}$. 

 When this is achieved, the reduced dynamics is given just by lowest-order terms; it is exactly known after the lowest orders have been derived, without computing the reduction at higher orders, which are useful only to determine the transformation. 

 In the procedure working on the Lagrangian $1$-form, the "components" are not the ones of the reduced equations of motion, but the ones of the reduced Lagrangian
\bq
 	\overline\Ga_n^j=0
 	\,,
\label{RequirPrior3}
\eq
for higher $n$. For differential forms, we use the same convention as for vectors: the index $n$ indicates the order in $\ep^n$ and the exponent $j$ refers to the component. It departs from the usual notation $\overline\Ga_n=(\overline\Ga_n)_j d\bfz^j$, but it avoids excessive use of parentheses. 

 Equation (\ref{RequirPrior3}) gives additional (i.e. non minimal) requirements for the reduced Lagrangian $\overline\Ga$ and it is used to determine the averaged transformation generators. When it cannot be obtained completely, the goal is to obtain it for as many components $j$ as possible, to get what can be considered as the maximal reduction. \\

 C) 
 When studying the Lie transform of the Lagrangian $1$-form, one variable, namely the \textsl{magnetic moment} $\overline\mu$, plays a key role, as the variable conjugated to the gyro-angle. 
 
 Basically, including $\overline\mu$ among the reduced coordinates is a way to obtain a more efficient reduction process by making one of the components of the reduced Lagrangian trivial $\overline\mu:=\overline\Ga^\theta$. Indeed, when the derivation is performed with the variable $p$, then the $\th$-component of the Lagrangian is given by a whole series, which is the magnetic moment. Changing the variable $p$ in such a way that the new variable absorbs this series is a way to have the reduced $\th$-component trivial, just given by a coordinate. This also simplifies the reduction algorithm, by providing a simpler expression for $\overline\om$, which will play a key role in the derivation. \\

 Second, the resulting variables $\overline\th$ and $\overline\mu$ are conjugated, which implies that the magnetic moment is a constant of motion besides the norm of the momentum $p$. And this conserved quantity is preferable to the variable $p$ because, unlike $p$, it remains an adiabatic invariant in the presence of a wide class of electric fields. 
 
 The variable $\overline\mu$ is a whole series in the Larmor radius, and its lowest-order term is the well-known adiabatic invariant $\mu$ often confounded with $\overline\mu$
\bq
	\overline\mu\approx\mu:=\tfrac{(p\sin\varphi)^2}{2mB}
	\,.
	\label{MagMomOrdMin}
\eq

 So, an interesting additional requirement is to include the magnetic moment in the reduced coordinates. For the equation to be solved, this requirement is expressed by 
\bq
 	\overline\Ga^\theta:=\overline \mu
 	\,.
\label{RequirPrior2}
\eq
Indeed, then the reduced Poisson bracket verifies 
\bq
	\overline\bbj^{\mu\theta }= 1, 
	\text{ and } 
	\overline\bbj^{\mu i }= 0
	\,,
\label{QcanCondition}
\eq
for $i\neq\theta$, which means that the dynamics of $\overline\mu$ is zero
$$
	\dot{\overline\mu}
	=
	\{\overline H,\overline\mu\}
	= -\p_\th \overline H =0
	\,,
$$
since the Hamiltonian $\overline H$ does not depend on $\overline\theta$, as a consequence of (\ref{RequirPrior1}).  \\

 There is a third reason for including the magnetic moment among the reduced coordinates and requiring $\overline\Ga^\th=\overline\mu$. It is concerned with the Hamiltonian structure of guiding-center dynamics. The $6$-dimensional reduced motion $\dot{\overline \bfz}$ is Hamiltonian, since it is just the transform of the Hamiltonian motion $\dot\bfz$. But the true reduced guiding-center motion is the $4$-dimensional slow motion $(\dot{\overline\bfq},\dot{\overline\varphi})$. It is the truncation of the full dynamics $\dot{\overline\bfz}$, but truncations of a Hamiltonian dynamics are in general not Hamiltonian. However, in some cases, truncations are automatically Hamiltonian, and a special case is the quarter canonical structure of the Poisson bracket, defined by conditions (\ref{QcanCondition}). This can be seen by imposing Dirac's constraints $(\overline\mu,\overline\th)$ to the reduced dynamics, or by verifying that the truncated bracket is actually just given by starting from the initial Lie algebra of all functions of the phase-space $f(\overline\bfz)$, and taking the sub-algebra of functions that do not depend on the gyro-angle $f(\overline\bfq,\overline\varphi,\overline\mu)$. Thus, including $\overline\mu$ in the reduced coordinates is a way to guarantee the reduced slow motion to be Hamiltonian.\\

 The requirement on the magnetic moment fixes one of the freedoms involved in the average components of the coordinate change. The other freedoms are used to make the reduced dynamics as simple as possible, by putting to zero as many average components of $\overline\Ga_n$ as possible for higher $n$, as indicated by formula (\ref{RequirPrior3}). \\

 D) 
 To sum it up, the guiding-center reduction involves a hierarchy of requirements: The primary requirement (minimal reduction, with an averaged reduced dynamics) is to remove the fast time-scale by averaging the Lagrangian over the gyro-angle; the corresponding equation is (\ref{RequirPrior1}). The secondary requirement (intermediate reduction, with a constant of motion and a Hamiltonian slow reduced motion) is to include the magnetic moment among the reduced coordinates by the quarter-canonical structure; the corresponding equation is (\ref{RequirPrior2}). The third optional requirement (maximal reduction, with a simplified reduced dynamics) is to use the remaining freedoms to make the reduced Lagrangian as simplified as possible; the corresponding equation is (\ref{RequirPrior3}). 
 
 This makes seven requirements at each order in $\ep$, one for each component of the reduced Lagrangian $\overline\Ga_n$ in formulae (\ref{ReducLagr}), and seven freedoms are needed. For a time-independent transformation, those are $\bfmg^\bfz_n$ and $S_n$, as announced in the previous subsection.

	\section{Derivation of the reduction}
 	\label{p2}

 Let us now turn to the guiding-center reduction. The details of the procedure as well as the practical computations may seem intricate and they hide somehow that the basic ideas of the reduction are very elementary. It is why the principles and general lines of the procedure are presented in the appendix, to give a clear view of the reduction process. 

 In this section, the three stages of the method presented in the appendix are shown to work with the coordinate $\sfc$ in a similar way as with the standard approach relying on a gyro-gauge. The transformation at lowest orders is computed for comparison with previous works, and it is shown how the reduction can be performed to arbitrary order in the Larmor radius by obtaining explicit induction relations. 

 Each order of the derivation can be given several numbers. For instance, what is usually called the first order is the order just after the lowest order. For the Lagrangian, it corresponds to $\Ga_0$ (since the lowest order corresponds to $\Ga_{-1}$), which is rather considered here as the order $0$. In addition, the order in the various quantities will be mixed up: for instance, in the derivation, the order involving $\Ga_2$ will be the equation for $\bfmg_3^{\sfa,\sfc}$, as well as for $\bfmg_2^{\sfb,\phi}$ and also for $\bfmg_1^{\th}$. For the sake of clarity, we will always consider the order $n$ as the one corresponding to $\Ga_n$ (or rather $\overline\Ga_n$), and we will often use the expression "at order $\Ga_n$", instead of "at the order corresponding to $\Ga_n$".

 \subsection{Preliminary transformation and initial setting}

 The goal is to solve equation (\ref{ReducLagrAlg}) for the guiding-center reduction, with the Lagrangian (\ref{LagrOrdering}), and with the requirements (\ref{RequirPrior1})-(\ref{RequirPrior2}) for the averaging reduction, for the magnetic moment reduction, and for the maximal reduction. \\

 First of all, the change of coordinates from the norm of the momentum $p$ to the magnetic moment $\overline\mu$ is not near identity, as is clear in equation (\ref{MagMomOrdMin}). Before beginning the reduction, a preliminary change of coordinates must be done, so that all the remaining transformation will be near identity. A suitable preliminary change of coordinates is
$$
	(\bfq,\varphi,p,\sfc)
	\longrightarrow
	(\bfq,\varphi,\mu,\sfc)
	\,,
$$
where $\mu$ is the zeroth-order magnetic moment 
$$
	\mu=\tfrac{(p\sin\varphi)^2}{2mB}
	\,,
$$
which is a well-known adiabatic invariant, and often confounded with $\overline\mu$. In the new variables, the Lagrangian becomes 
$$
	\Ga:=
	\big[
	e\bfma+\sqrt{2\mu mB}(\sfb\cot\varphi+\sfc)
	\big]
	\cdott d\bfq 
	- 
	\mu B (1+\cot^2\varphi) dt
	\,.
$$

 Interestingly, the pitch-angle $\varphi$ intervenes only through its cotangent, which was mentioned in \cite{GuilGCmin} as making all quantities polynomials. Here, this feature is obvious in the Lagrangian, and it is preserved by derivatives, so that it will be preserved throughout all of the derivation. Actually, the magnetic moment makes the polynomiality still more accurate than in \cite{GuilGCmin}, where the variables $\varphi$ and $p$ were used, and the Larmor-radius prefactor $r_L=p \sin\varphi$ was not polynomial in $\cot\varphi$. Here, the magnetic moment $\mu$ absorbs the $p \sin \varphi$ and all formulae will be purely polynomials in $\cot\varphi$ and monomials in $\sqrt\mu$, which is useful to simplify computations. 
 
So, we actually choose to change coordinates according to 
$$
	(\bfq,\varphi,p,\sfc)
	\longrightarrow
	(\bfq,\phi,\mu,\sfc)
	\,,
$$
with 
$$
	\phi:=\cot\varphi
$$
the variable that makes all formulae polynomials.

 Also, the structure of the Lagrangian shows that one can make the coefficients $e$ and $m$ disappear by noticing that the magnetic field $B$ appears only through $eB$, provided $\mu$ is considered as appearing only through $\mu m/e$, and $dt$ appears only through $dt/m$. 
 
 The particle charge $e$ is usually kept in guiding-center works because the order in $e^{-1}$ indicates the order in $\ep$ \cite{NortAdia}. Here, it is useless since the order in $\ep$ is already indicated by the order in other quantities: $\Ga_n$ and $\overline\Ga_n$ are of order $\tfrac{m\mu}{e}\left(\sqrt{\tfrac{m\mu}{e^2B}}\na\right)^{n}$. The reason is that all quantities will be series in $r_L\na=\sqrt{\tfrac{2m\mu}{e^2B}}\na$, as a result of the structure of the Lagrangian, and as will be confirmed by the derivation, e.g. formulae (\ref{LarmorRadiusOsc}),  (\ref{ResultsOrderGa11})-(\ref{ResultsOrderGa12}), (\ref{AvgG1mu})-(\ref{HamiltRedOrd2}), (\ref{ResultsDeparametrized1})-(\ref{ResultsDeparametrized2}), etc. Hence the order can be readily obtained by the overall order in $\sqrt B$ or $\na$. These last two quantities have the drawback of being an operator, or a space-dependent function; in addition, their order is only dimensional, e.g. $\tfrac{\na B\na B}{\sqrt B}$ is of order $\na^2B^{3/2}$. For a readily control of the order, it is useful to have a scalar parameter, which was considered as $e^{-1}$ in previous works. Here, this role can be played by $\sqrt\mu$, since all quantities will be monomial in it. Thus, keeping $e$ to indicate the order of expansion is indeed not necessary.
 
 Thus, we make the scaling 
\begin{align}
	A & \longrightarrow \underline A:=eA \ncr
	B & \longrightarrow \underline B:=eB 
\label{Scaling}
\\
	\mu & \longrightarrow \underline \mu 
	:= \tfrac{\mu m}{e} 
	= \tfrac{(p \sin\varphi)^2}{2eB} \ncr
	t & \longrightarrow \underline t:=\tfrac t m 
	\notag
	\,,
\end{align}
which avoids unnecessary coefficients in the derivation. It agrees with the physics, where the effect of the magnetic field on particle dynamics always includes the coupling constant $e$. For simplicity, we will drop the underline, e.g. we will write $B$ for $\underline B$. The Lagrangian becomes 
$$
	\Ga:=\big[\bfma+\sqrt{2\mu B}(\sfb\phi+\sfc)\big]\cdott d\bfq - \mu B (1+\phi^2) dt
	\,.
$$

\medskip
 The derivation starts from the Lagrangian with the expansion (\ref{LagrOrdering}) 
$$
	\Ga:=\Ga_{-1}+\Ga_0
	\,,
$$
with
\begin{align}
	\Ga_{-1}
	&=
	\bfma\cdott d\bfq
	\,,
\ncr
	\Ga_0
	&=
	\sqrt{2\mu B}~(\sfb\phi+\sfc)\cdott d\bfq 
	- \mu B(1+\phi^2) dt
	\,.
\label{LagrVarMu}
\end{align}
It can be divided into its average and fluctuating part 
\begin{align}
	\avg(\Ga_{-1})
	&=
	\Ga_{-1} = \bfma\cdott d\bfq
	\ncr
	\avg(\Ga_0)
	&=
	\sqrt{2\mu B}~\sfb\phi\cdott d\bfq 
	- \mu B (1+\phi^2) dt
	\ncr
	\osc(\Ga_0)
	&=
	\sqrt{2\mu B}~\sfc\cdott d\bfq 
	\,.
\label{LagrOrdVarMu}
\end{align}

\medskip
 The previous section showed that the process involves the Lagrange $2$-form $\om_n$. It can be split in three basic terms 
\begin{align}
	\om
	=d\Ga
	= \om_{-1} + \widetilde\om_0 + \overline\om_0
	\,,
\label{Om3terms}
\end{align}
with
\begin{align}
	\om_{-1}
	&
	:=d\Ga_{-1} 
	= d(\bfma)\cdott \wedge d\bfq
	\notag
	\\
	\widetilde\om_0
	&
	:= d \osc(\Ga_0)
	=d (\sqrt{2\mu B}\sfc) \cdott\wedge d\bfq 
	\ncr
	\overline\om_0
	&
	:= d\avg(\Ga_0)
	= 
	d(\sqrt{2\mu B}\sfb\phi)\cdott\wedge d\bfq 
	- d\big(\mu B (1+\phi^2)\big) \wedge dt
	\,,
	\notag
\end{align}
where the symbol $\wedge$ denotes the antisymmetry operator 
$$
	da .b. \wedge dc = da. b. dc- dc. b. da
	\,,
$$
for any matrix $b$ and any vectors $a$ and $b$.
 
The first term in (\ref{Om3terms}) is the lowest-order Lagrange $2$-form, related to Larmor gyrations.  The second (resp. third) term in (\ref{Om3terms}) is the exterior derivative of the oscillating (resp. average) zeroth-order Lagrangian. Be careful, this is not the oscillating (resp. average) zeroth-order Lagrange $2$-form; for instance
$$
	\widetilde\om_0
	:= d \big(\osc(\Ga_0)\big)
	\neq \osc(\om_0)=\osc(d\Ga_0)
	\,,
$$
because the exterior derivative does not preserve gyro-fluctuations.\\

The contributions (\ref{Om3terms}) to the Lagrange $2$-form are explicitly given by 
\begin{align}
	\om_{-1}
	:
	&
	=d\Ga_{-1} 
	= d\bfq \cdott \na (\bfma)\cdott \wedge d\bfq
	= d\bfq\cdott(-\bfmb)\x d\bfq
\label{FormOmegaI}
\\
	\overline\om_0
	:
	&
	= d\avg(\Ga_0)
	= 
	d(\sqrt{2\mu B}\sfb\phi)\cdott\wedge d\bfq 
	- d(\mu B (1+\phi^2)) \wedge dt
\ncr
	& = 
	\sqrt{2\mu B}
	\left\lbrace
	d\phi \sfb
	+d\bfq\cdott\na\sfb+
	\left[
	\tfrac{d\bfq\cdott\na B}{2B}
	+
	\tfrac{d\mu}{2\mu}
	\right]
	\phi\sfb
	\right\rbrace
	\cdott \wedge d\bfq
\ncr
	& \hspace{10.5ex}
	- \Big\lbrace
	(1+\phi^2)(B d\mu + \mu dB) + 2\mu B  d\phi 
	\Big\rbrace
	\wedge \tfrac{dt}{m}
\ncr
	\widetilde\om_0
	:
	&
	= d \osc(\Ga_0)
	=d (\sqrt{2\mu B}\sfc) \cdott\wedge d\bfq 
\ncr
	& = 
	\sqrt{2\mu B}
	\left\lbrace
	d\sfc
	+
	\left[
	\tfrac{d\bfq\cdott\na B}{2B}
	+
	\tfrac{d\mu}{2\mu}
	\right]
	\sfc
	\right\rbrace
	\cdott \wedge d\bfq
	\,.
	\notag
\end{align}

Now, $d\sfc$ involves two contributions: one corresponding purely to the gyro-angle, in which the variable $\sfc$ is changed at constant $\bfq$, and a second one coming from a change in the coordinate $\bfq$, which comes because the gyro-angle is a constrained coordinate, as mentioned about formulae (\ref{ConnectNatur}) and (\ref{ConnectBoth}). The second part is given by
$$
	d\bfq\cdott\na\sfc
	=
	- d\bfq\cdott\na\sfb\cdott\sfc\sfb  
	+ d\bfq\cdott\bfmr_g\sfa
	\,.
$$
The first one is then $(d\sfc-d\bfq\cdott\na\sfc)$, but it is written more precisely as $(d\sfc-d\bfq\cdott\na\sfc)\cdott\sfa\sfa$ since when $\bfq$ is constant, the variation of $\sfc$ can only be in the direction of $\sfa$, i.e. along the circle $\mathbb S^1(\bfq)$. On the whole 
$$
	d\sfc
	=
	- d\bfq\cdott\na\sfb\cdott\sfc\sfb  
	+ d\bfq\cdott\bfmr_g\sfa
	+
	(d\sfc-d\bfq\cdott\na\sfc)\cdott\sfa\sfa
	\,.
$$
This shows that, in the gauge-independent approach, the natural form corresponding to $d\th$ is 	
$$
	\de\th:=-(d\sfc-d\bfq\cdott\na\sfc)\cdott\sfa
	\,,
$$
where the minus sign comes to agree with the usual convention (\ref{DefThetaPerp}) on the orientation of the gyro-angle. Unlike $d\th$, this form is not closed in general; it is why it is written $\de\th$. In the gauge-dependent case, it is closed and indeed corresponds to $d\th$, as a consequence of (\ref{ConnectGauge}).\\

 As the reduction process mainly relies on inversions of $\om=d \Ga $ (or rather $\overline \om$), which can be viewed as a matrix inversion, a matrix notation is well suited and makes the discussion clearer. This implies to choose a basis of $1$-forms; the derivation of $\om$ above shows that a natural basis is
\bq
	\Big(
	\sfc\cdott d \bfq, 
	\sfa\cdott d \bfq, 
	\sfb\cdott d \bfq 
	~|~ 
	d\phi, 
	d\mu, 
	\de\th
	~|~
	-dt
	\Big)
	\,,
\label{Basis1Forms}
\eq
where a vertical dash $|$ is put to separate the space-, the momentum- and the time-components. The choice of $-dt$ makes the corresponding coordinate $H$ instead of $-H$. 

Those $1$-forms are not closed, unlike the standard $dz^i$, but it is not needed, provided one is careful of using intrinsic definitions for the operations involved in the procedure (see formulae (\ref{BasisVectField}) and (\ref{ExtDerivGene}) for instance). 

 In this basis, formulae (\ref{FormOmegaI}) become
\begin{align}
	\om_{-1}
	& =
	B
	\left(
	\begin{array}{ c  @{~}|@{~~} c  @{~~}|@{~} c }
     		\begin{smallmatrix}
			~~~0~~&~~~~~1~~~~~&~~0~~\\ -1&0&0\\0&0&0
			\end{smallmatrix}
     & ~~~~~~~~0~~~~~~~~ &~~~0~~~ \\ \hline
     0 & 0 &0\\ \hline
     0 & 0 &0
   \end{array}
   \right)
\label{MatrOm-1}
\\
	\overline\om_0
	& =
~~{\scriptsize
	\left(
	\begin{array}
				{ c@{~}c@{~}c@{\hspace{2ex}}  
				| @{~}c@{~}c@{~}c@{~~}  
				| @{~}c@{~} }
			0&- J& I
			&0&0&0
			& \sfc\cdott\na H_0 
		\\ 
			J&0&- K
			&0&0&0
			&\sfa\cdott\na H_0 
		\\
			- I& K&0
			&-\sqrt{2\mu B}&-\tfrac{\phi\sqrt{2\mu B}}{2\mu}&0
			&\sfb\cdott\na H_0 
	\\ \hline
			0&0&\sqrt{2\mu B}
			&&&
			&\p_\phi H_0
		\\ 
			0&0&\tfrac{\phi\sqrt{2\mu B}}{2\mu}
			&&\raisebox{-1ex}[0ex][0ex]{\large 0}&
			&\p_\mu H_0
		\\
			0&0 &0
			&&&
			&0
	\\ \hline
			-\sfc\cdott\na H_0 
			& -\sfa\cdott\na H_0  
			& -\sfb\cdott\na H_0 
		&
			-\p_\phi H_0
			& -\p_\mu H_0&0
			&0 
   \end{array}
   \right)
}
\ncr
	\widetilde\om_0
	& =
~~{\scriptsize
	\left(
	\begin{array}
				{ @{\hspace{1ex}}
				c@{\hspace{3ex}}
				c@{\hspace{4ex}}
				c@{\hspace{3ex}}  
				| @{\hspace{5ex}}
				c@{\hspace{2ex}}
				cc@{\hspace{1ex}}  
				| @{\hspace{3ex}}c@{\hspace{2ex}} }
			0&-\tilde J&\tilde I
			&0&-\tfrac{\sqrt{2\mu B}}{2\mu}&0
			&
		\\ 
			\tilde J&0&-\tilde K
			&0&0&\sqrt{2\mu B}
			&\raisebox{-1ex}[0ex][0ex]{\large 0}
		\\
			-\tilde I&\tilde K&0
			&0&0&0
			&
	\\ \hline
			0&0&0
			&&&
			&
		\\ 
			\tfrac{\sqrt{2\mu B}}{2\mu}&0&0
			&&\raisebox{-1ex}[0ex][0ex]{\large 0}&
			&\raisebox{-1ex}[0ex][0ex]{\large 0}
		\\
			0&-\sqrt{2\mu B} &0
			&&&
			&
	\\ \hline
			&\raisebox{-1.2ex}[0ex][0ex]{\large 0}&
			&&\raisebox{-1.2ex}[0ex][0ex]{\large 0}&
			&\raisebox{-1.2ex}{\large 0}
   \end{array}
   \right)
}
\,,
\notag
\end{align}
where for visual purpose, vertical and horizontal lines are used to separate the position-, momentum- and time-components. The spatial part of the matrix $\overline\om_0$ (resp. $\tilde\om_0$) is just a vector product $\mathbf v \times$, with the vector $\mathbf v:=I\sfa+J\sfb+K\sfc$ (resp. $\mathbf v:=\tilde I\sfa+\tilde J\sfb+\tilde K\sfc$), where the coefficients are 
\begin{align}
	I  
	&
	:= 
	-  \sfa\cdott \na\x(\sqrt{2\mu B}\phi\sfb)
	=
	\sqrt{2\mu B}~
	\phi
	\Big(
	\tfrac{\sfc\cdott\na B}{2B}
	-
	\sfb\cdott\na\sfb\cdott\sfc
	\Big)
\ncr
	J 
	&
	:= 
	-  \sfb\cdott \na\x(\sqrt{2\mu B}\phi\sfb)
	=
	\sqrt{2\mu B}~
	\phi
	\Big(
	\sfa\cdott\na\sfb\cdott\sfc
	-
	\sfc\cdott\na\sfb\cdott\sfa
	\Big)
\ncr
	K 
	&
	:= 
	- \sfc\cdott \na\x(\sqrt{2\mu B}\phi\sfb)
	=
	\sqrt{2\mu B}~
	\phi
	\Big(
	\sfb\cdott\na\sfb\cdott\sfa
	-
	\tfrac{\sfa\cdott\na B}{2B}
	\Big)
\ncr
	\tilde I 
	&
	:= 
	- \sfa\cdott \na\x(\sqrt{2\mu B}\sfc)
	=
	\sqrt{2\mu B}~
	\tfrac{
	\sfa\cdott\na\sfb\cdott\sfa
	-
	\sfc\cdott\na\sfb\cdott\sfc
	}
	{2}
\ncr
	\tilde J 
	&
	:= 
	- \sfb\cdott \na\x(\sqrt{2\mu B}\sfc)
	=
	\sqrt{2\mu B}~
	\left(
	\tfrac{\sfa\cdott\na B}{2B}
	-
	\sfc\cdott\bfmr_g
	\right)
\ncr
	\tilde K 
	&
	:= 
	- \sfc\cdott \na\x(\sqrt{2\mu B}\sfc)
	=
	\sqrt{2\mu B}~
	\left(
	\sfa\cdott\na\sfb\cdott\sfc
	+
	\sfb\cdott\bfmr_g
	\right)
	\,.
	\notag
\end{align}

\subsection{Immediate orders $-1$ and $0$}

 At lowest order $n=-1$, the reduced Lagrangian (\ref{ReducLagrAlg}) writes 
$$
	\overline \Ga_{-1}
	=
	\Ga_{-1} + dS_{-1}
	\,.
$$

 The change of variable has no effect at this order, since it is near-identity. But the averaging condition is trivially verified; it is actually a condition for the near-identity Lie transform to remove the fast time-scale, which is possible only when at lowest order, the Lagrangian is already gyro-averaged. Now, the only freedom involved $S_{-1}$ cannot be useful and is set to zero
$$
	S_{-1}=0
	\,.
$$

\medskip
 At the following order $n=0$, the reduced Lagrangian is 
\begin{align}
	& \overline \Ga_{0}
	=
	\bfmg_1\cdott\om_{-1}+\Ga_0 + dS_0
	\,.
\label{LagrRedOrdZeroEqZ}
\end{align}
To write vectors in matrix form, a basis has to be chosen for vector fields also. It is most convenient to choose the dual of the basis (\ref{Basis1Forms}) for $1$-forms, so that the coupling $\bfmg_1\cdott\om_{-1}$ is computed as a standard matrix product. The desired basis is easily identified as (using the natural isomorphism between vector fields and differential operators) 
\bq
	\big(
		\sfc\cdott\na ,
		\sfa\cdott\na ,
		\sfb\cdott\na 
		~|~
		\p_\phi ,
		\p_\mu ,
		-\sfa\cdott\p_\sfc 
		~|~
		-\p_t
	\big)
	\,,
\label{BasisVectField}
\eq
because $\na$ can be written $\p_{\bfq|\sfc}+\p_\bfq\sfc\cdott\p_{\sfc|\bfq}$. The operator $-\sfa\cdott\p_\sfc$ is the generator of Larmor gyrations, as shown in \cite{GuilMagMom}. It just an intrinsic definition of the usual $\p_\th$. The set (\ref{BasisVectField}) is actually the natural basis for vector fields, which confirms the relevance of the basis (\ref{Basis1Forms}) for $1$-forms. 

 Be careful that the chosen bases are not averages and they must be taken into account when computing averages or fluctuations of a quantity. For instance, it could seem that a vector (resp. a $1$-form) with components $(1,0,0~|~,0,0,0~|~0)$ is averaged, whereas it is not, since it is equal to $\sfc\cdott\na$ (resp. $\sfc\cdott d\bfq$).\\

 Then, equation (\ref{LagrRedOrdZeroEqZ}) is easily computed in matrix form
\begin{align}
	\overline \Ga_{0}
	& =
	\bfmg_1\cdott\om_{-1}+\Ga_0 + dS_0
\label{LagrRedOrdZeroEq}
	\\
	&~
{\footnotesize	
\raisebox{4ex}{=~}
\begin{array}{			r@{~~}r@{~~}c@{~~~}c@{~~~}c@{~~}
						c@{~~}c@{~~}c@{~~}c@{~~}
						c@{~~}c@{~~}l}
			&
			( &
			- B \bfmg_1^\sfa &
			B \bfmg_1^\sfc &
			0 &
			| &
			0 &
			0 &
			0 &
			| &
			0 &
			)
			\\
	+\rule{0ex}{3ex} &
	( & 	
			\sqrt{2\mu B} &
			0 &
			\sqrt{2\mu B} \phi &
			| &
			0 &
			0 &
			0 &
			| &
			\mu B (1+\phi^2) &
	)
	\\
	+\rule{0ex}{3ex} &
	( & 	
			\sfc\cdott\na S_0 &
			\sfa\cdott\na S_0 &
			\sfb\cdott\na S_0 &
			| &
			\p_\phi S_0 &
			\p_\mu S_0 &
			-\sfa\cdott\p_\sfc S_0 &
			| &
			0 &
	)
	\end{array}
}	
\,,
	\notag
\end{align}
where, as usual, $1$-forms are written as $1*7$ matrices: each column is an equation to be solved for the freedoms $\bfmg_1$ and $S_0$ in such a way that $\overline\Ga_0$ satisfies the desired requirements.

 The lowest-order Lagrange $2$-form $\om_{-1}$ is linked to Larmor gyration, as appears in (\ref{FormOmegaI}). It is not invertible. Only two components of $\bfmg_1$ are involved in the equation: $\bfmg_1^\sfc$ and $\bfmg_1^\sfa$; and only two components of $\overline\Ga$ can be controlled by these freedoms: $\overline\Ga_1^\sfc$ and $\overline\Ga_1^\sfa$. As announced in the appendix, the inversion is possible only under some conditions on the term $\sfmr_0=	- \overline \Ga_{0} + \Ga_0 + dS_0$, and the solution, if it exists, is not unique. \\

 More precisely, for the averaging requirement (\ref{RequirPrior1}), the condition of the right-hand side of equation (\ref{EquaG1Matrix}) being in the range of $\om_{-1}$ is satisfied, since the only fluctuating terms are in $\overline\Ga_1^{\sfc,\sfa}$, which can precisely be controlled by the freedoms $\bfmg_1^{\sfc,\sfa}$. The solution imposes the fluctuating part of $\bfmg_1^{\sfc,\sfa}$, which is given by 
\begin{align}
	\osc(\bfmg_1^\sfc\sfc+\bfmg_1^\sfa\sfa)
	&
	:=
	\tfrac{\sqrt{2\mu B}}{B}\sfa
	\,.
	\label{LarmorRadiusOsc}
\end{align}
Notice that the oscillating and averaged parts of the components $\bfmg_1^\sfc$ and $\bfmg_1^\sfa$ must be dealt with together through the combination $\bfmg_1^\sfc\sfc+\bfmg_1^\sfa\sfa$. This is easily illustrated with an example vector $\bfmx:=\sfb\cdott\na\sfb\cdott\sfc\sfc\cdott \nabla$: it is not a pure fluctuation, whereas its $\sfc$-component $\bfmx^\sfc:=\sfb\cdott\na\sfb\cdott\sfc$ is a pure fluctuation, and its fluctuating part is $\tfrac 12\sfb\cdott\na\sfb\cdott(\sfc\sfc-\sfa\sfa)\cdott \nabla$, which mixes up the components $\bfmx^\sfc$ and $\bfmx^\sfa$. 

 At this point, the average part of $\overline\Ga_1^{\sfc,\sfa}$ remains free and must be identified by the secondary and tertiary requirements (\ref{RequirPrior3}) and (\ref{RequirPrior2}). The magnetic moment requirement (\ref{RequirPrior2}) is not concerned here, since $\overline\Ga^\th$ is automatically zero. Last, for the requirement for the maximal reduction (\ref{RequirPrior3}), only the components $\overline\Ga_1^{\sfc,\sfa}$ can be controlled, and setting them to zero imposes the average part of $\bfmg_1^{\sfc,\sfa}$
\begin{align}
	\avg(\bfmg_1^\sfc\sfc+\bfmg_1^\sfa\sfa)
	&
	:=
	0
	\,.
	\label{LarmorRadiusAvg}
\end{align}

 Formulae (\ref{LarmorRadiusOsc})-(\ref{LarmorRadiusAvg}) give the traditional (lowest-order) Larmor radius, which is usually confused with the exact Larmor radius $r_L:=\bfq-\bar\bfq$:
$$
(r_L)_1:=(\bfq-\bar\bfq)_1
=\bfmg_1^\sfc\sfc+\bfmg_1^\sfa\sfa+\bfmg_1^\sfb\sfb
= \tfrac{\sqrt{2\mu B}}{B}\sfa
=\tfrac{p\sin\varphi}{B}\sfb\x\sfc
=\tfrac{\bfmb\x\bfp}{B^2}
\,,
$$
where the index denotes the order, and formulae (\ref{TransfCoord}) and (\ref{ResultsOrderGa11}) were implicitly used. 

 In a similar way, what is usually called guiding center in the literature actually corresponds to the first-order guiding center:
$$
\bar q := q - \tfrac{\bfmb\x\bfp}{B^2} + O(\epsilon^2)
$$
where we recover the traditional formula for the (first-order) guiding center, since we remind that here $\bfmb$ stands for $\underline\bfmb=e\bfmb$ (see formula (\ref{Scaling})).\\

 All other components of the first-order transformation generator $\bfmg_1^{\sfb,\phi,\mu,\th}$ still remain undetermined. They embody the non-uniqueness of the matrix inverse $(\om_{-1})\mun$ and the corresponding freedom will be useful for the solvability conditions at the next order. As a consequence, there will be some order mixing: the components $\sfb,\phi,\mu,\th$ of $\bfmg_1$ will be determined at higher order, at the same time as the components $\sfc,\sfa$ of $\bfmg_2$.

 Last, the gauge function $S_0$ cannot be useful and is set to zero  
$$
	S_{0}=0
	\,.
$$

\medskip
The final zeroth-order reduced Lagrangian (\ref{LagrRedOrdZeroEq}) writes 
\begin{align}
	\overline\Ga_0
	& =
	\sqrt{2\mu B}\sfb\phi\cdott d\bfq 
	- \mu B (1+\phi^2) dt
	\,.
\label{LagrRedOrdZeroSol}	
\end{align}
It is just the average of the zeroth-order initial Lagrangian $\Ga_0$: $\overline\Ga_0=\avg(\Ga_0)$. The zeroth-order reduced $2$-form $\overline\om_0$ is then
\bq
	d\overline\Ga_0=\overline\om_0
	\notag
	\,,
\eq
which justifies the notation $\overline\om_0$ introduced in (\ref{Om3terms}). 

 It is easy to see that the matrix $\overline\om_{-1}+\overline\om_0$ is not invertible, for instance all its gyro-angle components are zero. So, the pivotal matrix for high orders $\sfmm_\infty$ mentioned in the appendix (see page \pageref{DefinitMInfty}) is not identified yet.

\subsection{Turning point: Order $1$}

 At the following order $n=1$, the reduced Lagrangian is given by (\ref{ReducLagrAlg}), which is rather written with the unknown $\bfmg_{1,2}$ on the left-hand side
$$
	\bfmg_2\cdott\om_{-1}
	+ \tfrac{\bfmg_1}{2}\cdott(\widetilde\om_0+2\overline\om_0)
	=
	\overline \Ga_{1}
	- dS_1
	\,.
$$

 As announced in the appendix, the pivotal matrix is the set of $\om_{-1}$ and $(\widetilde\om_0+2\overline\om_0)/2$, acting on the set of unknown components of $(\bfmg_2,\bfmg_1)$. Again, it is not invertible; this corresponds to the case where the inversion is possible only under some integrability conditions on the right-hand side $\overline \Ga_{1}- dS_1$, to which one must add the set of $-\om_{-1}$ and $-(\om_0+\overline\om_0)/2$ acting on the set of components of $(\bfmg_2,\bfmg_1)$ that are already known, i.e. on $\bfmg_1^{\sfc,\sfa}$. And the solution, if it exists, is not unique. 
 
 More precisely, using a matrix notation and grouping in the left-hand side only the terms with unknown components of $(\bfmg_2,\bfmg_1)$ gives
\begin{align}
	& 
	\hspace{-3ex}
	~~~
	\left(
	     	\begin{smallmatrix}
			\bfmg_2^\sfc \\
			\bfmg_2^\sfa \\
			\bfmg_2^\sfb \\ \hline
			\bfmg_2^\phi \\
			\bfmg_2^\mu \\
			\bfmg_2^\th \\ \hline
			0\raisebox{-0.5ex}{\rule{0ex}{2ex}}
			\end{smallmatrix}
	\right)^T
	\cdott
	\left(
	\begin{array}
				{ c  | c  | c }
     		\begin{smallmatrix}
			~~\,0\,~~&~~~~B~~~~&~~~0~~~\\ -B&0&0\\0&0&0
			\end{smallmatrix}
     & ~~~~~~~~~~~~~0
     \raisebox{-2ex}{\rule{0ex}{6ex}}~~~~~~~~~~~~~ 
     &~~~0~~~ 
  \\ \hline
     0\raisebox{-2ex}{\rule{0ex}{6ex}} & 0 &0
  \\ \hline
     0\raisebox{-1ex}{\rule{0ex}{3ex}} & 0 &0
   \end{array}
   \right)
\ncr
	& 
	\hspace{-3ex}
	+
	\left( 
	     	\begin{smallmatrix}
			0\raisebox{-0.5ex}{\rule{0ex}{2ex}} \\
			0\raisebox{-0.5ex}{\rule{0ex}{2ex}} \\
			\bfmg_1^\sfb \\ \hline
			\bfmg_1^\phi \\
			\bfmg_1^\mu \\
			\bfmg_1^\th \\ \hline
			0\raisebox{-0.5ex}{\rule{0ex}{2ex}}
			\end{smallmatrix}
	\right)^T
	\cdott
{\scriptsize
	\left(
	\begin{array}
				{ @{\,}c@{~~\,}c@{~~\,}c@{~\,}  
				| @{~}c@{~~\,}c@{~~}c@{~~}  
				| @{~}c@{~} }
			0&- J_{21}& I_{21}
			&0&-\tfrac{\sqrt{2\mu B}}{4\mu}&0
			& \p_c H_0 
		\\ 
			J_{21}&0&- K_{21}
			&0&0&\tfrac{\sqrt{2\mu B}}{2}
			&\p_a H_0 
		\\
			- I_{21}& K_{21}&0
			&-\sqrt{2\mu B}
			&-\tfrac{\phi\sqrt{2\mu B}}{2\mu}&0
			&\p_b H_0 
	\\ \hline
			0&0&\sqrt{2\mu B}
			&&&
			&\p_\phi H_0
		\\ 
			\tfrac{\sqrt{2\mu B}}{4\mu}
			&0&\tfrac{\phi\sqrt{2\mu B}}{2\mu}
			&&\raisebox{-1ex}[0ex][0ex]{\large 0}&
			&\p_\mu H_0
		\\
			0&-\tfrac{\sqrt{2\mu B}}{2} &0
			&&&
			&0
	\\ \hline
			-\p_c H_0 
			& -\p_a H_0  
			& -\p_b H_0 
		&
			-\p_\phi H_0
			& -\p_\mu H_0&0
			&0 
   \end{array}
   \right)
}
\ncr
	& 
	\hspace{-3ex}
	\makebox[0ex][r]{=}
	- 
	\left( 
	     	\begin{smallmatrix}
			\bfmg_1^\sfc \\
			\bfmg_1^\sfa \\
			0\raisebox{-0.5ex}{\rule{0ex}{2ex}} \\ \hline
			0\raisebox{-0.5ex}{\rule{0ex}{2ex}} \\
			0\raisebox{-0.5ex}{\rule{0ex}{2ex}} \\
			0\raisebox{-0.5ex}{\rule{0ex}{2ex}} \\ \hline
			0\raisebox{-0.5ex}{\rule{0ex}{2ex}}
			\end{smallmatrix}
	\right)^T
	\cdott
{\scriptsize
	\left(
	\begin{array}
				{ @{\,}c@{~~\,}c@{~~\,}c@{~\,}  
				| @{~}c@{~~\,}c@{~~}c@{~~}  
				| @{~}c@{~} }
			0&- J_{21}& I_{21}
			&0&-\tfrac{\sqrt{2\mu B}}{4\mu}&0
			& \p_c H_0 
		\\ 
			J_{21}&0&- K_{21}
			&0&0&\tfrac{\sqrt{2\mu B}}{2}
			&\p_a H_0 
		\\
			- I_{21}& K_{21}&0
			&-\sqrt{2\mu B}
			&-\tfrac{\phi\sqrt{2\mu B}}{2\mu}&0
			&\p_b H_0 
	\\ \hline
			0&0&\sqrt{2\mu B}
			&&&
			&\p_\phi H_0
		\\ 
			\tfrac{\sqrt{2\mu B}}{4\mu}
			&0&\tfrac{\phi\sqrt{2\mu B}}{2\mu}
			&&\raisebox{-1ex}[0ex][0ex]{\large 0}&
			&\p_\mu H_0
		\\
			0&-\tfrac{\sqrt{2\mu B}}{2} &0
			&&&
			&0
	\\ \hline
			-\p_c H_0 
			& -\p_a H_0  
			& -\p_b H_0 
		&
			-\p_\phi H_0
			& -\p_\mu H_0&0
			&0 
   \end{array}
   \right)
}
\ncr
	& 
	\hspace{-3ex}
	+
	\hspace{11ex}
	\left( 
	     	\begin{smallmatrix}
			~~\overline\Ga_1^\sfc~~~ &
			~~\overline\Ga_1^\sfa~~~~ &
			~~\,\overline\Ga_1^\sfb~~ &
			| &
			\,~~\overline\Ga_1^\phi~~~~ &
			~~~\overline\Ga_1^\mu~~ &
			~~\,\overline\Ga_1^\th~~ &
			| &
			\,~~\overline H_1~
			\end{smallmatrix}
	\right)
\ncr
	&
	\hspace{-3ex}
	-
	\hspace{11ex}
			\left(
	     	\begin{smallmatrix}
			~\p_c S_1~~ &  
			~\p_a S_1~~ &
			~~\p_b S_1~ &
			| &
			~~\p_\phi S_1~~ &
			~~\p_\mu S_1~ &
			~\p_\th S_1~ &
			| &
			~~0~~
			\end{smallmatrix}
	\right)
	\,,
\label{LagrRedOrd1Z}
\end{align}
where the exponent $^T$ indicates matrix transpose. The coefficient $I_{21}$ is defined by $I_{21}:= \tfrac{2I+1\tilde I}{2}$, and $J_{21}$ and $K_{21}$ are defined the same way. Last, for shortness, we used the short-hands
\begin{align}
	\p_c&:=\sfc\cdott\na\,,\ncr
	\p_a&:=\sfa\cdott\na\,,\ncr
	\p_b&:=\sfb\cdott\na\,,\ncr
	\p_\th&:=-\sfa\cdott\p_\sfc\,.\notag
\end{align}
Be careful, $\p_c$ is different from $\p_\sfc$.

 Let us have a word on the graphical presentation for these matrix products, because it may seem surprising at first glance and it will be frequently used. In principle, $1$-forms are row-matrices, vectors are column-matrices, so that the pairing between them is just given by the matrix product. Now, if $2$-forms are presented as a $2*2$ matrix, then the vector implied in the left-pairing must be written as a row-matrix, for the pairing to be just the usual matrix product. In this paper, in order to save room for the editor, we wrote this row-matrix as the transpose of a column-matrix. By the way, it makes formulae easier to read, because each component of the vector is facing precisely the row of the matrix which it multiplies: for instance in the second term of equation (\ref{LagrRedOrd1Z}), the third row $\bfmg_1^\sfb$ of the vector $\bfmg_1$ multiplies the third row 
 $(- I_{21} ,  K_{21} , 0 ~|~ 
 -\sqrt{2\mu B} , -\tfrac{\phi\sqrt{2\mu B}}{2\mu} , 0 ~|~ 
 \p_b H_0 )$ 
of the matrix $2\overline\om_0+\widetilde\om_0$ which is on its right. So, this graphical presentation seems to be well-suited.\\

 Equation (\ref{LagrRedOrd1Z}) can be simplified by removing the components that do not contribute
\begin{align}
	&~~~~
	\left( 
	     	\begin{smallmatrix}
			\bfmg_2^\sfc \\
			\bfmg_2^\sfa \\
			\end{smallmatrix}
	\right)^T
	\cdott
{\scriptsize
	\left(
	     	\begin{array}{
	     					@{~~}c@{~~~~~~}c
	     					@{~~~~~~~~~\,}c@{~~~~~} 
	     					|@{~~~~~} 
	     						c@{~~~~~~~~~~~~}
	     						c@{~~~~~~~~~~}
	     						c@{~~~~\,} 
	     					| @{~~~~~}c@{~~~~}
	     				}
				0 & B & 0 
			& 0 & 0 & 0 &
			0
		\\
				-B & 0 & 0 
			& 0 & 0 & 0 &
			0
			\end{array}
   \right)
}
	\ncr
	&
	+ 
	\left( 
	     	\begin{smallmatrix}
			\bfmg_1^\sfb \\
			\bfmg_1^\phi \\
			\bfmg_1^\mu \\
			\bfmg_1^\th
			\end{smallmatrix}
	\right)^T
	\hspace{-2ex}
	\cdott
{\scriptsize
	\left(
	\begin{array}{ 
					@{~}c@{~}c@{~~}c@{~ }  
					| @{~}c@{~}c@{~~~~~}c@{~~~~\,}  
					| @{~}c@{~} 
				}
				-I_{21} &K_{21} &0
     &
			-\sqrt{2\mu B} 
			& -\tfrac{\phi\sqrt{2\mu B}}{2\mu} &0
	&
			H_0\tfrac{\p_b B}{B}
	\\ \hline
			0&0&\sqrt{2\mu B}
			&&&
			&2\mu B\phi 
		\\
			\tfrac{\sqrt{2\mu B}}{4\mu}
			&0&\tfrac{\phi\sqrt{2\mu B}}{2\mu}
	&& \raisebox{-1ex}[0ex][0ex]{\large 0} &
	&
			\tfrac{H_0}{\mu} 
		\\
			0&-\tfrac{\sqrt{2\mu B}}{2}&0
			&&&
			&0
   \end{array}
   \right)	
}
\ncr
	=
	&
	- 
	\left( 
	     	\begin{smallmatrix}
			\bfmg_1^\sfc \\
			\bfmg_1^\sfa 
			\end{smallmatrix}
	\right)^T
	\cdott
{\scriptsize
	\left(
	\begin{array}{ 
					@{~~~}c@{~~~~}c@{~~~}c@{~~~}  
					| @{~~~~~}c@{~~~~~~}c@{~~~}c@{~}  
					| @{~~}c@{~} 
				}
			0&-J_{21}&~~I_{21}
			&0&-\tfrac{\sqrt{2\mu B}}{4\mu}&0
			&H_0\tfrac{\p_c B}{B}
		\\
			J_{21}&0&-K_{21}
			&0&0&\tfrac{\phi\sqrt{2\mu B}}{2}
			&H_0\tfrac{\p_a B}{B}
   \end{array}
   \right)	
}
\ncr
	& 
	+
	\hspace{9ex}
	\left( 
	     	\begin{smallmatrix}
			~~~\overline\Ga_1^\sfc~~~ &
			\overline\Ga_1^\sfa~~~~ &
			\overline\Ga_1^\sfb~~~ &
			|~~ &
			\overline\Ga_1^\phi~~~~~ &
			\overline\Ga_1^\mu~~~~~~ &
			\overline\Ga_1^\th~ &
			| ~~&
			\overline H_1~~
			\end{smallmatrix}
	\right)
\ncr
	&
	-
	\hspace{9ex}
			\left(
	     	\begin{smallmatrix}
			~~\p_c S_1~ &  
			\p_a S_1~~ &
			\p_b S_1~ &
			|~ &
			\, \p_\phi S_1~~~ &
			\p_\mu S_1~~~\, &
			\p_\th S_1\, &
			| ~~~&
			0~~~
			\end{smallmatrix}
	\right)
	\,.
\label{LagrRedOrd1}
\end{align}
This formula illustrates the typical form for $\overline\Ga_n$ announced in the appendix. In the left-hand side, a matrix product involves the unknown components of $(\bfmg_{n+1},\bfmg_n,...)$. In the right-hand side, there are two  kinds of terms: the terms with the known components of $(\bfmg_n,\bfmg_{n-1},...)$, and the terms involving the freedoms to be determined by the integrability conditions, i.e. $\overline\Ga_n$ and $S_n$. 

The requirements concern the seven components of the reduced Lagrangian $\overline\Ga_n^i$, which are ideally put to zero. The freedoms are embodied in six components of $(\bfmg_{n+1},\bfmg_n,...)$ and in the gauge function $S_n$. More precisely, the primary requirement (averaging reduction) means that the fluctuating part of the reduced Lagrangian must be zero: $\osc(\overline\Ga_n^i)=0$; actually, the chosen basis for $1$-forms is not averaged, so that each component cannot be averaged separately, e.g. neither $\sfb\cdott\na\sfb\cdott\sfa~\sfa\cdott d\bfq$ nor $\sfb\cdott\na\sfb\cdott\sfc~\sfc\cdott d\bfq$ are averages, but the sum of them $\sfb\cdott\na\sfb\cdott d\bfq$ is an average. The secondary requirement (magnetic moment reduction) means that the gyro-angle component of the Lagrangian must be the magnetic moment: $\overline\Ga_1^\th=\overline\mu$, and $\overline\Ga_n^\th=0$ for $n\neq1$. The tertiary requirement (maximal reduction) means that the average part of the reduced Lagrangian should be zero as well: $\avg(\overline\Ga_n^i)=0$ (except $\overline\Ga_1^\th$). When integrability conditions cannot be solved for so strong requirements, the tertiary requirement is released, but as little as possible, and one sets to zero as many components $\overline\Ga_n^i$ as possible. \\

 Let us solve equation (\ref{LagrRedOrd1}). The pivotal matrix is not invertible. It has six columns and seven rows; the columns $\phi$ and $\mu$ are not linearly independent; in addition, its column for $\th$ is zero, which means that $\overline\Ga_1^\th$ cannot be controlled by the freedoms $(\bfmg_2,\bfmg_1)$. So, the existence of solution is submitted to integrability conditions (by which the freedoms $S_1$ and $\overline\Ga_1^i$ can be constrained), and then the existing solutions are not unique. \\

 The integrability conditions are automatically satisfied. The overall equation for the column $\phi$ will be automatically zero, because the procedure will imply $\bfmg_1^\sfb=0$, and $S_1=0$. 
 
 As for the column $\th$, it writes:
$$
	0=
	-\tfrac {\sqrt{2\mu B}}{2} \bfmg_1^\sfa 
	+ \overline\Ga_1^\th 
	- \p_\th S_1
	\,.
$$
The averaging requirement means that $\p_\th S_1 = 0 $, which implies
$$
	\osc(S_2)=0
	\,.
$$
The average part of the equation then writes 
\bq
	\overline\Ga_1^\th =
	\tfrac {\sqrt{2\mu B}}{2} \bfmg_1^\sfa
	= \mu 
	\,,
	\label{MagMomSucceeded}
\eq
which makes the automatically verified for this component. In fact, if $\mu$ had not been chosen as a preliminary coordinate, it is here that it could be identified. 

 As a side comment, it might seem that the right-hand side of equation (\ref{MagMomSucceeded}) should be $\overline\mu$ instead of $\mu$, because of formula (\ref{RequirPrior2}). However, in the derivation, all expressions for $\overline\Ga$ are functions of $\mu$. In the reduced system, the corresponding expressions will be the same functions evaluated on the reduced coordinate $\overline\mu$, as is well emphasized in \cite{Little81}. So, formula (\ref{MagMomSucceeded}) means that the first-order reduced Lagrangian $\overline\Ga_1$ will actually have its $\overline\th$ component equal to $\overline\mu$. It is why this result verifies the requirement (\ref{RequirPrior2}) on the magnetic moment. We will not insist more on this point. 

 With the result (\ref{MagMomSucceeded}), the reduced Lagrangian will contain the term $\mu\de\th$, implying the presence of the vector $\bfmr_g$, which is not determined. This feature already appeared in the computation of $\om_0$, but the vector $\bfmr_g$ involved in $\de\th=-(d\sfc- d\bfq\cdott\na\sfc)\cdott\sfa$ came from a term $-d\sfc\cdott\sfa$, so that the overall contribution of the vector $\bfmr_g$ actually cancelled. Just the same way now, the term $\mu \de\th$ is required to come from a total contribution 
$$
	-\mu \sfa\cdott d\sfc
	=
	\mu (\de\th - d\bfq\cdott\bfmr_g )
	\,.
$$
This imposes a non-zero contribution for $\overline\Ga^\bfq$:
$$
	\overline\Ga^\bfq
	:=
	{\overline\Ga'}^\bfq
	- \mu d\bfq\cdott\bfmr_g
	\,.
$$
For a maximal reduction, minimizing $\overline\Ga^\bfq$ then means minimizing $\overline\Ga{'}^\bfq$ (and ideally setting it to zero).\\

 For the remaining $5$ columns, the pivotal $5*5$ matrix is invertible for the five unknowns $\Big( \bfmg_2^\sfc,\bfmg_2^\sfa,\bfmg_1^\sfb,\bfmg_1^\phi,\bfmg_1^\mu \Big)$, as is clear through the following argument: 
 
 - The freedom $\bfmg_1^\sfb$ controls the column for $\overline\Ga_1^\phi$, since the coefficient $-\sqrt{2\mu B}=-p\sin\varphi$ is invertible and no other unknown component of $(\bfmg_2,\bfmg_1)$ appears in this column. This does not determines fully $\bfmg_1^\sfb$, since the freedom $\avg(S_1)$ appears in the column, and $\avg(S_1)$ is to be identified by the column $\overline\Ga_1^\mu$, in which $\bfmg_1^\sfb$ appears again. Thus, the set of  $\Big( \overline\Ga_1^\phi,\overline\Ga_1^\mu \Big)$ can be considered as a coupled set of equations for $\Big( \bfmg_1^\sfb,\avg(S_1)\Big)$. But it is solvable; it implies that 
\bq
	\avg(S_1)
	=
	\mathcal K(\phi\sqrt\mu, \bfq)
	\,,
\label{S1gene}
\eq 
is an arbitrary function of $\phi\sqrt\mu$ and $\bfq$, 
and that
\bq
	\bfmg_1^\sfb
	=
	\tfrac{1}{\sqrt{2\mu B}}\p_\phi \avg(S_1)
	\,.
\label{G2bgene}
\eq

 - In a similar way, the set of  $\Big( \overline\Ga_1'^\sfb,\overline\Ga_1^t \Big)$ is a coupled set of equations for $\Big( \bfmg_1^\phi,\bfmg_1^\mu\Big)$, which is solvable. 

 - Indeed, the freedom $\bfmg_1^\phi$ controls the column for $\overline\Ga_1'^\sfb$, since the coefficient $\sqrt{2\mu B}$ is invertible. The solution for $\bfmg_1^\phi$ is then parametrized by $\bfmg_1^\mu$, which is still unknown, but appears in this column. 
 
 - Then the freedom $\bfmg_1^\mu$ controls the column for $\overline\Ga_1^t=\overline H_1$, because when inserting the solution for $\bfmg_1^\phi$, the coefficient of $\bfmg_1^\mu$ becomes just $B$, which is invertible. 
 
 - The freedom $\bfmg_2^\sfa$ controls the column for $\overline\Ga_1'^\sfc$, since the coefficient $-B$ is invertible and no other unknown component of $(\bfmg_2,\bfmg_1)$ appears in this column (because now $\bfmg_1^\sfb$ and $\bfmg_1^\mu$ are not unknowns any more). 
 
 - The freedom $\bfmg_2^\sfc$ controls the column for $\overline\Ga_1'^\sfa$, since the coefficient $B$ is invertible. The solution for $\bfmg_2^\sfc$ is then parametrized by $\bfmg_1^\th$, which is still unknown, but appears in this column. 
 
 - Then, all the components of $\Ga_1'$ remain free. They can be used to explore the various guiding-center representations \cite{BurbySquir13}. Here, we are interested in a maximal reduction, which means to set them to zero, so that the optimal requirements are fulfilled. The freedom $\mathcal K$ in $\avg(S_1)$ (which is a parameter in the formulae obtained for $\bfmg_1^\sfb$, $\bfmg_1^\phi$, $\bfmg_1^\mu$, $\bfmg_2^\sfa$, and $\bfmg_2^\sfc$) cannot be useful to improve anything and is set to zero:
\bq
		\avg(S_1)=0
		\,.
\label{AvgS1settoZero}
\eq

 - The freedom $\bfmg_1^\th$ is still undetermined, it embodies the non-uniqueness of the solution implied by the pivotal matrix being not invertible. This freedom will be useful for the requirements at the following order. As a consequence, the order mixing will not include only two orders, but three of them, and the pivotal matrix will act on components of $(\bfmg_{n+1}, \bfmg_n, \bfmg_{n-1})$. In addition, the orders will not be solved independently, since the unknown $\bfmg_1^\th$ will be identified at order $\overline\Ga_2$, whereas it already appeared in the equations at order $\overline\Ga_1$, so that it is a parameter in the expression computed for $\bfmg_2^\sfc$. \\

 This procedure shows that the average and fluctuating parts of the equations are dealt with the same way, because the equations for $\bfmg_n$ is algebraic. This is very different from the minimal guiding-center reduction by Lie transforming the equation of motion, whose equation relies on the operator $\p_\th$, and which easily controls the fluctuating part of the equation, but involves secular differential equations for the average part of the equation \cite{GuilGCmin}. \\

 At that point, it seems that the order $\overline\Ga_1$ has been completed: it is indeed completely satisfactory in itself, since all of the requirements are perfectly fulfilled, with the resulting reduced Lagrangian $\overline\Ga_1=-\mu \sfa\cdott d\sfc$. However, the procedure will have to be slightly changed, because at the following order the secondary requirement for $\overline\Ga_2^\th$ can be controlled by no higher-order freedom; it can be controlled only by $\avg(\bfmg_1^\mu)$, which was already determined above by the tertiary requirement for $\overline\Ga_1^t$. These two requirements cannot be simultaneously fulfilled and one of them has to be dropped. 
 
 It is here that the requirements are not dealt with in the same way: as one of them must be dropped, the choice is imposed by the hierarchy and the secondary requirement must be preferred to the tertiary one. 
 
 As a consequence, $\avg(\bfmg_1^\mu)$ must be let free at first order. It remains a parameter in $\bfmg_2^\sfa$ and in $\overline H_1$. Another consequence is that the tertiary requirement for $\overline\Ga_1^t$ has been lost, and $\overline H_1$ has a non-zero value. Actually, the equation for $\overline H_1$ was coupled with the one for $\overline\Ga_1'^\sfb$, and the non-zero term can be put in either of these components of $\overline\Ga$. 
 
 One can consider recovering a zero value for this term by using the freedom $\mathcal K$ available in $\avg(S_1)$, which had been arbitrarily fixed to zero in the process above in equation (\ref{AvgS1settoZero}). But the corresponding equation for $\mathcal K$ has no solution. Indeed, requiring $\avg(\overline H_1)=0$ is an equation for $\mathcal K$, which is the only available freedom  
\bq
	\big[
	\phi \p_b
	+\tfrac{\na\cdott\sfb}{2}(1+\phi^2)\p_\phi
	\big]
	\mathcal K
	=
	\tfrac{B}{\sqrt{2\mu B}}\avg(\bfmg_1^\mu)
	.
\label{SecularDiffEquation}
\eq
With the expression (\ref{AvgG1mu}) for $\avg(\bfmg_1^\mu)$, it can be studied by expansion in series $\mathcal K(\phi\sqrt\mu)=\sum_k\mathcal K_k(\phi\sqrt\mu)^k$. Expanding the right- and left-hand side of the equation and equating the coefficient of the same orders in $\phi$ and $\mu$ gives a non-solvable equation. For instance, the coefficient of order $\phi^1\mu^1$ implies the following equation 
\bq
	\na\cdott\sfb ~\mathcal K_2 
	=
	-\sfb\cdott\na\x\sfb
	\,,
\label{Obstruction}
\eq
which has no solution for a general magnetic geometry. So, the only available freedom cannot be used to obtain the full reduction $\overline\Ga_1'^\sfb=0$ and $\overline H_1=0$. One of those components has to be non-zero.\\

 Computing explicitly the expressions for the solution at order $\overline\Ga_1$ according to the procedure identified above gives the following results 
\begin{align}
	\osc(S_1)
	& =
	\bfmg_1^\sfb 
	=
	0
\label{ResultsOrderGa11}
\\
	\osc(\bfmg_1^\phi)
	&=
	\tfrac{\sqrt{2\mu B}}{B} (1+\phi^2)
	\Big[  
	\tfrac{\cba+\abc}{4}
	+\phi\abb
	\Big]
\ncr
	\osc(\bfmg_1^\mu) 	
	&=
	\mu\tfrac{\sqrt{2\mu B}}{B}
	\Big[  
	-\tfrac{B\apos \sfa}{B}
	-\phi\tfrac{\cba+\abc}{2}
	-2\phi^2\abb
	\Big]
\label{OscG1mu}
\\
	\avg(\bfmg_1^\phi)
	&=
	\tfrac{1}{\sqrt{2\mu B}} 
	\Big[  
	\overline\Ga_1'^\sfb
	+ \mu \tfrac{\cba-\abc}{2}
	\Big]
	-\tfrac{\phi}{2\mu}\avg(\bfmg_1^\mu)
\ncr
	\overline H_1	
	&=
	\sqrt{2\mu B}\phi
	\Big[
	\overline\Ga_1'^\sfb
	+\mu\tfrac{\cba-\abc}{2}
	\Big]
	+
	B\avg(\bfmg_1^\mu)
\ncr
	\bfmg_2^\sfa 
	&=
	\tfrac{\mu}{B}
	\Big[
	\phi\tfrac{7\cba -9\abc}{4}
	-\phi^2\abb
	\Big]
	+\tfrac{1}{2\sqrt{2\mu B}}\avg(\bfmg_1^\mu)
\ncr
	\bfmg_2^\sfc	
	&=
	-\tfrac{\sqrt{2\mu B}}{B}\tfrac{\bfmg_1^\th}{2}
	-\tfrac{\mu}{B}\aca
	\,,
\label{ResultsOrderGa12}
\end{align}
where following Littlejohn, a condensate notation is used for gradients: the curved prime is used to indicate gradients of the magnetic field, and the short overbar over a vector $\sfc$ or $\sfa$ indicates the matrix transpose (for the euclidean scalar product), so that $\cba:=\sfa\cdott\na\sfb\cdott\sfc$ and $B\,\apos \sfa:=\sfa\cdott\na B$. This notation comes from \cite{Little83}, slightly adapted to make it more explicit, in order to fit in with higher-order expressions \cite{GuilGCmin}. Be careful, the straight prime and the long overline are different and do not indicate gradients or matrix transpose, e.g. in $\overline\Ga'$. 

 Finally, the first-order reduced Lagrangian writes 
\bq
	\overline\Ga_1
	=
	-\mu\sfa\cdott d\sfc
	+
	\overline\Ga_1'^\sfb \sfb\cdott d\bfq
	- \overline H_1 dt
	\,.
	\label{LagrRedOrd1Sol}
\eq

 In the results above, the term $\overline\Ga_1'^\sfb$ was kept free in order to include both of the choices considered above. As $\overline\Ga_1'^\sfb$ is free, the natural choice for a maximal reduction is $\overline\Ga_1'^\sfb=0$; then the reduced Hamiltonian is non-zero:
$$
	\overline H_1	
	=
	\mu
	\sqrt{2\mu B}\phi
	\tfrac{\cba-\abc}{2}
	+
	B\avg(\bfmg_1^\mu)
	\,.
$$
 Alternatively, $\overline H_1$ can be set to zero, by choosing 
\bq
	\overline\Ga_1'^\sfb
	=
	-\tfrac{B}{\sqrt{2\mu B}\phi}
	\avg(\bfmg_1^\mu)
	-
	\mu\tfrac{\cba-\abc}{2}
	\,.
\label{ChoiceH1=0}
\eq
This last choice is possible only if the inversion of $\phi$ does not cause a singularity, i.e. if $\avg(\bfmg_1^\mu)$ has no overall contribution of order zero in $\phi$, which will have to be verified when $\avg(\bfmg_1^\mu)$ is identified (see equation (\ref{AvgG1mu})).

 The first-order Lagrangian (\ref{LagrRedOrd1Sol}) induces the following first-order reduced Lagrange matrix, which will be part of the pivotal matrix at higher orders: 
\begin{align}
	&
	\overline\om_1=
{\scriptsize
	\left(
	\begin{array}
				{ c@{~}c@{~}c@{~}  
				| @{~}c@{~}c@{~}c@{~}  
				| @{~}c }
			&&
			&0&\bfmr_g\cdott\sfc&0
			&\p_c \overline H_1
		\\ 
			&\makebox[0pt][l]{\raisebox{0ex}[0pt][0pt]{\text{\large {$\overline\om_1^{\bfq;\bfq}$}}}}&
			&0&\bfmr_g\cdott\sfa&0
			&\p_a \overline H_1
		\\
			&&
			&-	\p_\phi\overline\Ga_1'^\sfb 
			&-	\p_\mu\overline\Ga_1'^\sfb+\bfmr_g\cdott\sfb &0
			&\p_b \overline H_1
	\\ \hline
			0&0& \p_\phi\overline\Ga_1'^\sfb 
			&0&0&0
			&\p_\phi \overline H_1
		\\ 
			-\bfmr_g\cdott\sfc&-\bfmr_g\cdott\sfa& -\bfmr_g\cdott\sfb+\p_\mu\overline\Ga_1'^\sfb 
			&0&0&1
			&\p_\mu \overline H_1
		\\
			0&0&0
			&0&-1&0
			&0
	\\ \hline
			-\p_c \overline H_1&
			-\p_a \overline H_1&
			-\p_b \overline H_1
	& 
			-\p_\phi \overline H_1 & 
			-\p_\mu \overline H_1 &
			0
	&
	0
   \end{array}
   \right)
}
	\,,
\label{BarOmega1}
\end{align}
where the matrix $\overline\om_1^{\bfq;\bfq}$ is defined by
\begin{align}
	d\bfq\cdott \overline\om_1^{\bfq;\bfq}\cdott d\bfq
	:=
	d\bfq\cdott\Big[
	-
	\mu\na\sfb\cdott\sfa 
	\wedge \bar\sfc\sfb\apos
			-\Big(
			\na\x 
			\big(\overline\Ga_1'^\sfb\sfb \big)
			\Big)
			\x
	\Big]
	d\bfq	
			\,.		
			\notag
\end{align}

 At the end of the first-order analysis, three freedoms remain: $\bfmg_1^\th$, which is a parameter in $\bfmg_2^\sfc$; $\bfmg_1^\mu$, which is a parameter in $\bfmg_1^\phi$ , in $\bfmg_2^\sfa$ and in $\overline H_1$; and either $\overline\Ga_1'^\sfb$ or $\overline H_1$, which is a parameter in the other one.\\

 The results (\ref{ResultsOrderGa11})-(\ref{ResultsOrderGa12}) have physical implications. For instance, $\bfmg_2^{\sfc,\sfa}$ determines (together with $\bfmg_1$) the perpendicular component of the second-order Larmor radius $(\bfq-\overline\bfq)_2=\bfmg_2^\bfq-\tfrac {1}{2}\bfmg_1^\bfz\cdott \p_\bfz \bfmg_1^\bfq$. Its averaged contribution will imply that the Larmor radius is not a pure fluctuation, and later on, a non-zero $\bfmg_2^\sfb$ will be obtained, which will imply that the Larmor radius is not purely transverse to the magnetic field. 
 
 Those features are well known in guiding-center works, and we do not insist on them. Here, we focus on the mechanism of the reduction, to show how it can be performed to arbitrary order in the Larmor radius using gauge-independent coordinates for the gyro-angle, and why it can be considered as a maximal reduction. \\

 This point is the turning point of all the reduction: the matrix $\overline\om_{-1}+\overline\om_0+\overline\om_1$ is invertible. 
  
 Thus, in the induction for high orders, the pivotal matrix $\sfmm_\infty$ will be the set of $\Big( \overline\om_{-1},  \overline\om_0, \overline\om_1 \Big)$ acting on the set of unknown components of $(\bfmg_{n+1},\bfmg_n,\bfmg_{n-1})$. 
 
 It means that with respect to the reduction procedure announced in \ref{Sect:Appendi}, the first stage of the reduction is achieved, and the order at which $\overline\om_{-1}+\overline\om_0+ ... + \overline\om_{n_b}$ becomes invertible is $n_b:=1$. So, the critical order at which the pivotal matrix becomes the same at each order is $n_c\leqslant 2n_b+2=4$ (see the appendix for a more detailed analysis of the reduction procedure, with especially the role of the orders $n_b$ and $n_c$, introduced in pages \pageref{DefinitNc} and \pageref{DefinitNb}). Accordingly, the algorithm for the derivation at high orders (third stage mentioned in the appendix) can be identified by now, but it will be efficient only for orders $n\geqslant 4$. The intermediate orders (named "second stage" in the appendix) must be studied separately. 
 
 In order to introduce the derivation order by order, we first go through the second stage and postpone the third stage, but it is important to notice that this last is independent of the second stage and could be studied before. Especially, all the high-orders algorithm relies on the matrix $\sfmm_\infty$, together with the differential operators involved in $dS_n$. They are already known by now, and are determined by the choices that have been made previously, and mainly by the choices at order $1$.

\subsection{Core of the second stage: order $2$}

 At the following order $n=2$, the reduced Lagrangian is given by (\ref{ReducLagrAlg})
\bq
	\overline \Ga_{2}
	=
	\bfmg_3\cdott\om_{-1}
	+ \bfmg_2\cdott\overline\om_0
	+ \tfrac{\bfmg_1}{6}\cdott d
	\big[\bfmg_1\cdott(3\overline\om_0+2\tilde\om_0)\big]
	+ dS_2
	\,.
\label{LagrRedOrd2Z}
\eq
 At the previous orders, it appeared that for the algebraic part of the equations, the requirements on the average Lagrangian were dealt with exactly the same way as the requirements on the fluctuating Lagrangian. So, they will not be studied separately. Only the integrability condition will restore a difference between them for some of the unknowns. 
 
 At this point, the scheme for the unknown $\bfmg_i$ is not purely algebraical: $\bfmg_1$ is still not completely known, and it is involved in a first-order differential non-linear equation because of the term $i_{\bfmg_1}d\lambda$, with 
$$
	\lambda:=\tfrac 16 \bfmg_1\cdott
	(3\overline\om_0+2\tilde\om_0)
	\,.
$$

\medskip 
 In equation (\ref{LagrRedOrd2Z}), the exterior derivative must be computed for the $1$-form $\lambda$ that is not explicitly known yet. Unlike in previous subsections, an explicit computation is not possible, and an abstract formula must be used; care must be taken that the usual formula (\ref{ExtDeriv}) for exterior derivative cannot be used, because it is valid only when the basis is composed of closed $1$-forms. Otherwise, it is replaced by the more general formula:
\begin{align}
	d\ga
	=
	d(\ga^j \bfe_j)
	&
	=
	d(\ga^j) \wedge \bfe_j
	+ \ga^j d\bfe_j
\label{ExtDerivGene}
\\
	& =
	\bfe_i(\p_i\ga^j-\p_j\ga^i)\bfe_j
	+ \ga^j d\bfe_j
	\,,
\notag
\end{align}
for any $1$-form $\ga$, with components $\gamma^j$ in the basis $\bfe_j$. 

 In formula (\ref{ExtDerivGene}), two operations are to be identified: the action of the exterior derivative on scalar functions $d\ga^j$ expressed in the chosen basis for $1$-forms, and the exterior derivative of the basis $d\bfe_j$.
 
 For the exterior derivative on scalar functions $d\ga^j$, it writes as usual   
$$
	d(\ga^j) \wedge \bfe_j
	=
	\bfe_i(\p_i\ga^j-\p_j\ga^i)\bfe_j
	\,,
$$ 
provided the differential operators $\p_i$ are given by the dual basis (\ref{BasisVectField}) to the chosen basis (\ref{Basis1Forms}) of $1$ forms $\bfe_i$. 

 As for the exterior derivatives of the basis $d\bfe_i$, they are easily computed as:
\begin{align}
	d(d\phi)
	&
	=d(d\mu)=0
\ncr
	d(\sfb\cdott d\bfq)
	&
	=
	-d\bfq\cdott(\na\x\sfb)\x d\bfq
\ncr
	d(\sfc\cdott d\bfq)
	&
	=
	-d\bfq\cdott(\na\x\sfc)\x d\bfq
	-\delta \th \wedge \sfa\cdott d\bfq
\ncr
	d(\sfa\cdott d\bfq)
	&
	=
	-d\bfq\cdott(\na\x\sfa)\x d\bfq
	+\delta \th \wedge \sfc\cdott d\bfq
\ncr
	d(\delta \th)
	&
	=
	-\bar\sfa\sfb\apos d\bfq\wedge\bar\sfc\sfb\apos d\bfq
	+ d(\bar\sfa\sfc\apos)\wedge d\bfq
	\notag
	\,.
\end{align}
In $d \delta \th$, the derivative $\sfa\cdott\sfc\apos $ appears. Such terms should be avoided, since they are not determined, as shown by (\ref{ConnectBoth}). But there is no trouble here, since in all the $1$-forms in this derivation, the term $\bfe_\th=-\sfa\cdott(d\sfc-d\bfq\cdott\na\sfc)$ comes from $\sfa\cdott d\sfc = -\bfe_1 + d\bfq\cdott\na\sfc\cdott\sfa$. The exterior derivative of the second term will generates $d(\sfa\cdott\sfc\apos )\wedge d\bfq$, which will automatically cancel the term $\bfmr_g$ coming from $d\delta\th$.

 As a result, there is no need to compute those terms. Computing the exterior derivative of a $1$-form $\ga$ can be made according to the following procedure: write $\ga=\ga'-\ga^\th\bfmr_g\cdott d\bfq$, where now the spatial components $\ga'^\bfq\cdott d\bfq$ do not involve $\sfa\sfc\apos $. Then apply formula (\ref{ExtDerivGene}) to $\ga'$ and to $-\ga'^\th\bfmr_g\cdott d\bfq$. The cancellation of the terms containing $d(\bfmr_g)\cdott\wedge d\bfq$ gives the resulting formula:
\bq
	d\ga 
	= 
	\bfe_i(\p_i\ga'^j-\p_j\ga'^i)\bfe_j
	+
	\ga'^i d'\bfe_i
	-
	\bfe_i\p_i\ga'^\th\wedge \bfmr_g\cdott d\bfq
	\,,
\label{ExtDerivPractical}
\eq
in which $d'\bfe_i=d\bfe_i- \delta^{i\th} d(\sfa\cdott\sfc\apos )\wedge d\bfq$, where $\delta$ is the Kronecker delta, which means that $d'\bfe_i$ is exactly $d\bfe_i$ but without the problematic term in $d(\delta\th)$. \\

 With formula (\ref{ExtDerivPractical}), the quadratic term in $\bfmg_1$ is found to have its momentum components linear in the unknowns $\Big( \bfmg_1^\th, \avg(\bfmg_1^\mu) \Big)$, and differential only for $\bfmg_1^\th$:
\begin{align}
	6 \big( i_{\bfmg_1}d\lambda \big)^\phi
	&=
	-\bfmg_1^\sfa \p_\phi 
	\big[
		2\bfmg_1^\sfa\bfmr_g\cdott\sfa
		-
		2\bfmg_1^\th
	\big]
	\label{QuadraticG1Momentum}
\\
	6 \big( i_{\bfmg_1}d\lambda \big)^\mu
	&=
	-\bfmg_1^\sfa \p_\mu 
	\big[
		2\bfmg_1^\sfa\bfmr_g\cdott\sfa
		-
		2\bfmg_1^\th
	\big]
	+4\bfmg_1^\sfa\bfmr_g\cdott\sfa
	-
	4
	\bfmg_1^\th
\ncr
	6 \big( i_{\bfmg_1}d\lambda \big)^\th
	&=
	-\bfmg_1^\sfa \p_\th 
	\big[
		2\bfmg_1^\sfa\bfmr_g\cdott\sfa
		-
		2\bfmg_1^\th
	\big]
	+
	6\bfmg_1^\mu
	+\bfmg_1^\sfa
	.6 J_{322}
	\,,
	\notag
\end{align} 
where the coefficient $J_{322}$ is defined by 
$$
	J_{322}
	:=
	\tfrac{\mu}{3}
	\big(
	3J+2\tilde J +2 \bfmr_g\cdott\sfa 
	\big)
	=
	\mu
	\left[ 
	\phi(\cba-\abc) 
	+
	\tfrac{B\apos \sfa}{3B} 
	\right]
	\,.
$$
Notice that it does not depend on $\bfmr_g$, precisely because the term  $2 \sfa\sfc'\sfa$ coming from formula (\ref{ExtDerivGene}) cancels the corresponding term in $\tilde J$.\\

 The differential equation for $\bfmg_1^\th$ does not make things much more complicated, since it can be easily solved, e.g. by expansion in $\sqrt\mu$ and $\phi$. Alternatively, a trick can be used to make the scheme purely algebraical \cite{BrizGC09}; it is not essential for the derivation, but we will use it because it simplifies much the explanations. 

 The idea is to notice that in this case, the only differential operators involved come from the second term of the exterior derivative $(\p_i\lambda^j-\p_j\lambda^i)~d\bfy^i\otimes d\bfy^j$, i.e. they are involved in expressions that write $-\bfmg_1^i \p_j\lambda^i d\bfy^j=-\bfmg_1^i d\lambda^i $. In addition, in the sum over the index $i$, only one of the terms involves a differential operator $d$ acting on the unknown $\bfmg_1^\th$, namely the term with $i=\sfa$. An integration by parts over this term can transfer the differential operator over the pre-factor $\bfmg_1^\sfa$, which is already known. This is a way to make the equation algebraic. 
 
 This integration by parts is justified by the gauge function. The equation for $\overline\Ga_2$ can be added a total derivative, which can be chosen $d (\bfmg_1^\sfa\lambda^\sfa)$ and extracted from $dS_2$ by $dS_2=dS_2'+d(\bfmg_1^\sfa\lambda^\sfa)$:
$$
	-\bfmg_1^\sfa d\lambda^\sfa 
	+dS_2
	= 
	-\bfmg_1^\sfa d \lambda^\sfa
	+dS_2'
	+d(\bfmg_1^\sfa\lambda^\sfa)
	=
	dS_2'
	+\lambda^\sfa d\bfmg_1^\sfa 
	\,,
$$
and this formula is not differential any more, but algebraic for $\bfmg_1^\theta$, which is contained in $\lambda^\sfa$. \\

 Then, equation (\ref{LagrRedOrd2Z}) becomes 
\bq
	\overline \Ga_{2}
	=
	\bfmg_3\cdott\om_{-1}
	+ \bfmg_2\cdott\overline\om_0
	+ \Lambda
	+ dS_2'
	\,,
\label{LagrRedOrd2ZZ}
\eq
with 
\begin{align}
	\Lambda 
	:
	= 
	&
	i_{\bfmg_1} d\lambda
	+
	d(\bfmg^\sfa\lambda^\sfa)
\ncr
	=
	&
	\hspace{11ex}
	\left(
	     	\begin{smallmatrix}
			\Lambda_\sfc &
			\Lambda_\sfa &
			\Lambda_\sfb &
			| &
			~0~ &
			~0~ &
			~0~ &
			| &
			\Lambda_t
			\end{smallmatrix}
    \right)
\ncr
	&
	+
		\left( 
	     	\begin{smallmatrix}
			\bfmg_1^\sfa \\
			\bfmg_1^\mu \\
			\bfmg_1^\th 
			\end{smallmatrix}
	\right)^T
	\cdott
{\scriptsize
	\left(
	\begin{array}
				{ c@{~}c@{~}c@{~~}  
				| @{~~}c@{~}c@{~}c@{~~}  
				| @{~~}c@{~} }
			&&
			&0& \bar\sfa\sfc\apos\sfa & J_{322}
			&
		\\ 
			~~&\raisebox{-1ex}[0ex][0ex]{\large 0}&~~
			&0&0&1
			&\raisebox{-1ex}[0ex][0ex]{\large 0}
		\\
			&&
			&0&-1&0
			&
   \end{array}
   \right)
}
	\notag
	\,,
\end{align}
where the momentum components are written in matrix form because they will determine the unknowns $\big(\bfmg_1^\th,\avg(\bfmg_1^\mu)\big)$.

 Now, the induction relation just relies on a pivotal matrix $\sfmm_2$, which appears in the set of 
$$
	\Big( \bfmg_3\cdott\om_{-1}, 
	\bfmg_2\cdott\overline\om_0, 
	\Lambda 
	\Big)
	\,,
$$ 
in which a linear algebraic operator (a matrix) acts on the unknown components of $(\bfmg_3,\bfmg_2,\bfmg_1)$. 

 It is invertible in the sense that it determines six unknown components of $(\bfmg_3,\bfmg_2,\bfmg_1)$, which is the maximum that can be done at each order. Remember the matrix cannot be fully invertible, since the transformation is time-independent, so that $\bfmg_n$ is $6$-dimensional, whereas the matrix has value in the $7$-dimensional space $(\bfq,\bfp,t)$; and by the way, the $7$-dimensional matrix is anti-symmetric, hence not invertible. The seventh requirement is to be provided by the gauge function $S_2'$, which is the only integrability condition involved at this order. 
 
 More precisely, removing all the coefficients that do not contribute, as was done in (\ref{LagrRedOrd1}), equation (\ref{LagrRedOrd2ZZ}) becomes 
\begin{align}
	&~~~~
	\left( 
	     	\begin{smallmatrix}
			\bfmg_3^\sfc \\
			\bfmg_3^\sfa \\
			\end{smallmatrix}
	\right)^T
	\hspace{1ex}
	\cdott
{\scriptsize
	\left(
	     	\begin{array}{
	     					c@{~~~}c@{~~~~~}c@{~~~~} 
	     					|@{~~~~~~} 
	     						c@{~~~~~~~~~~~}
	     						c@{~~~~~~\,}
	     						c@{~~} 
	     					| @{~~~~\,}c@{~~~\,}
	     				}
				0 & B & 0 
			& 0 & 0 & 0 &
			0
		\\
				-B & 0 & 0 
			& 0 & 0 & 0 &
			0
			\end{array}
   \right)
}
\ncr
	&
	+ 
	\left( 
	     	\begin{smallmatrix}
			\bfmg_2^\sfb \\
			\bfmg_2^\phi \\
			\bfmg_2^\mu \\
			\bfmg_2^\th
			\end{smallmatrix}
	\right)^T
	\hspace{0ex}
	\cdott
{\scriptsize
	\left(
	\begin{array}{ 
					c@{~~~}c@{~}c@{~ }  
					| @{~}c@{~}c@{~}c@{~\,}  
					| @{~}c 
				}
				-I &K &0
     &
			-\sqrt{2\mu B} 
			& -\tfrac{\phi\sqrt{2\mu B}}{2\mu} &0
	&
			H_0\tfrac{\p_b B}{B}
	\\ \hline
			0&0&\sqrt{2\mu B}
			&&&
			&2\mu B\phi 
		\\
			0&0&\tfrac{\phi\sqrt{2\mu B}}{2\mu}
	&& \raisebox{-1ex}[0ex][0ex]{\large 0} &
	&
			\tfrac{H_0}{\mu} 
		\\
			0&0&0
			&&&
			&0
   \end{array}
   \right)	
}
\ncr
	&
	+ 
	\left(
	     	\begin{smallmatrix}
			\bfmg_1^\mu \\
			\bfmg_1^\th 
			\end{smallmatrix}
	\right)^T
	\hspace{1ex}
	\cdott
{\scriptsize
	\left(
	     	\begin{array}{
	     			@{~~~}c@{~~~~~~}c@{~~~~~~}c@{~~~}
	     			|@{~~~~~}c@{~~~~~~~~}c@{~~~~~~}c@{~~~}
	     			|@{~~~~~\,}c@{~~~~}
	     				}
			0 &  
			0 &  
			0 & 
			0 &  
			0 &  
			1 &  
			0
	\\
			0 &  
			0 &  
			0 & 
			0 &  
			-1 &  
			0 &  
			0  
			\end{array}
   \right)
}
\ncr
	=
	&
	- 
	\left( 
	     	\begin{smallmatrix}
			\bfmg_2^\sfc \\
			\bfmg_2^\sfa \\
			\bfmg_1^\sfa 
			\end{smallmatrix}
	\right)^T
	\hspace{1ex}
	\cdott
{\scriptsize
	\left(
	\begin{array}{ 
					@{~~}c@{~~~}c@{~~~}c@{~~~}  
					| @{~~~~}c@{~~~~~~~}c@{~~~~}c@{~}  
					| @{~}c 
				}
			0&-J&I
			&0&0&0
			&H_0\tfrac{\p_c B}{B}
		\\
			J&0&-K
			&0&0&0
			&H_0\tfrac{\p_a B}{B}
		\\
			0&0&0
			&0& \bar\sfa\sfc\apos\sfa &J_{322}
			&0
   \end{array}
   \right)	
}
\ncr
	& 
	-
	\hspace{11ex}
	\left( 
	     	\begin{smallmatrix}
			~\Lambda_\sfc~ &
			\Lambda_\sfa~~\, &
			\Lambda_\sfb\, &
			|~ &
			~\,0\,~~~~~ &
			0~~~~~ &
			0~\, &
			|~ &
			~\, \Lambda_t~\,
			\end{smallmatrix}
	\right)
\ncr
	& 
	+
	\hspace{11ex}
	\left( 
	     	\begin{smallmatrix}
			~\overline\Ga_2^\sfc~~ &
			\overline\Ga_2^\sfa~~ &
			\overline\Ga_2^\sfb\, &
			|~ &
			~\overline\Ga_2^\phi~~~~ &
			\overline\Ga_2^\mu~~~ &
			\overline\Ga_2^\th~ &
			| &
			~\, \overline H_2~\,
			\end{smallmatrix}
	\right)
\ncr
	&
	-
	\hspace{11ex}
			\left(
	     	\begin{smallmatrix}
			\p_c S_2' &  
			\p_a S_2' &
			\p_b S_2' &
			| &
			\p_\phi S_2'~ &
			\p_\mu S_2'~ &
			\p_\th S_2' &
			| &
			~~\,0~~\,
			\end{smallmatrix}
	\right)
	\,,
\label{LagrRedOrd2}
\end{align}
where the left-hand side contains just the terms involved in the matrix inversion to determine the unknown components of $\big( \bfmg_3,\bfmg_2,\bfmg_1 \big)$.\\

 Formula (\ref{LagrRedOrd2}) is similar to (\ref{LagrRedOrd1}), and the same comments can be done as at the previous order, in the two paragraphs following formula (\ref{LagrRedOrd1}). 

 To solve the equation, an analysis similar to the one at the previous order leads to the procedure summarized in the following tabular, where each row corresponds to one of the equations. The component $\overline\Ga_2^i$ of the reduced Lagrangian involved in the corresponding equation is indicated in the first column; the unknown which controls the equation and permits $\overline\Ga_2^i=0$ is indicated in the second column; the coefficient to be inverted is indicated in the third column. 
$$
	\begin{array}{l|@{~~}l@{~}l}
		\text{Equation} 
		& \text{Unknown} 
		& \text{Coefficient}
	\\ \hline
		\avg(\overline\Ga_2^\th) & \avg(\bfmg_1^\mu) & 1
	\\
		\osc(\overline\Ga_2^\th) & \osc (S'_2) & \p_\th 
	\\
		\overline\Ga_2^\phi & \bfmg_2^\sfb & -\sqrt{2\mu B}
	\\
		\overline\Ga_2^\mu & \bfmg_1^\th & -1
	\\
		\overline\Ga_2^\sfb & \bfmg_2^\phi & \sqrt{2\mu B}
	\\
		\overline H_2 & \bfmg_2^\mu & B
	\\
		\overline\Ga_2^\sfc & \bfmg_3^\sfa & -B
	\\
		\overline\Ga_2^\sfa & \bfmg_2^\sfc & B
	\end{array}
$$
 
 A few comments are in place. As had been announced in the derivation of formula (\ref{OscG1mu}), the new feature is that the equation for $\avg(\overline\Ga_2^\th)$ can be controlled only by the average first-order magnetic moment $\avg(\bfmg_1^\mu)$. It is why it was not available at the previous order. 
 
 The coefficient $\p_\th$ is an operator, but it is invertible over gyro-fluctuations. 

 The set of  $\Big( \overline\Ga_2^\phi,\overline\Ga_2^t \Big)$ is a coupled set of equations for $\Big( \bfmg_2^\phi,\bfmg_2^\mu\Big)$, but each of the unknowns can be assigned to one of the equation because the system can be solved in the following way. The freedom $\bfmg_2^\phi$ controls the column for $\overline\Ga_2^\sfb$, since the coefficient $\sqrt{2\mu B}$ is invertible. The solution for $\bfmg_2^\phi$ is parametrized by $\bfmg_2^\mu$, which is still unknown, but appears in this equation. Then, the freedom $\bfmg_2^\mu$ controls the column for $\overline\Ga_2^t=\overline H_1$, because when inserting the solution for $\bfmg_2^\phi$, the coefficient of $\bfmg_2^\mu$ becomes $B$, which is invertible. 
 
 At the end, the reduced Lagrangian $\overline\Ga_2$ is free and can be set to zero, as required for the maximal reduction. The freedom $\avg(S_2')$ (which is a parameter in the formulae obtained for $\bfmg_1^\th$, $\bfmg_2^\sfb$, $\bfmg_2^\phi$, $\bfmg_2^\mu$, $\bfmg_3^\sfa$, and $\bfmg_3^\sfc$) cannot be useful and is set to zero. \\

 In a similar way as at the previous order, at that point, the order $\overline\Ga_2$ is completely satisfactory in itself, since the reduced Lagrangian has been fully reduced $\overline\Ga_2=0$. 
 
 However, the procedure will have to be changed, because at the following order the secondary requirement for $\overline\Ga_3^\th$ can be controlled only by $\avg(\bfmg_2^\mu)$, which is therefore not available to get the tertiary requirement for $\overline\Ga_2^t$. So, this last requirement has to be dropped. 
 
 Accordingly, $\avg(\bfmg_2^\mu)$ remains free at this order, and it is a parameter in $\bfmg_3^\sfa$ and in $\overline\Ga_2^t$. This last has a non-zero value, but its equation was coupled with the one for $\overline\Ga_2^\sfb$, and the non-zero term can be put in either of these components of $\overline\Ga_2$. \\

 One can consider recovering a zero value for this term by using the freedom $\avg(S_2')$, whose value had been arbitrarily fixed to zero in the process above. 
 
 Indeed, if $\avg(S_n)$ is let as a free parameter in $\bfmg_2^\sfb$, $\bfmg_1^\th$ and $\bfmg_2^\phi$, then when replacing these variables by their expression, the requirement $\avg(\overline H_2)=0$ becomes an equation for $\avg(S_2')$. Unfortunately, this equation is not easily studied. 
 
 As a first attempt, $S_2'$ can be chosen such that it is absent from $\bfmg_1^\th$, as it was done in equations (\ref{S1gene})-(\ref{G2bgene}). Then it is easy to see that the equation for $\avg(S_2')$ will have the same structure as equation (\ref{SecularDiffEquation}). This equation may not be integrable, as it was the case for equation (\ref{SecularDiffEquation}). In this case, one should relax the condition for $\avg(S_2')$ to be absent from $\bfmg_1^\th$, and the differential equation for $\avg(S_2')$ could be more difficult to study, because $\bfmg_1^\th$ is involved in the $1$-form $\Lambda$ in equation (\ref{LagrRedOrd2}) in a rather intricate way. \\

 Applying the procedure identified above gives
\begin{align}
	\avg(\bfmg_1^\mu)
	&= 
	\mu\tfrac{\sqrt{2\mu B}}{B}
	\phi
	(\abc-\cba)
\label{AvgG1mu}
\\
	\osc (S'_2)
	&=
	\mu\tfrac{\sqrt{2\mu B}}{B}
	\Big\lbrack
	-
	\tfrac{2B\apos \sfc}{3B}
	+\phi\tfrac{\aba-\cbc}{4}
	-2\phi^2\cbb	
	\Big\rbrack
\ncr
	\bfmg_2^\sfb
	&=
	\tfrac{\mu}{B}
	\Big\lbrack
	\tfrac{\aba-\cbc}{4}
	-4\phi\cbb
	\Big\rbrack
	+
	\tfrac{\p_\phi \avg(S_2')}{\sqrt{2\mu B}}
\ncr
	\bfmg_1^\th
	&=
	\tfrac{\sqrt{2\mu B}}{B}
	\Big\lbrack
	\sfa\sfc'\sfa
	-\tfrac{B\apos \sfc}{B}
	+
	\phi\tfrac{\aba-\cbc}{4}
	-
	\phi^2\cbb
	\Big\rbrack
\ncr
	& \hspace{27ex}
	+
	\p_\mu \avg(S_2')
\label{SolG1th}
\\
	\bfmg_3^\sfa
	&=
	\tfrac{1}{B}
	\Big\lbrack
	\sqrt{2\mu B}
	\Big\lbrace
	J \bfmg_2^\sfa
	-I \bfmg_2^\sfb
	\Big\rbrace
	+ S_2'\sfc
	+ \Lambda^\sfc
	\Big\rbrack
\ncr
	\bfmg_3^\sfc
	&=
	\tfrac{1}{B}
	\Big\lbrack
	\sqrt{2\mu B}
	\Big\lbrace
	J \bfmg_2^\sfc
	-K \bfmg_2^\sfb
	\Big\rbrace
	- S_2'\sfa
	- \Lambda^\sfa
	\Big\rbrack
\ncr
	\osc(\bfmg_2^\mu)
	&=
	\tfrac{\osc}{-B}
	\Big\lbrack
	\sfmr_2^t
	+
	\bfmg_2^\sfb
	(1+\phi^2)\mu B\apos \sfb
	-
	\sqrt{2\mu B} \phi 
	\sfmr_2^\sfb
	\Big\rbrack
\ncr
	\osc(\bfmg_2^\phi)
	&=
	\tfrac{\osc}{2\mu B}
	\Big\lbrack
	\phi\sfmr_2^t
	+
	(1+\phi^2)
	\Big\lbrace
	\bfmg_2^\sfb \phi\mu B\apos \sfb
	-
	\sqrt{2\mu B}\sfmr_2^\sfb
	\Big\rbrace
	\Big\rbrack
\ncr
	\avg(\bfmg_2^\phi)
	&=
	\tfrac{\avg}{2\mu B}
	\Big\lbrack
	-B\phi \bfmg_2^\mu
	+
	\sqrt{2\mu B}
	\big\lbrace
	\overline\Ga_2^\sfb
	-
	\sfmr_2^\sfb
	\big \rbrace
	\Big\rbrack
\ncr
	\overline H_2
	&=
	\avg
	\Big\lbrack
	\sfmr_2^t
	+ \bfmg_2^\sfb (1+\phi^2)\mu B\apos \sfb
\ncr
	& \hspace{10.6ex} +
	B \bfmg_2^\mu
	+
	\sqrt{2\mu B}\phi
		\Big\lbrace 
		\overline\Ga_2^\sfb - \sfmr_2^\sfb 
		\Big\rbrace
	\Big\rbrack
	\,,
\label{HamiltRedOrd2}
\end{align}
where 
\begin{align}
	\sfmr_2
	:=
	&
	\Lambda+dS_2'
	+
	\left( 
	     	\begin{matrix}
			\bfmg_2^\sfc \\
			\bfmg_2^\sfa 
			\end{matrix}
	\right)^T
	\cdott
{\small
	\left(
	\begin{array}{ ccc  | @{~}c@{~}c@{~}c@{~}  | @{~} c }
				0&-J &I 
				& 0 & 0 & 0
				&H_0\tfrac{\p_c B}{B} 
			\\
				J &0&-K 
     &
			0 & 0 & 0
			& H_0\tfrac{\p_a B}{B}
   \end{array}
   \right)	
}
	\notag	
	\,.
\end{align} 

 With the results (\ref{AvgG1mu})-(\ref{HamiltRedOrd2}), the parameters involved in formulae (\ref{ResultsOrderGa11})-(\ref{ResultsOrderGa12}) can be made explicit:
\begin{align}
	\avg(\bfmg_1^\phi)
	&=
	\tfrac{1}{\sqrt{2\mu B}} 
	\bigg[  
	\overline\Ga_1^\sfb
	+ \mu \tfrac{1+2\phi^2}{2} (\cba-\abc)
	\Bigg]
\label{ResultsDeparametrized1}
\\
	\overline H_1	
	&=
	\sqrt{2\mu B}\phi
	\Big[
	\overline\Ga_1^\sfb
	-\mu\tfrac{\cba-\abc}{2}
	\Big]
\ncr
	\bfmg_2^\sfa 
	&=
	\tfrac{\mu}{B}
	\Big[
	\phi\tfrac{5\cba -7\abc}{4}
	-\phi^2\abb
	\Big]
\ncr
	\bfmg_2^\sfc	
	&=
	\tfrac{\mu}{B}
	\Big[
	-\tfrac{B\apos \sfc}{B}
	+
	\phi\tfrac{\aba-\cbc}{4}
	-
	\phi^2\cbb
	\Big]
	+
	\tfrac{\sqrt{2\mu B} }{B}
	\tfrac{\p_\mu \avg(S_2')}{2}
	\,.
\label{ResultsDeparametrized2}
\end{align}
The reduced first-order Hamiltonian for the choice $\overline\Ga_2^\sfb=0$ is
$$
	\overline H_1 = \mu \sqrt{2\mu B} \phi\tfrac{\abc-\cba}{2}
	\,.
$$
The reverse choice $\overline H_1=0$ is possible, since its existence condition in equation (\ref{ChoiceH1=0}) is satisfied, as is clear in (\ref{AvgG1mu}). It corresponds to a component $\overline\Ga_1^\sfb$ of the reduced Lagrangian given by  
\bq
	\overline\Ga_1'^\sfb=
	\mu\tfrac{\cba-\abc}{2}
	\,,
	\label{LagrRedOrd1Alain}
\eq
which is regular in $\phi=0$, as expected.

 In formulae (\ref{AvgG1mu})-(\ref{ResultsDeparametrized2}), the components $\bfmg_1$ and $\bfmg_2^\bfq$ have been completely computed and simplified for comparison with previous results, because it is the point where usual derivations stop. 
 
 For the other components $\bfmg_2^\phi$, $\bfmg_3^\sfc$, $\bfmg_3^\sfa$, and $\overline H_3$, formulae (\ref{AvgG1mu})-(\ref{ResultsDeparametrized2}) are explicit solutions. Their right-hand side involves only known quantities (or quantities that are free parameters for these relations), but it has not been not expanded and simplified. This can can be done in a straightforward way just by computing explicitly the terms involved, but we will not pursue in that direction, since the calculation for the $\Lambda_{\sfc,\sfa,\sfb}$ is lengthy, and useless for our purpose, which is just to show how the procedure can be performed to arbitrary order. In addition, they are the topic of a work by the authors of \cite{BrizTron12}, and were already partly introduced in \cite{ParrCalv11}.\\

 In the results, the term $\overline\Ga_2^\sfb$ was kept free to include both of the choices considered above: setting the non-zero term in $\overline H_2$ just means choosing $\overline\Ga_2^\sfb=0$. The other choice $\overline H_2=0$ corresponds to 
$$
	\overline\Ga_2^\sfb
	=
	\avg
	\Big\lbrack
	\sfmr_2^\sfb
	-
	\tfrac{1}{\sqrt{2\mu B}\phi}
	\Big\lbrace
	B\bfmg_2^\mu
	+
	\sfmr_2^t
	+
	\bfmg_2^\sfb(1+\phi^2)\mu B\apos \sfb
	\Big\rbrace
	\Big\rbrack
	\,.
$$
This last choice is possible only if the term inside the curled parentheses has no overall contribution of order zero in $\phi$. 

 Also $\avg(S_2')$ was kept free, because one can consider using it to obtain the full reduction with both $\overline\Ga_2^\sfb=0$ and $\overline H_2=0$: it will imply a differential equation for $\avg(S_2')$, which might not be integrable, just as happened at the previous order in equation (\ref{SecularDiffEquation}), but it might be integrable, or partly integrable, and then provide the full reduction, or at least a stronger reduction. Otherwise, it can be set to zero. 

 At the end of the analysis at order $n=2$, it remains one unknown, one freedom, and a binary choice. The unknown is $\avg(\bfmg_2^\mu)$; it is a parameter in $\bfmg_2^\phi$ and in $\overline H_2$ (or $\overline\Ga_2^\sfb$); it will be determined at the following order. The freedom is $\avg(S_2')$, which remains free, but could not be used to improve the reduction. The binary choice is that either $\overline\Ga_2^\sfb$ or $\overline H_2$ is set to zero, the other is computed accordingly. This is very similar to what had occurred at the previous order, but now, the unknown $\bfmg_2^\th$ does not appear as a parameter, since it is not at all involved at this order, as is clear in equation (\ref{LagrRedOrd2}). \\

 \subsection{End of the second stage: order $3$}

 Let us turn now to the following order $\overline\Ga_3$. The equation writes (\ref{ReducLagrAlg})
\begin{align}
	\overline \Ga_{3}
	=
	\bfmg_4\cdott\om_{-1}
	+ \bfmg_3\cdott\overline\om_0
	&
	+ \bfmg_2\cdott\overline\om_1
	- \tfrac{(\bfmg_2\cdott d)^2}{2}\Ga_{-1}
\ncr
	&
	+ \tfrac{(\bfmg_1\cdott d)^2}{24}
	\bfmg_1\cdott(3\om_0+\overline\om_0)
	+ dS_3
	\,.
\notag
\end{align}
The unknowns are components of $(\bfmg_4,\bfmg_3,\bfmg_2)$. The pivotal matrix is not $\sfmm_\infty$, i.e. not just given by $(\om_{-1},\overline\om_0,\overline\om_1)$, because of the correcting term $- \tfrac{(\bfmg_2\cdott d)^2}{2}\Ga_{-1}$. This term might make the derivation more difficult, since it is not algebraic, but non-linear and differential for $\bfmg_2$
$$
	- \tfrac{\bfmg_2}{2}\cdott d(\bfmg_2\cdott d\Ga_{-1})
	=
	- \tfrac{\bfmg_2}{2}\cdott 
	\big[
	d
		\big\lbrace
		-B(\bfmg_2^\sfa \sfc) 
		+B(\bfmg_2^\sfc \sfa) 
		\big\rbrace
	\cdott\wedge d\bfq
	\big]
	\,.
$$
It can be written in matrix form
$$
	-\tfrac{1}{2}
	\left(
	\begin{array}{ c}
	\bfmg_2^{\sfc,\sfa,\sfb}    
     \\ \hline
	\bfmg_2^{\phi,\mu,\th}    
     \\ \hline
     0
   \end{array}
   \right)^T
\cdott
	\left(
	\begin{array}{ c  @{~}|@{~} c  @{~}|@{~} c }
	M_{11}     
	& M_{12} 
     & 0 
     \\ \hline
     M_{21} & 0 & 0
     \\ \hline
     0 & 0 & 0
   \end{array}
   \right)
   \,,
$$
where the matrices $M_{ij}$ have obvious definitions, and are independent of $\bfmg_2^{\mu,\th}$. 

 This is enough to show that this additional term can be transferred into the right-hand side, i.e. it involves only terms that are already known at each step of the computation. 
 
 When computing $\overline\Ga_3^\bfp$ for the unknowns $\bfmg_2^\th$, $\avg(\bfmg_2^\mu)$, $\bfmg_3^\sfb$ and $\osc (S_3)$, the only components of $\bfmg_2$ involved in the correcting term are in $\bfmg_2^\bfq$, which is already known at that point. Then, when computing $\overline\Ga_3^{\bfq,t}$ for the unknowns $\bfmg_3^\phi$, $\bfmg_4^\sfc$, $\bfmg_4^\sfa$, and $\osc(\bfmg_3^\mu)$ all the components of $\bfmg_2$ are involved in the correcting term, but they are all known at that point. 

 As a consequence, the correcting term can be put in the right-hand side of the equation, and the pivotal matrix is actually $\sfmm_\infty$, i.e. the set of $( \om_{-1},  \overline\om_0, \overline\om_1 )$ acting on the unknown components of $(\bfmg_4,\bfmg_3,\bfmg_2)$. This means that the critical order at which the pivotal matrix becomes the same at each order is $n_c:=3$, and the order $\overline\Ga_3$ can be included in the third stage, with all higher orders, which is studied in the following section.

\subsection{Third stage: algorithmic orders $4$ and higher}

 Now, the second stage of the method mentioned in appendix is ended and the third stage is beginning, which means that the matrix to be inverted is always the same at any order $n\geqslant 3$, and it is indeed invertible. So, the reduction can be performed to arbitrary order. The only possible complication comes from the integrability condition for the gauge-function $S_n$ (and possibly $\overline\Ga_n$), but after settling it, the reduction process becomes fully algorithmic and unique.

\subsubsection{Equation and algorithm}

 This is proven by induction. Let us suppose that at some order $n\geqslant 3$, the set of unknowns are 
\begin{align}
	\bfg_n:=\Big( \overline\Ga_n, S_n, 
	\bfmg_{n-1}^\th,\avg(\bfmg_{n-1}^\mu),
	&
	\osc(\bfmg_{n}^\mu), 
\label{DefVectBfgText}
\\
	&
	\bfmg_n^\sfb,\bfmg_n^\phi,
	\bfmg_{n+1}^\sfc,\bfmg_{n+1}^\sfa \Big)
	\,,
	\notag 
\end{align}
which means that before that order, all lower-order quantities $\bfg_{i<n}$ are already determined, and that after that order, all higher-order quantities $\bfg_{i>n}$ will remain free parameters. This assumption is verified at order $n=3$, which initializes the induction. As announced in the appendix, we have included the reduced Lagrangian $\overline\Ga_n$ in the vector $\bfg_n$, because some components of $\overline\Ga_n$ cannot be set to zero and have to be computed in the process. 

 The reduced Lagrangian is given by equation (\ref{MatrPivot})
$$
	\bfmg_{n+1}\cdott\om_{-1}
	+ \bfmg_n\cdott\overline\om_0
	+ \bfmg_{n-1}\cdott\overline\om_1
	=
	\overline \Ga_{n}
	-\sfmr_n 
	- dS_n
	\,,
$$
where $\sfmr_n$ indicates all terms of $... e^{\bfmg_2}e^{\bfmg_1}\Ga$ that are of order $n$ but do not involve $\bfmg_{n+1}$, $\bfmg_{n}$ or $\bfmg_{n-1}$:
$$
	\sfmr_n
	:= [... e^{\bfmg_2}e^{\bfmg_1}\Ga]_n 
	- 	[\bfmg_{n+1}\cdott\om_{-1}
	+ \bfmg_n\cdott\overline\om_0
	+ \bfmg_{n-1}\cdott\overline\om_1]
	\,,
$$
in which the index $n$ indicates the term of order $n$.\\

 Then, the pivotal matrix $\sfmm_\infty$ is the set of 
$$\
	\Big( \om_{-1}, 
	\overline\om_0, 
	\overline\om_1
	\Big)
	\,,
$$ 
acting on the unknown components of $(\bfmg_{n+1},\bfmg_n,\bfmg_{n-1})$. 

 It is invertible in the sense that it determines six unknown components of $(\bfmg_{n+1},\bfmg_n,\bfmg_{n-1})$, which is the maximum that can be done at each order. The last requirement is provided by the integrability condition for the gauge function $S_n$. 
 
 More precisely, grouping as usual in the left-hand side only the terms with unknown components of $(\bfmg_{n+1},\bfmg_n,\bfmg_{n-1})$, and removing all the coefficients that do not contribute, the induction relation for $\overline\Ga_n$ writes
\begin{align}
	&~~
	\left( 
	     	\begin{smallmatrix}
			\bfmg_{n+1}^\sfc \\
			\bfmg_{n+1}^\sfa \\
			\end{smallmatrix}
	\right)^T
	\hspace{5ex}
	\cdott
{\scriptsize
	\left(
	     	\begin{array}{
	     					c@{~~~}c@{~~~~~}c@{~~~~} 
	     					|@{~~~~~~} 
	     						c@{~~~~~~~~~~}
	     						c@{~~~~~~}
	     						c@{~~} 
	     					| @{~~~~}c@{~~~~}
	     				}
				0 & B & 0 
			& 0 & 0 & 0 &
			0
		\\
				-B & 0 & 0 
			& 0 & 0 & 0 &
			0
			\end{array}
   \right)
}
	\ncr
	&
	+ 
	\left( 
	     	\begin{smallmatrix}
			\bfmg_n^\sfb \\
			\bfmg_n^\phi \\
			\bfmg_n^\mu 
			\end{smallmatrix}
	\right)^T
	\hspace{5ex}
	\cdott
{\scriptsize
	\left(
	\begin{array}{ 
					c@{~~}c@{~~}c@{~ }  
					| @{~}c@{~}c@{~}c@{~}  
					| @{~}c 
				}
				-I &K &0
     &
			-\sqrt{2\mu B} 
			& -\tfrac{\phi\sqrt{2\mu B}}{2\mu} &0
	&
			H_0\tfrac{\p_b B}{B}
	\\ 
			0&0&\sqrt{2\mu B}
			&0&0&0
			&2\mu B\phi \\
			0&0&\tfrac{\phi\sqrt{2\mu B}}{2\mu}
	&0& 0 &0
	&
			\tfrac{H_0}{\mu} 
   \end{array}
   \right)	
}
\ncr
	&
	+ 
	\left(
	     	\begin{smallmatrix}
			\avg(\bfmg_{n-1}^\mu) \\
			\bfmg_{n-1}^\th 
			\end{smallmatrix}
	\right)^T
	\cdott
{\scriptsize
	\left(
	     	\begin{array}{
	     			@{~~}c@{~~~~}c@{~~~~}c@{~~~}
	     			|@{~~~~~}c@{~~~~~~~}c@{~~~~~}c@{~~}
	     			|@{~~}c@{~~}
	     				}
			0 &  
			0 &  
			\p_\mu\overline\Ga_1'^\sfb & 
			0 &  
			0 &  
			1 &  
			\p_\mu\overline H_1
	\\
			0 &  
			0 &  
			0 & 
			0 &  
			-1 &  
			0 &  
			0  
			\end{array}
   \right)
}
\ncr
	=
	&
	- 
	\sfmr'_n
\label{LagrRedOrdN}
\\
	& 
	+
	\hspace{15ex}
	\left( 
	     	\begin{smallmatrix}
			~\overline\Ga_n^\sfc~~ &
			\overline\Ga_n^\sfa~~ &
			\overline\Ga_n^\sfb~ &
			|~ &
			\overline\Ga_n^\phi~~~ &
			\overline\Ga_n^\mu~~~ &
			\overline\Ga_n^\th &
			|~ &
			~\overline H_n~
			\end{smallmatrix}
	\right)
\ncr
	&
	-
	\hspace{15ex}
			\left(
	     	\begin{smallmatrix}
			\p_c S_n &  
			\p_a S_n &
			\p_b S_n &
			| &
			\p_\phi S_n &
			\p_\mu S_n &
			\p_\th S_n &
			| &
			~~0~~~
			\end{smallmatrix}
	\right)
	\notag
	\,,
\end{align}
where the terms involving known components of $(\bfmg_{n+1},\bfmg_n,\bfmg_{n-1})$ have been grouped with $\sfmr_n$:
\begin{align}
	&
	\sfmr'_n
	=
	\sfmr_n
\label{DefSfmrPrime}
\\
	&
	-
	\hspace{3ex}
	\left( 
	     	\begin{smallmatrix}
			\bfmg_n^\sfc \raisebox{-0.5ex}{\rule{0ex}{2.2ex}}\\
			\bfmg_n^\sfa \raisebox{-0.5ex}{\rule{0ex}{2.2ex}}
			\end{smallmatrix}
	\right)^T
	\hspace{3ex}
	\cdott
{\scriptsize
	\left( 
     		\begin{array}{
     						@{~~~~}c@{~~~~~~~}c@{~~~~~~~}c@{~~~}
     						|
     						@{~~~~~}c
     						@{~~~~~~~~~~}c
     						@{~~~~~~}c@{~~~}
     						|
     						@{~~~}c@{~~}
     					}
			0&-J&I
			&0&0&0
			& H_0 \tfrac{B\apos \sfc}{B}\\ 
			J&0&-K
			&0&0&0
			& H_0 \tfrac{B\apos \sfa}{B}
			\end{array} 
   \right)
}
\ncr
	&
	- 
	\left( 
	     	\begin{smallmatrix}
			\bfmg_{n-1}^\sfc \\
			\bfmg_{n-1}^\sfa\\
			\bfmg_{n-1}^\sfb \\ \hline
			\bfmg_{n-1}^\phi \\
			\osc(\bfmg_{n-1}^\mu) \\
			0 \\ \hline
			0
			\end{smallmatrix}
	\right)^T
	\cdott
{\scriptsize
	\left(
	\begin{array}
				{ c@{~}c@{~}c@{~}  
				| @{~}c@{~~}c@{~~}c@{~~~}
     			|
     			@{~~~}c@{~~}
     			}
			&&
			&0&0&0
			&\p_c \overline H_1
		\\ 
			&\makebox[0pt][l]{\raisebox{0ex}[0pt][0pt]{\text{\large {$\overline\om_1^{\bfq;\bfq}$}}}}&
			&0&0&0
			&\p_a \overline H_1
		\\
			&&
			&-	\p_\phi\overline\Ga_1'^\sfb 
			&-	\p_\mu\overline\Ga_1'^\sfb &0
			&\p_b \overline H_1
	\\ \hline
			0&0& \p_\phi\overline\Ga_1'^\sfb 
			&0&0&0
			&\p_\phi \overline H_1
		\\ 
			0&0& \p_\mu\overline\Ga_1'^\sfb 
			&0&0&1
			&\p_\mu \overline H_1
		\\
			0&0&0
			&0&-1&0
			&0
	\\ \hline
			-\p_c \overline H_1&
			-\p_a \overline H_1&
			-\p_b \overline H_1
	& 
			-\p_\phi \overline H_1 & 
			-\p_\mu \overline H_1 &
			0
	&
	0
   \end{array}
   \right)
}
   \,.
	\notag
\end{align}
 Again, the same comments as the ones after equation (\ref{LagrRedOrd1}) are in place. Also, when computing $\sfmr_n$, formula (\ref{ExtDerivPractical}) is to be used to account for the derivative of the chosen basis of $1$-forms and for the cancellation of the derivatives of $\bfmr_g$. \\

 When solving the induction relation, the mechanism is the same as at order $\overline\Ga_0$ for $\overline\Ga_n^{\sfc,\sfa}$, and the same as at order $\overline\Ga_1$ for $\overline\Ga_n^{\sfb,\phi,t}$ and $\osc(\overline\Ga_n^\th)$. In addition, the new feature is the presence of $\overline\om_1$ for $\overline\Ga_n^{\mu}$ and $\avg(\overline\Ga_{n}^\th)$, but the mechanism is similar to what happens at order $\overline\Ga_2$, in the sense that the pivotal coefficients are the same. Mainly, the procedure relies on three conjugation-like relations: $(\sfc, \sfa)$ are conjugated for $\om_{-1}$; $(\mu, \th)$ are half-conjugated for $\overline\om_{1}$, in the sense that the structure is quarter-canonical; and $(\sfb, \phi)$  are half-conjugated for the symplectic part of $\overline\om_{0}$.
 
 The same procedure can be applied, which is reminded in the following tabular.
$$
	\begin{array}{l|@{~~}l@{~}l}
		\text{Equation} 
		& \text{Unknown} 
		& \text{Coefficient}
	\\ \hline
		\avg(\overline\Ga_n^\th) & \avg(\bfmg_{n-1}^\mu) & 1
	\\
		\osc(\overline\Ga_n^\th) & \osc (S'_n) & \p_\th 
	\\
		\overline\Ga_n^\phi & \bfmg_n^\sfb & -\sqrt{2\mu B}
	\\
		\overline\Ga_n^\mu & \bfmg_{n-1}^\th & -1
	\\
		\overline\Ga_n^\sfb & \bfmg_n^\phi & \sqrt{2\mu B}
	\\
		\overline H_n & \bfmg_n^\mu & B
	\\
		\overline\Ga_n^\sfc & \bfmg_{n+1}^\sfa & -B
	\\
		\overline\Ga_n^\sfa & \bfmg_{n+1}^\sfc & B
	\end{array}
$$

 Then, all the components of $\Ga_n$ remain free. This can be used to explore the various guiding-center representations at higher orders \cite{BurbySquir13}. Here, we are interested in a maximal reduction, which means to set them to zero, so that the optimal requirements are fulfilled. The freedom $\avg(S_n)$, which is a parameter in the formulae obtained for $\bfmg_{n-1}^\th$, $\bfmg_n^\sfb$, $\bfmg_n^\phi$, $\bfmg_n^\mu$, $\bfmg_{n+1}^\sfa$, and $\bfmg_{n+1}^\sfc$, cannot be useful and can be set to zero. \\

 In the same way as at orders $1$ and $2$, $\avg(\bfmg_n^{\mu})$ must remain free at that order, because it will be needed to solve $\overline\Ga_{n+1}^\th$ at the following order, just as $\avg(\bfmg_{n-1}^\mu)$ is needed here to solve $\overline\Ga_{n}^\th$. So, one cannot have the reduced Lagrangian fully simplified $\overline\Ga_n=0$. One of its component remains uncontrolled, either $\overline\Ga_n^t$ or $\overline\Ga_n^\sfb$.  
 
 One can consider recovering a zero value for this component by using the freedom $\avg(S_n)$, whose value had been arbitrarily  fixed to zero in the process above. Then, when computing $\bfmg_{n-1}^\th$, $\bfmg_n^\sfb$ and $\bfmg_n^\phi$, the average gauge function $\avg(S_n)$ remains a free parameter. When replacing these variables by their expressions, the equation $\avg(\overline H_n)=0$ becomes a differential equation for $\avg(S_n)$, whose structure is 
\bq
	\big[
	\phi \p_b
	+\tfrac{\na\cdott\sfb}{2}(1+\phi^2)\p_\phi
	\big]
	\avg(S_n)
	=
	o.t.
	\,,
	\label{SecularEqOrdN}
\eq
where $o.t.$ means other terms that can be explicitly computed. This resembles equation (\ref{SecularDiffEquation}), but here, the condition (\ref{S1gene}) has not been required in the process, contrary to what happened in previous orders. The reason is that now $\overline\Ga_n^\mu$ is controlled by $\bfmg_{n-1}^\th$, which has no effect on $\overline H_n$. The integrability of equation (\ref{SecularEqOrdN}) will depend on the right-hand side and must be studied at each order; a priori, it is not guaranteed, since obstructions such as (\ref{Obstruction}) are possible. 

 So, a systematic procedure cannot use the freedom $\avg(S_n)$ to get the additional requirement $\overline H_n=0$, which must be dropped. Then the freedom $\avg(S_n)$ is useless and can be set to zero. \\

 At the end of the $n$-th-order analysis, exactly all of the unknowns $\bfg_n$ have been determined. All the components of $(\bfmg_{n+1},\bfmg_n,\bfmg_{n-1})$ that remain unknown are in $\bfg_{i>n}$. Yet, this does not allow us to conclude that the induction is proven, because the unknown $\avg(\bfmg_n^\mu)$ already appeared as a parameter in $\bfmg_{n+1}^\sfa$ and in $\overline H_1$; hence it is not completely free, whereas the induction assumes it is free (independent of the quantities $\bfg_{i\leqslant n}$), since it is in $\bfg_{n+1}$; it will be determined at the next order, and this could imply coupled equations, whose solvability is to be verified. 
 
 However, $\avg(\bfmg_n^\mu)$ will be computed in equation for $\avg(\overline\Ga_{n+1}^\th)$, which corresponds to the column $\overline\Ga_n^\th$ in equation (\ref{LagrRedOrdN}) at the next order, and does not involve any of the parameter-dependent quantities. Thus, there is no coupled equations, and the solutions are indeed explicit. This terminates the proof of the induction: the reduction can be performed to arbitrary order in $\ep$.
 
 Notice that here, the induction relation is considered from the point of view of $\overline\Ga_n$; this caused an interlocking between the orders, where in the solution at each order, a parameter is involved, which will be identified at the next order, when computing $\overline\Ga_{n+1}^\bfp$. To avoid this interlocking phenomenon, it is possible to consider the induction relation from the point of view of $\Big( \overline\Ga_{n}^{\bfq,t}, \overline\Ga_{n+1}^\bfp \Big)$. The drawback would be that when solving the equations for $\overline\Ga_{n+1}^\bfp$, one would begin the heavy computations for $\sfmr'_{n+1}$, which are involved in $\overline\Ga_{n+1}^\bfq$, hence at the next order. In computations by hands, this can be a trouble, but when using computer-assisted computations, this is no trouble and it would probably be a more relevant choice.

\subsubsection{Explicit induction relations}

 The argument above emphasizes the distinction to be made between four kinds of quantities. First, some of the quantities are already known before the computation at order $n$, namely $\bfg_{k<n}$. 
 
 Second, for $i\not\in\{\sfb,t\}$ the components $\overline\Ga_n^i$ have not been computed yet, but they can be excluded both from the unknowns and from the parameters, since the algorithm shows that before any computation, they are already known to be zero for all $n \geqslant 3$, to fulfil the requirements (\ref{RequirPrior1})-(\ref{RequirPrior2}) for $\overline\Ga$. 
 
 Third, some quantities are not known yet, and will be determined after the matrix inversion, namely 
\begin{align}
(\bfg_{n})_\infty:=
			\Big(
				\avg(\bfmg_{n-1}^\mu) ,
				&
				\osc(S_n) 
			~|~ 
\label{DefVectBfgInftyText}
\\
				&
				\bfmg_n^\sfb ,
				\bfmg_{n-1}^\th ,
				\bfmg_{n+1}^\sfa ,
				\bfmg_{n+1}^\sfc 
			~| ~
\ncr
				&
				\avg(\bfmg_n^\phi) ,
				\osc(\bfmg_n^\phi) ,
				\osc(\bfmg_n^\mu) ,
				\overline H_n
			\Big)
			\,,
			\notag
\end{align}
in which a vertical dash $|$ was written at the places where a vertical line will be written in the matrix $\sfmm_\infty$ below. 

 Last, other quantities are not known and will remain free after the matrix inversion, namely $\overline\Ga_n^\sfb$, $\avg(S_n)$ and $\bfg_{k>n}$; the variables $\bfg_{k>n}$ will be determined at higher order, but one of its component, $\avg(\bfmg_n^\mu)$, is already involved in the equations at order $n$ and behaves as a parameter in this matrix inversion. So, the parameters are 
\begin{align}
(\bfg_{n})_\al:=
			\Big(
				&
				\avg(\bfmg_{n}^\mu) ,
				\avg(S_n) ,
				\overline\Ga_n^\sfb
			\Big)
			\,.
\label{DefVectBfgAlpha}
\end{align}
Notice that $\avg(\bfmg_{n}^\mu)$ is included in the parameters $(\bfg_{n})_\al$ even if it is not an element of $\bfg_n$ but of $\bfg_{n+1}$. \\

 With the procedure above, the left-hand side of equation (\ref{LagrRedOrdN}) can be written as just a matrix product, provided the pivotal matrix is extended, to act on all the quantities $(\bfg_{n})_\infty$ to be computed at this order, even the reduced Hamiltonian $\overline H_n$ and the gauge function $S_n$. To include the gauge function $S_n$ in the vector which is acted upon by the matrix, some coefficients in the matrix must be operators, and the equation will be transposed, so that the operators act on their right. For clarity, the order of the columns is chosen to fit with the steps of the algorithm
\begin{align}
	\Big(
	\avg(\overline\Ga_n^\th) ,
	\osc(\overline\Ga_n^\th) 
			~|~ 
	&
	\overline\Ga_n^\phi ,
	\overline\Ga_n^\mu ,
	\overline\Ga_n^\sfc ,
	\overline\Ga_n^\sfa 
			~|~ 
	\ncr
	&
	\avg(\overline\Ga_n^\sfb) ,
	\osc(\overline\Ga_n^\sfb) ,
	\osc(\overline\Ga_n^t) ,
	\avg(\overline\Ga_n^t) 
	\Big)
	\,,
	\notag
\end{align} 
so that the equations are solved one after the other in order. A vertical dash $|$ was written at the places where a horizontal line will be written in the matrix $\sfmm_\infty$ below. 

 With this order for the rows and for the columns, the equation becomes 
\bq
	\sfmm_\infty
	\cdott(\bfg_n)_\infty ^T
	+
	\sfmm_\al
	\cdott(\bfg_n)_\al ^T	
	+
	\sfmr'^T=0
	\,.
\label{InductionOrdN}
\eq
Here, the rows are the equations to be solved, corresponding to the (re-ordered) columns of equation (\ref{LagrRedOrdN}). The first term involves exactly the unknown quantities to be identified at this order. The second term involves exactly the parameters involved at this order but which will remain free at the end of this order. The third term involves only quantities that are already known at the beginning of this order. 

 The matrices are given by 
\bq
	\hspace{-8ex}
	\sfmm_\infty:=
	\left(
	\begin{array}{ c@{~}c  | cccc  | cccc }
		1 & 
		&  
		& 
		& 
		& 
		& 
		& 
		& 
		& 
	\\
		& \p_\th
		&  & 
		&  & 
		&  & 
		& & 
\\ \hline
		& \p_\phi
		& -\sqrt{2\mu B} & 
		&& 
		&&
		&& 
	\\
		& \p_\mu
		& -\phi\sqrt{\tfrac{B}{2\mu}}  & -1
		&&
		&&
		&&
	\\
		&\p_c
		& -I &
		&-B &
		&&
		&&
	\\
		&\p_a
		& K &
		&&B
		&&
		&&
\\ \hline
		\avg\p_\mu\overline\Ga_1^\sfb
		& \p_b
		&&
		&&
		&\sqrt{2\mu B}&
		&&
	\\
		\osc\p_\mu(\overline\Ga_1^\sfb)
		& \p_b
		&&
		&&
		&&\sqrt{2\mu B}
		& \phi\sqrt{\tfrac{B}{2\mu}} &
	\\
		\osc\p_\mu(\overline H_1) 
		&
		& -H_0 \na\cdott\sfb &
		&&
		&& 2\mu B\phi
		&\tfrac {H_0} \mu&
	\\
		\avg\p_\mu(\overline H_1)
		&
		& -H_0 \na\cdott\sfb &
		&&
		& 2\mu B\phi &
		&&-1
	\end{array}
	\right)
	\,,
	\label{MatrixExtPivotOrdN}
\eq
and
$$
	\sfmm_\al:=
	\left(
	\begin{array}{ c@{~}c@{~}c }
		0
		&0
		&0
		\\
		0
		&0
		&0
	\\ \hline
		&\p_\phi
		&
		\\
		&\p_\mu
		&
		\\
		&\p_c
		&
		\\
		&\p_a
		&
	\\ \hline
		\phi\sqrt{\tfrac{B}{2\mu}}
		&\p_b
		& -1
		\\
		&\p_b
		&
		\\
		0
		&0
		&0
		\\
		\tfrac{H_0}{\mu}
		&
		&
	\end{array}
	\right)
	\,,
$$
in which the zeros were written only in the empty rows, for clarity, and we used that $\na\cdott\bfmb=0$ implies $\tfrac{\p_b B}{B}=-\na\cdott\sfb$. \\

 In the matrix $\sfmm_\infty$, grouping together the eighth and ninth rows and columns produces a $9*9$ lower triangular matrix, whose coefficients on the diagonal are invertible, since the operator $-\sfa\cdott\p_\sfc$ is invertible on gyro-fluctuations. The eighth and ninth rows and columns have been grouped together because they constitute an invertible $2*2$ matrix on the diagonal. A convenient way to invert the resulting $9*9$ matrix is to separate its diagonal terms:
$$
	\sfmm=\sfmm'+\mathsf D
	\,,
$$
where $\sfmm'$ has null diagonal and $\mathsf D$ is purely diagonal. Then the equation can be written
$$
	-\mathsf D
	\cdott(\bfg_n)_\infty ^T
	=
	\sfmm_\infty'
	\cdott(\bfg_n)_\infty ^T
	+
	\sfmm_\al
	\cdott(\bfg_n)_\al ^T	
	+
	\sfmr'^T
	\,.
$$
The solution of equation (\ref{InductionOrdN}) is then 
\bq
	(\bfg_n)_\infty ^T
	=
	 (-\mathsf D)\mun\cdott
	 \Big[ 
	\sfmm_\infty'
	\cdott(\bfg_n)_\infty ^T
	+
	\sfmm_\al
	\cdott(\bfg_n)_\al ^T	
	+
	\sfmr'^T
	\Big]
	\,,
\label{InductionOrdNSol}
\eq
where $(-\mathsf D)\mun$ is a diagonal matrix with coefficients
\begin{align}
	(-\mathsf D)\mun
	:=
	\text{Diag}
	\Big(
	-1,-(\p_\th)\mun
	~|~
	&
	\tfrac{1}{\sqrt{2\mu B}},
	1, 
	\tfrac1B,-\tfrac 1B
	~|~ 
	\ncr
	&
	-\tfrac {1}{\sqrt{2\mu B}},
	-\mathsf D_2\mun, 
	1
	\Big)
	\,,
	\notag
\end{align}	
in which $\mathsf D_2\mun$ is the inverse matrix for the coupled system (the eighth diagonal term of the 9*9 matrix mentioned above) 
$$
	-\mathsf D_2\mun:=
	-
	\left(
     		\begin{smallmatrix}
		\sqrt{2\mu B}
		& \phi\sqrt{\tfrac{B}{2\mu}} 
	\\
		2\mu B\phi
		&B(1+\phi^2)
			\end{smallmatrix}
	\right)\mun
	=
	\left(
     		\begin{smallmatrix}
			-\tfrac{1+\phi^2}{\sqrt{2\mu B}}
			&\tfrac{\phi}{2\mu B}
			\\ 
			\phi\sqrt{\tfrac{2\mu }{B}}
			&
			-\tfrac 1B
			\end{smallmatrix}
	\right)
	\,.
$$

\medskip
 The operator $\p_\th\mun$ is the gyro-integral operator. It can be computed without introducing any gyro-gauge, with the intrinsic calculus introduced in \cite{GuilMagMom}, or with the matrix calculus introduced in \cite{GuilGCmin}. Also, the coordinate $\theta$ can be used as an intermediate quantity for this computation, which is made at constant $\bfq$, so that the presence of a gauge (only for the intermediate computation) is of no consequence; then $\p_\th\mun$ is the primitive with respect to $\th$ such that its gyro-average is zero. Over the Fourier modes $k\neq0$ (i.e. over gyro-fluctuations), it is the operator $\tfrac 1{ik}$. \\

 Even if $(\bfg_n)_\al$ appears in its right-hand side, formula (\ref{InductionOrdNSol}) is an explicit solution for the induction relation: in the matrix $\sfmm_\infty'$, all the coefficients on the diagonal or above it are zero, so that when computing the unknowns one after the other starting from the left, each of them is computed as a function of previously computed quantities, i.e. the right-hand side contains only known quantities or parameters, but none of the remaining unknowns. Alternatively, the induction can be solved using a standard matrix inverse 
\bq
	(\bfg_n)_\infty ^T
	=
	-
	\sfmm_\infty\mun\cdott
	\Big[ 
	\sfmm_\al
	\cdott(\bfg_n)_\al ^T	
	+
	\sfmr'^T
	\Big]
	\,,
\label{InductionOrdNSol2}
\eq
with $\sfmm_\infty\mun$ easily computed from (\ref{MatrixExtPivotOrdN}),  but the coefficients are more complicated, and practical computations usually follow the procedure of formula (\ref{InductionOrdNSol}).

 For the solution $\bfg_n$, formula (\ref{InductionOrdNSol}) or (\ref{InductionOrdNSol2}) must be completed by the following relations for the trivial components of $\overline\Ga_n$:
\begin{align}
	\overline\Ga_n^{i}&=0 
	\text{~~for all $i\not\in\{\sfb,t\}$}
	\notag
	\,,
\end{align}
and by the determination of the parameters 
$(\bfg_{n})_\al$.\\

 The first parameter $\avg(\bfmg_{n}^\mu)$ will be determined at the following order, in an equation that does not involve $\bfmg_n^\phi$ and $\overline H_n$, so that there are not coupled equations between the orders. 
 
 The second parameter $\overline\Ga_n^\sfb$ is either put to zero or adjusted so as to make the reduced Hamiltonian $\overline H_n$ zero by the relation
\begin{align}
	\overline\Ga_n^\sfb
	=
	&
	\avg\Big[
	\sfmr_n'^\sfb
	+
	\bfmg_{n-1}^{\mu} \p_\mu\overline\Ga_1'^\sfb
	+
	\p_b S_n
\label{ChoiceHn=0}
\\
	&
	-
	\tfrac{1}{\sqrt{2\mu B}\phi}
	\Big\lbrace
		\bfmg_{n-1}^{\mu} \p_\mu\overline H_1
		-
		\bfmg_n^\sfb H_0 \na\cdott\sfb 
		+
		B\bfmg_n^\mu
		+
		\sfmr_n'^t
	\Big\rbrace
	\Big]
	\,.
	\notag
\end{align}
This last choice is possible only if the term inside the parentheses has no overall contribution of order zero in $\phi$.

 The last parameter $\avg(S_n)$ is determined by equation (\ref{SecularEqOrdN}), in order to make both $\overline\Ga_n^\sfb$ and $\overline H_n$ zero, when the equation is integrable. Otherwise, it is determined to cancel as many terms of $\overline\Ga_n^\sfb$ and $\overline H_n$ as possible, or it can be set to zero for simplicity. 

 Accordingly, at the end of each order $n\geqslant 3$, the situation is the same as at the end of order $2$, with the presence of one parameter $\avg(\bfmg_{n}^\mu)$, of one binary choice between $\overline\Ga_n^\sfb$ and $\overline H_n$, and of one free variable $\avg(S_n)$. When the integrability condition can be satisfied, the reduction of the Lagrangian $\overline\Ga_n$ is complete and the solution is defined to within an element in the kernel of the operator (\ref{SecularEqOrdN}). Otherwise, $\avg(\bfmg_{n}^\mu)$ is generally set to zero, and after the binary choice the transformation is unique, but on the whole there are two maximal reductions. As announced in the appendix, the unicity of the transformation is determined by the integrability condition for $S_n$, and possibly by an additional criterion for simplicity.

	\section{Comparison with previous works}

\subsection{Agreeing results}
\label{SectElecField}

 Computations of the previous section for the guiding-center transformation and reduced Hamiltonian can be summarized by formulae (\ref{LarmorRadiusOsc}),  (\ref{ResultsOrderGa11})-(\ref{ResultsOrderGa12}) and (\ref{AvgG1mu})-(\ref{ResultsDeparametrized2}). As for the reduced symplectic Lagrangian, it is exactly known, and is given by formulae (\ref{LagrRedOrdZeroSol}) and (\ref{LagrRedOrd1Sol}), together with the prescription that all other terms $\overline\Ga_n^j$ are zero, except $\overline\Ga_{n\geqslant 2}^\sfb$ (as well as $\overline\Ga_1'^\sfb$), which can be freely chosen, for instance it can be chosen zero, or such that it makes the reduced Hamiltonian $\overline H_n$ zero by formula (\ref{ChoiceHn=0}). 
 
 These results agree with the standard results of the literature, provided the connection vector is defined as $\bfmr_g:=\bfmr$, which corresponds to the traditional gauge-dependent framework. For instance, in the paper \cite{Little83} Littlejohn made the choice $\overline \Ga_2=0$, and accounting for this choice, our formulae agree with his ones. In the paper \cite{CaryBriz}, the choice is $H_2=0$, and again, accounting for this choice through formula (\ref{LagrRedOrd1Alain}) for $\overline\Ga_1'^\sfb$, our results agree with the ones of \cite{CaryBriz}. Thus, the procedure of the previous section succeeds in obtaining the standard guiding-center reduction without introducing any gyro-gauge and using purely intrinsic coordinates. \\

 The explicit induction relation (\ref{InductionOrdNSol}) shows that the reduction is possible to arbitrary orders, but it also gives an explicit formula to practically compute the transformation generator order by order. At any order in the Larmor radius, all that remains to do is to develop the Lie derivatives involved in the term $\sfmr_n'$. Only the number of terms generated by the Leibniz rule makes the process difficult to compute by hand at higher order, but the explicit induction involves few basic operations (just exterior derivatives and matrix products) and can easily be implemented to higher orders on a computer. Actually, as the series is a polynomial, the derivation does not rely on formal calculus but just on symbolic calculus, which is still easier to implement. \\

 Once the generators $\bfmg_n$ are obtained, the guiding-center coordinate transformation is given by 
\begin{equation}
	\bfz\longrightarrow \overline \bfz:=...e^{-\bfmg_2}e^{-\bfmg_1}\bfz 
	\,.
\label{TransfCoord}
\end{equation}

 The reduced Lagrangian $1$-form is 
\begin{align}
	\overline\Ga=\avg(\Ga)
	&
	-\mu\sfa\cdott d\sfc
	+\Big[
	\overline\Ga_1'^\sfb
	+ \sum_{n\geqslant2}\overline\Ga_n^\sfb
	\Big]
	\sfb\cdott d\bfq
	-
	\sum_n\overline H_n dt
\label{LagrRedTotalSolution}
\\
	&=
	(\bfma+\overline\Ga^\sfb \sfb)\cdott d\bfq
	- \mu\sfa\cdott d\sfc
	- \overline H dt
	\,,
\notag	
\end{align}
where the Hamiltonian terms $\overline H_n$ are provided by (\ref{InductionOrdNSol}), whereas the parallel Lagrangian terms $\overline\Ga_1'^\sfb$ and $\overline\Ga_n^\sfb$ are chosen freely at each order $n$, and can be chosen zero.

 The reduced dynamics is obtained the usual way, by computing the Lagrange matrix $\overline\om_s:=d\overline\Ga_s$, then inverting it to get the Poisson matrix $\overline\bbj:=\overline\om_s\mun$, and last computing Hamilton's equations $\dot{\overline\bfz^i}:=\bbj^{ij}\p_j \overline H$. Alternatively, the reduced equations of motion can be obtained by Lie transforming directly the velocity vector field   
$$
	\dot\bfz\longrightarrow \dot{\overline\bfz}
	:=...e^{\sfml_2}e^{\sfml_1}\dot\bfz 
	\,.
$$

 Here, performing these computations is useless, since the reduced Lagrangian (\ref{LagrRedTotalSolution}) completely agrees with previous results in the literature in the gauge-dependent case, and so will the reduced motion. \\

 We only indicate the guiding-center Poisson bracket, in which the effects of the gauge independence and of the higher-order corrections are interesting. It is computed from the Lagrange $2$-form, easily obtained from Eq.~(\ref{LagrRedTotalSolution}) in a matrix form:
$$
	\overline\om:=d\overline\Ga
	=
	\left(
	\begin{array}
				{ c@{\hspace{1ex}}
				c@{\hspace{1ex}}
				c@{\hspace{2ex}}  
				| @{\hspace{2ex}}
				c@{\hspace{1ex}}
				c@{\hspace{2ex}}
				c@{\hspace{2ex}}  
				| @{\hspace{2ex}}c@{\hspace{1ex}} }
			0&-\mathtt B&\mathtt A
			&0&\mathtt x&0
			&\sfc\cdott\na\overline H
		\\ 
			\mathtt B &0&-\mathtt C
			&0&\mathtt y&0
			&\sfa\cdott\na\overline H
		\\
			-\mathtt A&\mathtt C&0
			&-\mathtt D&\mathtt z-\mathtt E&0
			&\sfb\cdott\na\overline H
	\\ \hline
			0&0&\mathtt D
			&0&0&0
			&\partial_\phi \overline H
		\\ 
			-\mathtt x&-\mathtt y& \mathtt E-\mathtt z 
			&0&0&1
			&\partial_\mu \overline H
		\\
			0&0 &0
			&0&-1&0
			&0
	\\ \hline
			-\sfc\cdott\na\overline H&-\sfa\cdott\na\overline H&-\sfb\cdott\na\overline H
			&-\partial_\phi \overline H&-\partial_\mu \overline H&0
			&0
   \end{array}
   \right)
\,,
$$
with 
\begin{align}
	\mathtt A &:= -\sfa\cdott\na\x 
					( \bfma + \overline\Ga^\sfb\sfb)
				-\mu \sfc\cdott\na\sfb\cdott\sfb\x\sfb\apos\sfb
	= I + O(B^0)
	\,,
	~~~~~~
	& \mathtt x := \bfmr_g\cdott\sfc
	\,,
\ncr
	\mathtt B &:= -\sfb\cdott\na\x 
					( \bfma + \overline\Ga^\sfb\sfb)
				-\mu \sfa\cdott\na\sfb\cdott\sfb\x\sfb\apos\sfc
	= - B + J + O(B^0)
	\,,
	~~~~~~
	& \mathtt y := \bfmr_g\cdott\sfa 
	\,,
\ncr
	\mathtt C &:= -\sfc\cdott\na\x 
					( \bfma + \overline\Ga^\sfb\sfb)
				-\mu \sfb\cdott\na\sfb\cdott\sfb\x\sfb\apos\sfa
	= K + O(B^0)
	\,,
	~~~~~~
	& \mathtt z := \bfmr_g\cdott\sfb 
	\,,
\ncr
	\mathtt D &:= \partial_\phi \overline\Ga^\sfb
	=\sqrt{2\mu B}+O(B^0)
	\,,
\label{CoeffLagrMatrRed}
\\
	\mathtt E &:= \partial_\mu \overline\Ga^\sfb 
	= \phi\tfrac{\sqrt{2\mu B}}{2\mu}+O(B^0) 
	\,,
\notag
\end{align}
where we used the fact that the order in $\sqrt B$ indicates the expansion order, as mentioned about Eq.~(\ref{Scaling}). 

 Then the Poisson bracket in matrix form writes:
$$
	\overline{\mathbb J}:= (\overline\omega^{\bfz;\bfz})^{-1}
	=
	\left(
	\begin{array}
				{ @{\hspace{1ex}}
				c@{\hspace{2ex}}
				c@{\hspace{2ex}}
				c@{\hspace{2ex}}  
				| @{\hspace{2ex}}
				c@{\hspace{3ex}}
				c@{\hspace{3ex}}
				c@{\hspace{2ex}}  
				}
			0&\frac{1}{\mathtt B}&0
			&\frac{\mathtt C}{\mathtt B\mathtt D}&0
			&\frac{\mathtt y}{\mathtt B}
		\\ 
			-\frac{1}{\mathtt B} &0&0
			&\frac{\mathtt A}{\mathtt B\mathtt D}&0
			&-\frac{\mathtt x}{\mathtt B}
		\\
			0&0&0
			&\frac{\mathtt B}{\mathtt B\mathtt D}&0&0
	\\ \hline
			-\frac{\mathtt C}{\mathtt B\mathtt D}
			&-\frac{\mathtt A}{\mathtt B\mathtt D}
			&-\frac{\mathtt B}{\mathtt B\mathtt D}
			&0&0&-\alpha
		\\ 
			0&0& 0
			&0&0&-1
		\\
			-\frac{\mathtt y}{\mathtt B}
			&\frac{\mathtt x}{\mathtt B}
			&0
			&\alpha&1&0
   \end{array}
   \right)
\,,
$$
with
$$
	\alpha:=
		-\frac{\mathtt E}{\mathtt D}
		+	
		\frac
		{\mathtt C\mathtt x + \mathtt A \mathtt y 
							+ \mathtt B \mathtt z}
		{\mathtt B\mathtt D} 
		=
		-\frac{\mathtt E}{\mathtt D}
		+	
		\frac
		{\bfmb_*\cdott\bfmr_g}
		{\mathtt B\mathtt D} 
		\,,
$$
where the reduced magnetic field $\bfmb_*$ is defined as usual by formula (\ref{Bstar}). The coefficients $\mathtt B$ and $\mathtt D$ are invertible since Eq.~(\ref{CoeffLagrMatrRed}) shows that they are small corrections from $-B$ and $\sqrt{2\mu B}$, which are invertible. 

 As a result, between two arbitrary functions of the reduced phase space $f (\overline z)$ and $g (\overline z)$, the Poisson bracket is
\begin{equation}
	\lbrace f, g\rbrace 
	=
	- \na_* f \cdot 
	\frac{\sfb}{\mathtt B} \x \na_* g
	+
	\frac{\bfmb_*}{\mathtt B\mathtt D} \cdot \na_* f \wedge
	\partial_\phi g
	+
	\frac{\mathtt E}{\mathtt D} \partial_\phi f \wedge \partial_\theta g
	+
	\partial_\theta f \wedge \partial_\mu g
	\,,
\label{PoissBrackRed}
\end{equation}
where the symbol $\na_*$ is a shorthand for the operator
\begin{equation}
	\na_*:=\na+\bfmr_g\partial_\theta
	\,.
\label{NablaStar}
\end{equation}
This operator does not depend of the free function $\bfmr_g$, since for any choice of $\bfmr_g$ the definition (\ref{NablaStar}) gives the same result, which is equal to the covariant derivative $\na$ when the gauge vector is chosen zero, i.e. $(\bfmr_g)_*=\na_*\sfc\cdott\sfa=0$.

 The Poisson bracket (\ref{PoissBrackRed}) agrees with the literature, e.g. \cite{CaryBriz, BrizHahm07, Little81, Little83, ParrCalv11}. Especially, the traditional ordering is patent: the last two terms are of order $\mu^{-1}$, and correspond to the fast gyro-angle dynamics; the second term is of order   $\tfrac{\na}{\sqrt{\mu B}}=\mu^{-1}\epsilon$, and corresponds to the intermediate motion along the  magnetic field lines; as for the first term, it is of order $B^{-1}\na^2=\mu^{-1}\epsilon^2$     , and corresponds to the slow drifts across the magnetic field lines. Remind $\epsilon:=r_L\na=\sqrt{\tfrac{\mu}{B}}\na$, as defined in (\ref{DefinitEpsilon}), is the usual small parameter of guiding-center theory. 
 
 In previous works, the reduced Poisson bracket often had only three coefficients (the ones for $\bfmb_*$, namely $\mathtt A$, $\mathtt B$, and $\mathtt C$), either because of incidental lowest-order simplifications, or because of the choice they had performed for $\overline\Ga^\sfb$. For a general result about maximal reductions at arbitrary order, the Poisson bracket is given by Eq.~(\ref{PoissBrackRed}), where the higher-order correction $\overline \Ga_{n\geqslant 1}^\sfb$ to the Lagrangian impacts five coefficients $\mathtt A$, $\mathtt B$, $\mathtt C$, $\mathtt D$, and $\mathtt E$ (and hence $\bfmb_*$), through the definitions (\ref{CoeffLagrMatrRed}), i.e. as an effect of the five derivative operators $\partial_\sfc$, $\partial_\sfa$, $\partial_\sfb$, $\partial_\phi$, $\partial_\mu$, acting on the seminal term $\overline \Ga_{n\geqslant 1}^\sfb$. \\

 The only difference compared to previous results, besides the coordinate system and the term $\overline\Ga_{n\geqslant 1}^\sfb$ being let free, is that the gauge vector $\bfmr=\na\sfe_1\cdott\sfe_2$ is replaced by the general connection $\bfmr_g=\na\sfc\cdott\sfa$. This difference could impact the reduced magnetic field $\bfmb_*$, given by 
\bq
 \bfmb_*
	:=
	-\mathtt A\sfa - \mathtt B\sfb - \mathtt C \sfc
	=
	\na\x
 \Big(
 	\bfma
 	+
 	\sfb
 	\overline\Ga^\sfb
 \Big)
 +
 \mathbf V
 \,,
\label{Bstar}
\eq
where the vector $\mathbf V$ is generated by the term $-\mu\sfa\cdott d\sfc$ in the Lagrangian $\overline\Ga$, which implies for the Lagrange $2$-form the presence of the term
\begin{align}
	-\mu ~ d\sfa\cdott \wedge d\sfc 
	&
	=
	-\mu ~ d\sfa\cdott
	(\sfa\sfa+\sfb\sfb+\sfc\sfc) \cdott
	\wedge d\sfc
\ncr
	&
	=
	-\mu ~ d\sfa\cdott
	\sfb
	\wedge
	\sfb \cdott
	d\sfc
\ncr
	&
	=
	-\mu ~ d\sfb\cdott
	\sfa
	\wedge
	\sfc \cdott
	d\sfb
\ncr
	&
	=
	-\mu ~ 
	d\bfq\cdott\na\sfb
	\cdott
	\sfa
	\wedge
	\sfc\cdott
	\sfb\apos d\bfq
\ncr
	&
	=
	-\mu~ 
	d\bfq\cdott
	\na\sfb\cdott\sfb\x\sfb\apos
	d\bfq
\ncr
	&
	=:
	-d\bfq\cdott\mathbf V\x d\bfq
	\,.
\label{CurlGaugeVectIntrin}
\end{align} 
The first equality comes by inserting the identity $(\sfa\sfa+\sfb\sfb+\sfc\sfc)$ beside the wedge symbol. The second and third equalities come because $(\sfa,\sfb,\sfc)$ is orthonormal. The fact that the vector $\sfb$ depends only on $\bfq$ implies the fourth equality, which can be rewritten to get the final answer. 

 Eq.~(\ref{CurlGaugeVectIntrin}) shows why the vector $\mathbf V$ (and hence also the reduced Poisson bracket) is indeed independent of the free function $\bfmr_g$. In the reduced Lagrange $2$-form, the concerned term (\ref{CurlGaugeVectIntrin}) is the counterpart in the gauge-independent approach of the term $d\bfq\cdot(\mu\na\x\bfmr)\x d\bfq$ involving the gauge vector in the traditional approach. Actually, the curl of the gauge vector $\na\x\bfmr$ is also gauge-independent, and the corresponding term is explicitly given by  
\begin{align}
	-\mu  ~ d(d\bfq\cdott\na\sfe_1\cdott\sfe_2)
	&
	=
	-\mu~ d\bfq\cdott\na\sfe_2\cdott\wedge\sfe_1\apos d\bfq
	\ncr
	&
	=
	-\mu~ d\bfq\cdott\na\sfb\cdott\sfb\x\sfb\apos d\bfq
	\,,
	\label{CurlGaugeVect}
\end{align}
where again, the second equality comes by inserting the identity $(\sfb\sfb+\sfe_1\sfe_1+\sfe_2\sfe_2)\cdott$ beside the wedge symbol and by using the fact that $(\sfb,\sfe_1,\sfe_2)$ is orthonormal. 

 So, in the reduced magnetic field $\bfmb_*$, the term with $\bfmr_g$ exactly fits with the corresponding term with $\bfmr$ of previous results. This is a good illustration of how the approach using the coordinate $\sfc$ completely agrees with the gauge-independent part of the usual results, whereas it proceeds in a different way and never introduces the gauge $\sfe_1$. Indeed, in the usual approach, $\na\sfe_1\cdott\sfe_2$ is assumed to depend only on the position, which implies the formula above. On the contrary, in the gauge-independent approach, the velocity is present in the first lines of (\ref{CurlGaugeVectIntrin}), but it comes out from the computation that the result is naturally just a purely spatial term in the Lagrange matrix.
 
 By the way, the argument above shows that $\mathbf V$ is indeed the curl of $-\mu\bfmr_g=-\mu\na\sfc\cdott\sfa$, but it is not surprising since the spatial part of a term $df\wedge dg$ is given by the cross product with the curl of $-f \na g $: 
\begin{align}
	d\bfq\cdott(df\wedge dg)^{\bfq;\bfq}\cdott d\bfq
	&
	=
	d\bfq\cdott\na f \wedge g\apos d\bfq
	\ncr
	&
	=
	d\bfq\cdott(\na(f
	g\apos)) \wedge d\bfq
	\ncr
	&
	=
	-
	d\bfq\cdott(\na\x(f g\apos)) \x d\bfq
	\,.
\notag
\end{align} 

 Thus, formula (\ref{Bstar}) can be rewritten 
$$
 \bfmb_*
:=\na\x
 \Big[
 	\bfma
 	+
 	\sfb
 	\Big\lbrace
 		\sqrt{2\mu B}\phi
 		+
 		\overline\Ga_1'^\sfb
 		+
 		\sum_{n\geqslant2} \overline\Ga_n^\sfb
 		-
 		\mu\bfmr_g
 	\Big\rbrace
 \Big]
 \,,
$$
which is manifestly divergenceless.\\

 Here, for clarity and following previous works, we considered no electric field, in order to focus on the reduction mechanism. The extension for a non-zero electric field is straightforward, because the Lagrangian is changed only by the addition of \label{LagrElecField}$-e\Phi dt$, with $\Phi(\bfq)$ the electric potential \cite{CaryBriz, BrizHahm07, Little83}. It affects the spatial part of $\p_\bfz H$, which plays no pivotal role in the derivation: it always appears in the right-hand side of the equations, i.e. in terms that are already known. As a consequence, the presence of this term does not change the procedure at all. In the results, each term of order $n$ becomes a polynomial in $\Phi$ (or rather $\na\Phi$), which introduces a second parameter $\ep_E$ in the theory, which corresponds to $\tfrac{mE}{pB}$ or to $\tfrac{mE\na}{eB^2}$. A more detailed study shows that the momentum in denominators is only the perpendicular momentum $\norm{\bfp_\perp}$, and that at the lowest orders the perpendicular electric field can be one order higher than the parallel electric field \cite{BrizHahm07, Little81, Little83}. The series remains perturbative provided the associated parameter is small $\ep_E<<1$, as usual in guiding-center reductions.

\subsection{Polynomiality in the momentum coordinates}

 In the results above, the usual expansions are recovered, but the choice of the coordinates $\sfc$ and $\phi$ makes all quantities polynomial in the used coordinates and monomial in $\sqrt \mu$ and $\sqrt B$. This is useful to simplify the derivation, which can be considered as a symbolic-calculus algorithm based on just two operations acting on words (polynomials) composed from a very restricted alphabet. Such structures can also be useful when going beyond the formal expansions and considering them as asymptotic series. For instance, the polynomiality in the cotangent of the pitch-angle is important to control the loss of accuracy of the expansion in the domains where the direction of the particle momentum is close to the direction of the magnetic field. 
 
 Previous results were obviously polynomial in the variable $\sfc$, but they did not use it as a coordinate, since they replaced it by the variable $\th$. In addition, the expansion in the usual small parameter $\ep:=\sqrt{2\mu /B}\na:=r_L\na$ was also present, but it did not correspond to an expansion in $\mu$ nor in $B$, but only in $\na$, which is not a scalar quantity; it is why keeping the quantity $e^{-1}$ was useful to indicate the expansion order. Here, the order is directly indicated by the power in $\sqrt\mu$ or $\sqrt B$, since each order is a monomial in those quantities. 
 
 The monomiality in $\sqrt \mu$ is especially interesting, because the orders in the operator $\na$ and in the function $B$ have only a dimensional meaning: the term of order $\ep^n$ will involve terms like $\left(\tfrac{\na B}{B}\right)^n$, but also terms like $\tfrac{\na^{n-k} B}{B}\tfrac{\na^k B}{B}$, or $\na^n \sfb$. In a similar way, the order in $B$ is given by the order of the prefactor $r_L^n=\sqrt{2\mu /B}^n$, but the variable $B$ appears in other places when it is acted upon by gradients; then, it is compensated by a $B$ in the denominator, which means that gradients act only on the variable $\ln B$. \\

 The use of the variable $\phi$ instead of the usual $v_\parallel$ is crucial for the results at each order to be both polynomial in the coordinates and monomial in $\sqrt\mu$ and $\sqrt B$. It is a generalization of \cite{GuilGCmin}, which considered only the minimal guiding-center reduction. It seems it had not been noticed in previous works on the full guiding-center reduction \cite{CaryBriz, Little81, Little83, ParrCalv11}. 
 
 When using the standard variable $v_\parallel$ instead of $\phi$, the monomiality in $\sqrt\mu$ and $\sqrt B$ is not verified: for instance, in $\bfmg_1^\th$, the term of order $\phi^2$ writes $-\tfrac{v_\parallel^2}{B\sqrt{2\mu B}}\cbb$, which is not a polynomial in $\sqrt \mu$ and which is not of the same order in $\sqrt\mu$ nor in $\sqrt B$ as the term of order $\phi^1$, since this last writes $\tfrac{v_\parallel}{B}\tfrac{\aba-\cbc}{4}$, as is confirmed in \cite{CaryBriz, Little83}, for instance. 
 
 On the contrary, when using the variable $\phi$, the polynomiality is verified, because of the structure of the Lagrangian, of the action of derivatives, especially (\ref{ConnectBoth}), and of the coefficients to be inverted for the matrix inversions involved in the derivation. Notice that the induction procedure also guarantees that all formulae will be polynomial in the variable $\sfb$. 
 
 As for the monomiality, it is easily understood from a dimensional analysis: only three independent dimensional quantities are involved in the results, e.g. $\na$, $\mathbf B$ and $\mathbf p$. When using the momentum coordinates $(\phi,p,\sfc)$ or $(\phi,\mu,\sfc)$, two of them are dimensionless, and only one of them can generate the dimension of $p$, namely $p$ or $\sqrt\mu$. For an expansion in $r_L\na=\sqrt{\tfrac{\mu}{B}}\na$, the variable $\mu$ can be involved only in the pre-factor of each order, otherwise, it could not be compensated to generate a dimensionless quantity. 
 
 On the other hand, when using the momentum coordinates $(v_\parallel,p,\sfc)$ or $(v_\parallel,\mu,\sfc)$, there is a redundancy in dimension between $v_\parallel$ and $p$ (or $\sqrt{\mu B}$), which means that ratios of them are expected in order to get dimensionless quantities. In an expansion in $\ep$, if the term of order $(r_L\na)^n$ is a polynomial in $v_\parallel$, it has to be actually a polynomial in $\tfrac{v_\parallel}{p}$ or $\tfrac{v_\parallel}{\sqrt{\mu B}}$. As a consequence, formulae will be sums of terms $(\tfrac{p\na}{B})^n\left(\tfrac{v_\parallel}{p}\right)^j$, which is not a monomial in $p$ (or equivalently in $\sqrt\mu$ and $\sqrt B$); by the way, it is not a polynomial either, because of the terms where $j>n$.
 
 Last, the polynomiality in $\phi$ means that each term writes $(\tfrac{p\sin\varphi\na}{B})^n\left(\tfrac{p\cos\varphi}{p\sin\varphi}\right)^j=\left(\tfrac{p\na}{B}\right)^n\cos^j\varphi\sin^{n-j}\varphi$, which agrees with the idea that the entities $\cos\varphi$ and $\sin\varphi$ come from expansions of the momentum $\bfp$, or rather the corresponding dimensionless vector $\tfrac{\bfp}{p}$.

\subsection{A two-fold maximal reduction}

 In addition to the averaging reduction and the inclusion of the magnetic moment among the reduced coordinates, the goal was to obtain a reduced dynamics as strongly reduced as possible. So, a complete achievement is obtained when all the components of $\overline \Ga_n$ are zero, at least for higher orders. For the procedure, it means using the seven unknowns to solve the seven corresponding equations, or rather twice these numbers, if one considers the average and the fluctuating parts as different variables. It did work for all the requirements but one, which could not be satisfied and had to be dropped. 

 The obstruction for a complete reduction $\overline\Ga_{n \geqslant  3}=0$ comes from the requirement (\ref{RequirPrior2}): it imposes to obtain  $\overline\Ga_n^\th$ by fixing the freedom $\avg(\bfmg)$, which was the only freedom available for the equation $\overline\Ga_t=0$ and is no more available for it. It is why one of the requirements has to be dropped. Then, it remains more unknowns than requirements. So, the transformation is not unique. Especially, while it naturally appeared in $\overline H$, the non-zero component of $\overline\Ga$ can be transferred to $\overline \Ga^\sfb$. Thus two maximal reductions can be considered. \\

 The first alternative (called the \textsl{Hamiltonian} representation in \cite{BrizTron12}) sets $(\overline\Ga_s)_{n\geqslant 2}=0$. Then, the reduced Poisson bracket is completely known before computing the transformation to higher orders; it is given by the lowest three orders of the symplectic Lagrangian $(\overline\Ga_s)_{n\leqslant1}$. The reduced Hamiltonian is not exactly known; it is given by a whole series in $\ep$ and must be computed order by order. The reduced dynamics is a Hamiltonian perturbation of the guiding-center equations of motion at order $2$. 
 
 The second alternative (called the \textsl{symplectic} representation in \cite{BrizTron12}) is to set $\overline H_{n \geqslant 1}=0$. Then, the Hamiltonian is completely known, and the structure of the Poisson bracket is also known. The only unknown information on the reduced dynamics is concentrated in the component of the reduced Lagrangian parallel to the magnetic field $\overline\Ga_n^\sfb$, which is given by a whole series in $\ep$. The Poisson bracket includes a kind of reduced magnetic field $\bfmb_*$, induced by the higher-order terms of the Lagrangian. 
 
 The choice of symplectic or Hamiltonian representation can be made at each order in the derivation, but it seems more convenient to be consistent and to make the same choice for all orders, as suggested in \cite{BrizTron12}. \\

 These two maximal reductions give a unified view of various choices that can be found in the literature, and they anticipate what will happen at higher orders. Indeed, even in the standard non-canonical Hamiltonian approach of the guiding-center reduction introduced by Littlejohn, several transformations can be found, often related by differences of choice related to this two-fold maximal reduction. For instance, Littlejohn's initial guiding-center reduction \cite{Little81} corresponded to the second possibility above at order $n=1$, but at higher order, it is unclear whether the procedure provided a maximal reduction, or if some terms could remain both in the Hamiltonian and in the Poisson bracket. The seminal reduction by Lie transforming the Lagrangian \cite{Little83} corresponded to a maximal reduction with the first choice ("Hamiltonian representation"). 
 
 Later papers by Lie transforming the Lagrangian turned to the other choice, e.g. \cite{CaryBriz} used a maximal reduction (at order $n=1$) but with the second possibility. Recently, while improvements in the second order were addressed \cite{BrizTron12, ParrCalv11}, interest was renewed in the first possibility. The paper \cite{ParrCalv11} actually corresponds to a mixed choice, where the second possibility is used at order $n=1$ and the first one is used at order $n=2$. The work \cite{BrizTron12} introduced the designation of \textsl{Hamiltonian} and \textsl{symplectic} representation to differentiate between the two choices, and studied an "equivalence relation" between the two choices (when the same choice is made for all orders), which is a way to go from one representation to the other by a redefinition of the reduced parallel momentum $\overline p_\parallel$ (or equivalently of $\overline\phi$).\\

 The algorithm at higher order, with the condition (\ref{ChoiceHn=0}) for the symplectic representation, had not been studied in detail. In the previous section, the Hamiltonian representation appeared as indeed guaranteed at arbitrary order, and as naturally provided by the procedure as a maximal guiding-center reduction, whereas the symplectic representation appeared as submitted to a condition at each order in $\ep$: in formula (\ref{ChoiceHn=0}), the term inside the bracket must have no overall contribution of order $0$ in $\phi$. This could explain why first papers addressing both the first and the second order in $\ep$ systematically used the Hamiltonian representation; the symplectic representation was used only later, when the condition was observed as verified.
 
 The equivalent relation introduced in \cite{BrizTron12} relies on a relationship between these representations, which allows to go from one to the other. Hence it might seem that it guarantees the existence of the symplectic representation, but it is not the case.
 
 The underlying idea (see equation (17) in \cite{BrizTron12}) is the following. Start from the reduced Lagrangian written in symplectic representation (we use here the variable $p_\parallel$ instead of $\phi$, in order to agree with the notation used in \cite{BrizTron12}):
\begin{align}
	\overline H
	&
	=
	\mu B + \tfrac{\overline p_\parallel^2}{2}
	\,,
\ncr
	\overline\Ga^\sfb
	&
	=
	\overline p_\parallel 
	+
	\sum_{n\geqslant 1} (\Pi_\parallel)_n 
	\notag
	\,.
\end{align}
 Then, redefine the reduced  parallel momentum so as to absorb all the series $\overline\Ga^\sfb$ in it:
\bq
	\overline p_\parallel'
	:=
	\overline p_\parallel 
	+
	\sum_{n=1} (\Pi_\parallel)_n
	\,.
	\label{TransformSymplectToHamil}
\eq
With this coordinate, the symplectic part of the Lagrangian is fully reduced: 
$$
	\overline\Ga^\sfb
	=
	\overline p_\parallel'
	\,.
$$
To obtain the reduced Hamiltonian with this coordinate, one just inverts the series (\ref{TransformSymplectToHamil}):
\bq
	\overline p_\parallel
	:=
	\overline p_\parallel' 
	-
	\sum_{n\geqslant 1} (\Pi_\parallel)_n 
	=
	\overline p_\parallel' 
	+
	\sum_{n\geqslant 1} (\Pi_\parallel')_n 
	\,,
	\label{SeriesInverseRegular}
\eq
with $(\Pi_\parallel')_n$ some coefficients easily obtained by inserting iteratively the first equality  in the occurrences of $\overline p_\parallel$ in $\sum_{n=1} (\Pi_\parallel)_n$, as is standard to invert a near-identity series. 

 Then the reduced Hamiltonian in the new coordinate writes 
$$
	\overline H
	=
	\mu B + \tfrac{1}{2}
	\big[
		\overline p_\parallel' 
		+
		\sum_{n\geqslant 1} (\Pi_\parallel')_n 
	\big]^2
	\,.
$$
It is a full series in $\ep$, which corresponds to the Hamiltonian representation. This is a constructive procedure showing that when the symplectic representation exists, then the Hamiltonian representation exists and is easily obtained.\\

 Now, what is actually needed is to go in the reverse direction, since the derivation of the guiding-center reduction shows that the Hamiltonian representation is natural and guaranteed to exist, whereas the symplectic is suspected of having existence conditions.

 It turns out that the procedure in the reverse direction can break down. Start from the reduced Lagrangian written in Hamiltonian representation:
\begin{align}
	\overline H
	&
	=
	\mu B + \tfrac{\overline p_\parallel^2}{2}
	+
	\sum_{n\geqslant 1}
	\overline H_n
	\,,
\ncr
	\overline\Ga^\sfb
	&
	=
	\overline p_\parallel 
\label{PparSympRepr}
	\,.
\end{align}
 Then, redefine the reduced  parallel momentum so as to absorb all the higher-order terms $\sum_{n\geqslant 1}\overline H_n$ in the term with $\overline p_\parallel$:
\bq
	\overline p_\parallel'^2
	:=
	\overline p_\parallel^2 
	+
	2 \sum_{n\geqslant 1} \overline H_n
	\,.
	\label{TransformHamiltToSymplect}
\eq
With this coordinate, the Hamiltonian part of the Lagrangian is fully reduced: 
$$
	\overline H
	=
	\mu B + \tfrac{\overline p_\parallel'^2}{2}
	\,.
$$
To obtain the reduced symplectic Lagrangian with this coordinate, one just inverts the near-identity transformation (\ref{TransformHamiltToSymplect}) by writing it as:
\bq
	\overline p_\parallel'
	:=
	\pm 
	\sqrt
	{
	\overline p_\parallel^2 
	+
	2 \sum_{n\geqslant 1} \overline H_n
	}
	=
	\overline p_\parallel
	\sqrt 
	{ 
	1
	+
	\tfrac{2 \sum_{n\geqslant 1} \overline H_n}{\overline p_\parallel^2}	
	}
	\label{SeriesInverseSingular}
	\,,
\eq 
and then by expanding the term $\sqrt{1+\ep}$. This assumes that the ratio $\tfrac{2 \sum_{n=1} \overline H_n}{\overline p_\parallel^2}$ is small. The point is that this condition is not guaranteed a priori, even if the series is near-identity. 
 
 Indeed, the derivations here are formal. "Near-identity" has only a dimensional meaning. It means that the ratio between the first and zeroth-order term is of order $\ep=r_L\na $, but only in dimension, its value might not to be small if it is multiplied by a large dimensionless factor such as $1/\cos\varphi$, as in (\ref{SeriesInverseSingular}). Thus, a division by $\overline p_\parallel$ can cause a singularity. 

 When going from the symplectic to the Hamiltonian representation, no such a division was needed, since the series inversion (\ref{SeriesInverseRegular}) just consisted in composing series. On the contrary, when starting from the Hamiltonian representation, the series inversion (\ref{SeriesInverseSingular}) involves a division by $\overline p_\parallel^2$. This causes a singularity if $\overline H_n$ contains a term of order $0$ or $1$ in $\overline p_\parallel$. 
 
 It is interesting to see that difficulties arise here at $\overline p_\parallel=0$ (or equally at $\phi=0$), which is precisely where they appeared in the guiding-center reduction in the previous section. This suggests that singularities in $\phi=0$ are indeed a difficulty for the symplectic representation. Accordingly, at each order, it can be used only when the absence of singularity in (\ref{ChoiceHn=0}) is verified.

\subsection{Gyro-gauge independence}

 The intrinsic formulation of the guiding-center reduction was motivated by questions about the traditional gyro-angle variable $\th$. The derivation with the coordinate $\sfc$ shows that it does succeed in shedding light on those questions.
 
 First, the traditional coordinate was a detour. In all guiding-center works, the variable $\th$ never appears in itself (except in its own definition and subsequent relations); for instance, it does not appear explicitly in one of the components of $\bfmg$ or $\overline\Ga$, which all depend on $\th$ only through the corresponding physical quantity $\sfc$ (or $\sfa:=\sfb\x\sfc$); even the gyro-angle component of the generator $\bfmg^\th$ verifies this statement. The detour is not given by the physics, since it imposes to fix arbitrarily a gauge $\sfe_1(\bfq)$, which is not related to the physics of the problem. The role of the intrinsic approach was to avoid this detour, and it achieves its goal since it obtains the full guiding-center results without introducing any gauge and by working purely with $\sfc$. 

 From a mathematical point of view also, the use of the variable $\th$ was not completely satisfactory, because the gyro-angle corresponds to a circle bundle \cite{Little88,BurbQin12,Krus62}. The traditional coordinate $\th$ makes this structure somehow disappear, because the manifold trivially becomes $\mathbb R^3\x\mathbb S^1$. It is why the variable $\th$ does not have a global existence in a general magnetic geometry \cite{BurbQin12}. On the contrary, when using the physical variable $\sfc$, the circle bundle naturally arises: as $\sfc$ is defined on a space-dependent circle, spatial displacements imply a variation of $\sfc$, so that a covariant derivative is involved, which encodes the circle-bundle geometry for the gyro-angle \cite{BurbQin12} and does not imply some restricted class of circle bundle. A more detailed study of the coordinate system is outside the scope of the present paper, and will be reported in \cite{GuilAnholon}.\\

 In some way, the relevance of this coordinate is obvious, since it just results from keeping the initial coordinate, in which all the circle-bundle picture was included. From this point of view, performing the derivation with this variable is a way to see how it globally agrees with the physics and the mathematics of the problem, and to make intrinsic definitions arise naturally for all the quantities involved in the process. 
 
 Indeed, the previous section shows that the reduction follows the same procedure with the vectorial constrained coordinate $\sfc$ as with the scalar coordinate $\th$, but that there are slight changes in the quantities used. The gauge vector $\bfmr$ came as naturally replaced by the connection $\bfmr_g$ for the covariant derivative. The generator of Larmor gyrations $\p_\th$ came with an intrinsic definition $-\sfa\cdott\p_\sfc$. The basic $1$-form for the gyro-angle $d\th$ appeared as replaced by a non-closed $1$-form $\de\th$, which agrees with the fact that $\th$ considers the circle bundle trivial, whereas it should not. This implied to use more intrinsic definitions for the operations used, such as (\ref{ExtDerivGene}) and (\ref{ExtDerivPractical}) for exterior derivatives. Also, this implied to be careful on how the basis of $1$-forms and of vector fields are chosen, but the natural ones were found to agree with each other. 

 Thus, the formalism with $\sfc$ is slightly more involved, but it perfectly fits both with the physics and the mathematics of the problem, which correspond to a non-trivial circle bundle. \\

 From a formal point of view, the results with the physical variable $\sfc$ mainly correspond to replacing the gauge vector $\bfmr$ by the connection term $\bfmr_g$. Thus, they include the standard gauge-dependent results as a special case, but they are more general: in the usual approach $\bfmr(\bfq)$ depends only on the position and cannot be chosen freely (e.g. $\bfmr=0$ is not possible \cite{Little81, Little88, BurbQin12}); here, $\bfmr_g(\bfq,\bfp)$ can be any function of the position and momentum. Especially, the physical definition of $\sfc$ corresponds to the function $\bfmr_g:=-\phi\na\sfb\cdott\sfc$, which depends also on the momentum and preserves the polynomiality in $\phi$. \\

 Other gauge-dependent quantities are interesting to consider. In previous works, the coordinate transformation $\th\longrightarrow\overline\th$ was gauge dependent (see for instance equation (30c) in \cite{Little83}, or in \cite{BrizHahm07} the solution for the generator $\bfmg_1^\zeta$ below equation (5.45)), as well as the definition of the coordinate $\th$, and also the gradient $\p_{\bfq|\th}$. It is why the gauge vector $\bfmr$ was involved in some of the resulting formulae, e.g. the Poisson bracket, in such a way as to make all the physical or geometrical (intrinsic) quantities gauge-independent. For instance, in the Poisson bracket, gradients appear only in the combination $\na_*:=\na+\bfmr\p_\th$ \cite{BrizHahm07}. It would be interesting to interpret it as the gradient corresponding to a special gauge, because it would remove the appearance of the gauge vector in all the derivation, and would simplify computations. The issue is that it is not possible because it would correspond to fix the gauge in such a way that $\na\sfe_1\cdott\sfe_2=0$, which is not possible even locally \cite{Little81, Little88}.

 In the gauge-independent approach, all the coordinates, including the gyro-angle $\sfc$, are gauge-independent, as well as the transformation $\sfc\longrightarrow\overline\sfc $: at first order, it is not transformed by $\sfa\bfmg_1^\th$, because the covariant derivative must be taken into account, which means that it is given by $\sfa\bfmg_1^\th-\bfmg_1^\bfq\cdott\na\sfc$. This last quantity is indeed independent of the connection vector $\bfmr_g:=\na\sfc\cdott\sfa$, as can be verified in (\ref{SolG1th}). In the same way, at higher order, all the transformed coordinates $\overline\bfz = ...e^{-\bfmg_2}e^{-\bfmg_1}\bfz$ will be independent of $\bfmr_g$, where $\bfz:=(\bfq,\phi,\mu,\sfc)$ are physical coordinates. 
 
 Gradients are also involved in combinations involving $\bfmr_g$ (see Eq.~(\ref{PoissBrackRed}), for instance). This is no surprise, since the connection on the fibre bundle involves some arbitrariness, but the combinations can always be written $\na_*:=\na+\bfmr_g\p_\th$, which is connection-independent. In addition, it can be interpreted as the covariant derivative associated with the trivial connection $\bfmr_g=0$. Thus, when working with the coordinate $\sfc$, this choice can be used to simplify computations and to make them connection-independent.

\subsection{Maximal vs. minimal reduction}

 The derivation procedure confirms the respective interests of Lie transforming the velocity vector field and the Lagrangian $1$-form. 
 
 As with concerns the minimal requirements for the guiding-center transformation, working on the equation of motion is much more efficient, since it systematically obtains the fluctuating part of the reduced motion just by inverting the operator $\p_\th$. 
 
 The procedure with the Lagrangian is much more involved, as can be seen in previous sections, especially because the order mixing makes the scheme more elaborated and because the algorithmic stage begins only at higher order: the induction matrix mixes up the orders, changes at each order for $n\leqslant 3$, and involves differential operators in some coefficients. It is why in this paper, as in previous works, only a part of $\bfmg_2$ is explicitly computed, whereas the work \cite{GuilGCmin} directly obtained the full second-order generator $\bfmg_2$.

 In addition, the minimal guiding-center reduction can hardly be obtained by working on the Lagrangian, because going from the Lagrangian to the motion mixes the components up. To guarantee an averaged slow reduced motion for the four coordinates $(\overline\bfq,\overline\phi)$, one would need to average all of the seven components of the Lagrangian, which is not a minimal transformation. \\

 As with concerns the additional requirements for the slow dynamics, working on the equation of motion is not efficient, because the equations to be solved are secular differential equations that are not simple to deal with \cite{GuilGCmin}. Working on the Lagrangian is more efficient, because it essentially consists in algebraic equations, which deals the same way with gyro-averages as with gyro-fluctuations. This makes it easy to identify good choices for the averaged transformation generator $\avg(\bfmg_n)$ in order to obtain a reduced Lagrangian as strongly reduced as possible. Thus, it provides a maximal guiding-center reduction almost as simply as the minimal one. 
 
 Also, working on the Lagrangian $1$-form makes it easy to impose requirements on the reduced Hamiltonian structure, for instance to obtain a quarter-canonical structure for the coordinates $(\overline\mu,\overline\th)$, which both provides a constant of motion $\overline\mu$ and a Hamiltonian sub-dynamics for the $4$-dimensional reduced motion $(\dot{\overline\bfq},\dot{\overline\phi})$.

	\section{Conclusion}

 The full guiding-center reduction can be performed to arbitrary order in the Larmor-radius expansion by Lie transforming the Lagrangian $1$-form while keeping physical gyro-gauge-independent variables as coordinates, following the same procedure as when working with the standard gauge-dependent gyro-angle. 
 
 For higher orders, the procedure was shown to be completely algorithmic. The pivotal role is played by the inverse of the lowest-order Lagrange matrix $\overline\om_{-1}+\overline\om_0+\overline\om_1$, together with a differential equation for the function $S_n$. An extended matrix was defined and used to explicitly solve the induction equation to arbitrary order in the Larmor radius. \\

 The results exactly agree with previous works, but they were obtained without introducing any gyro-gauge, and working purely with the physical coordinate $\sfc$ as the gyro-angle coordinate. In addition, the choice of the cotangent of the pitch-angle as a coordinate for the parallel velocity made the results purely polynomial in the coordinates and monomial in $\sqrt \mu$ and $\sqrt B$. \\

 Compared to the method by Lie transforming the equations of motion, the process is much more elaborated, especially because of the order mixing, but it easily obtains a much stronger result. It does not rely on differential equations for the reduced motion, but on algebraic equations for the reduced Lagrangian. A quarter-canonical reduced Hamiltonian structure provides a constant of motion $\overline\mu$ and a Hamiltonian sub-dynamics for the $4$-dimensional slow reduced motion $(\dot{\overline\bfq}, \dot{\overline\phi})$. In addition, the procedure makes the reduced dynamics trivial not only in the gyro-fluctuating components of the Lagrangian, but also in six of the averaged components out of seven.

 As a result, all but one of the components of the reduced Lagrangian $1$-form are put to zero for all orders higher than two. Only one of them cannot be made exact, and is given by a whole series. The two canonical choices are recovered: either to enclose the series into the Hamiltonian (Hamiltonian representation), then the reduced Poisson bracket is exact, or to enclose the series into the spatial component of the Lagrangian parallel to the magnetic field (symplectic representation), then the Hamiltonian is exact and the uncertainty of the reduced motion is traduced by five coefficients, and especially a kind of reduced magnetic field $\bfmb_*$. 

 The Hamiltonian representation appeared as naturally induced by the reduction process, whereas the symplectic representation is subjected to a condition at each order, to avoid a singularity in $\phi=0$, i.e. at the bounce points of particle trajectories. 
 
 These representations make the reduction maximal because for a general magnetic field, the procedure cannot get a stronger reduction for which even the last component of the Lagrangian would be zero. The obstruction originates from the special role of the magnetic moment; in the Hamiltonian representation, this can be viewed because the magnetic-moment component of the transformation generator can remove only the fluctuating part of the reduced Hamiltonian function, since the averaged part is imposed by the requirement of adiabatic invariance. \\

 The use of gauge-independent coordinates had little effect on the reduction procedure. All the ingredients of the standard reduction with gauge-dependent coordinates were found to be present, but they naturally arose with an intrinsic definition or they were replaced by a different intrinsic object playing a similar role. 

 It was observed to fit in with both the physics and the mathematics of the system, by restoring the general circle-bundle framework, which practically disappeared with the coordinate $\th$, and by making the coordinates directly induced by the physical state of the system. 
 
 For instance, the gauge removal introduced a vectorial quantity $\sfc$ for the gyro-angle coordinate. This caused the coordinate system to be constrained and implied a connection for the covariant derivative on a space-dependent circle, which is directly linked to the circle-bundle structure underlying in the gyro-angle coordinate and which replaced the gauge vector of the gauge-dependent approach.
 
 The closed $1$-form $d\th$ was replaced by a non-closed $1$-form $\de\th$, which is related to the non-triviality of the circle bundle for a general magnetic geometry. 
 
 In previous works relying on the coordinate $\th$, the gauge-independence of the physical results implied that gradients were systematically involved in special expressions, which were not related to derivative operators because no gauge fixing were suited to them. These expressions were found to be related to covariant derivatives corresponding to suitable connections. \\

 Unlike the gauge fixing for the coordinate $\th$, the connection fixing for the coordinate $\sfc$ depends not only on the position, but on the momentum as well. This is all the more convenient as the physical definition of $\sfc$ and its associated connection depend on both the position and the momentum. In addition, this removes one of the assumptions causing the presence of anholonomy in the gyro-angle motion. 
 
 So, the intrinsic gyro-angle coordinate $\sfc$ is a way to tackle some of the questions involved in the guiding-center anholonomy and gauge- (or connection-) arbitrariness. These questions are outside the scope of this paper and will be reported elsewhere \cite{GuilAnholon}. \\

 In this paper, we focused on the formal derivation of perturbation series, as is usual in guiding-center works, and as is the standard first step in perturbation theory \cite{Little82}. A next step will be to relate these formal expansions with asymptotic series, in a similar way as what was done in \cite{BerkGard57} for Kruskal's work, and what is beginning being done about Littlejohn's works \cite{Lutz12}. Also, as usual in perturbation theory, convergence of the guiding-center series is an interesting question to investigate, probably with methods of accelerated convergence \cite{BogoMitr76}. In these attempts, the structures of the expansion series, such as the polynomiality induced by the cotangent of the pitch-angle, are expected to play a role.

	\section*{Acknowledgement}

\noindent 
We acknowledge financial support from the Agence Nationale de la Recherche (ANR GYPSI). This work was also supported by the European Community under the contract of Association between EURATOM, CEA, and the French Research Federation for fusion study. The views and opinions expressed herein do not necessarily reflect those of the European Commission. The authors also acknowledge fruitful discussions with Alain Brizard, Phil Morrison, Mathieu Lutz and with the \'Equipe de Dynamique Nonlin\'eaire of the Centre de Physique Th\'eorique of Marseille.

\appendix

\section{Mechanism of the reduction}
\label{Sect:Appendi}

 In this appendix, we introduce the mechanism at work when Lie transforming the Lagrangian $1$-form. Indeed, the basic ideas of the derivation are very elementary, but they are hidden by the details of the procedure, which are rather involved because of some order mixing and other subtleties between algebraic and differential integrability conditions. In addition, practical computations in the case of the guiding-center are somehow intricate. All the same, the method is very efficient and has quite a wider domain of application than just the guiding-center reduction. So, it seems useful to give a general overview of the method for people not familiar with it.

\subsection{Fundamental ingredients}

 A) 
 The goal is to solve equations (\ref{ReducLagr}) for the unknowns $\bfmg_n$ and $S_n$, with the requirements (\ref{RequirPrior1}), (\ref{RequirPrior3}) and (\ref{RequirPrior2}) identified above. Ideally, the maximal reduction sets $\overline\Ga_n=0$ for all higher orders $n$, as a result of (\ref{RequirPrior3}). 
 
 The solution is built order by order in the Larmor radius. Each order implies to solve the equation 
$$
	\overline\Ga_n
	=\bfmg_{n+1}\cdott \om_{-1}
	+\sfmr_n
	+d S_n
	\,,
$$
where $\sfmr_n$ is a shorthand for all other terms, that do not contain the highest-order generator $\bfmg_{n+1}$. 

 The very basic idea of the reduction is that $\bfmg_{n+1}$ is involved only through a matrix product. So, at any order, the solution is just given by a matrix inversion
\bq
	\bfmg_{n+1} 
	= 
	(\om_{-1})^{-1}\cdott
	\big[
	\overline\Ga_n-\sfmr_n - d S_n
	\big]
	\,,
\label{BasicIdea}
\eq
provided $\om_{-1}$ is invertible; then $\overline\Ga_n$ can be chosen zero, and $S_n$ is not useful and can also be set to zero. This idea of matrix inverse is the key ingredient of the underlying mechanism, even if the corresponding basic picture is not true at the lowest orders, and at higher orders, it is slightly complicated by some order mixing and integrability conditions, especially for $S_n$.\\

 B) 
 As a matter of fact, the matrix $\om_{-1}$ is usually not invertible, since it corresponds to the fast part of the dynamics, here the Larmor gyration, which does not concern all the phase-space coordinates. \\

 At zeroth order, under the requirement $\overline\Ga_n=0$, equation (\ref{ReducLagr}) writes 
$$
 	\bfmg_1\cdott\om_{-1}
	=
	-\Ga_0 - dS_0
	\,.
$$
It has a solution only if the right-hand side $-\Ga_0-dS_0$ is in the range of the matrix $\om_{-1}$ (solvability condition); this is a necessary condition for the corresponding reduction to exist. Usually, it is not verified for so strong a requirement as $\overline\Ga_0=0$, and the reduced Lagrangian $\overline\Ga_0$ has to be used as a softening parameter. Then equation (\ref{ReducLagr}) writes 
\bq
 	\bfmg_1\cdott\om_{-1}
	=
	\overline\Ga_0-\Ga_0 - dS_0
	\,.
\label{EquaG1Matrix}
\eq

 One has to check that, with the freedoms $S_0$ and $\overline\Ga_0$, the solvability condition can be satisfied at least for the minimal requirement (\ref{RequirPrior1}) and if possible for the intermediate requirement (\ref{RequirPrior2}); then, the reduction is possible, and the maximal requirement (\ref{RequirPrior3}) can be considered by trying to remain as close as possible to the condition $\overline\Ga_0=0$.

 At that point, the solution exists, but it is not unique; it is defined to within an element of the kernel of $\om_{-1}$. The choice of this element may be free at this stage of the reduction, but care must be taken that it may be constrained by the solvability conditions at the following order. \\

 At the next order $n=1$, equation (\ref{ReducLagr}) writes 
\bq
	\bfmg_2\cdott\om_{-1}
	+ \tfrac{\bfmg_1}{2}\cdott(\om_0+\overline\om_0)
	= \overline \Ga_{1}- dS_1
	\,.
\label{EqForX2}
\eq
Now, the pivotal matrix $\sfmm_1$ to be inverted is the set of $\om_{-1}$ and $\tfrac{\om_0+\overline\om_0}{2}$, acting on the set of unknown components of $(\bfmg_2,\bfmg_1)$. Its rank is greater than (or equal to) the rank of $\om_{-1}$. Care must be taken that some of the coordinates of $\bfmg_1$ are already determined. This introduces some order mixing, where some components of $\bfmg_n$ are determined at order $\overline\Ga_{n-1}$, others are computed at order $\overline\Ga_{n}$ at the same time as some of the components of $\bfmg_{n+1}$.

 Notice that $\sfmm_1$ cannot be invertible on the unknown components of $\bfmg_2$, because of the non-trivial kernel of $\om_{-1}$. In the same way as at zeroth order, this kernel has to be excluded when studying the invertibility of $\sfmm_1$, because it will be involved only at the following order. 

 Then, if the pivotal matrix $\sfmm_1$ is invertible for the ideal requirement $\overline\Ga_1=0$, then the solution exists and is unique. Otherwise, there is again both a solvability condition and a non-uniqueness of the solution. More precisely, the solvability condition means that there is a solution only if the right-hand side is in the range of the matrix to be inverted, and that in order to fulfil this condition, the reduced Lagrangian $\overline \Ga_1$ may have be chosen non-zero, but having as many null components as possible. The non-uniqueness means that the solution is determined only to within an element of the kernel of the pivotal matrix $\sfmm_1$. The choice of this element may be free, but care must be taken that it may be constrained by the solvability condition at the following order. \\

 At the next orders, the same process goes on. The pivotal matrix $\sfmm_n$ evolves at each order and its rank increases to determine more and more of the unknowns. At high orders, it becomes of constant rank, and actually it becomes the same at each order. The critical value of $n$ at which this occurs will be denoted by \label{DefinitNc}$n_c$, and the corresponding pivotal matrix will be denoted by \label{DefinitMInfty}$\sfmm_\infty$. So, for $n\geqslant n_c$, the pivotal matrix verifies $\sfmm_n=\sfmm_\infty$, whereas for $n=n_c-1$, it verifies $\sfmm_n\neq\sfmm_\infty$.
 
 This can be explained as follows: Equation (\ref{ReducLagrAlg}) generically (i.e. at high orders) writes 
\begin{align}
	\overline\Ga_n=
	& \left[ 
	(\bfmg_{n+1}\cdott d) 
	+(\bfmg_n\cdott d) (\bfmg_1\cdott d)
	+...
	+\tfrac{(\bfmg_1\cdott d)^{n+1}}{n+1 !}
	\right] 
	\Ga_{-1}
\ncr
	+
	& \left[ 
	(\bfmg_n\cdott d) +(\bfmg_{n-1}\cdott d) (\bfmg_1\cdott d)
	+...+\tfrac{(\bfmg_1\cdott d)^n}{n !}
	\right] 
	\Ga_{0}
	+ dS_n
	\,.
	\label{RedLagrEquaGeneric}
\end{align}
In this analysis, low orders are excluded because there would be some additional coefficients coming from the exponential series: for instance for $n=1$, the term $(\bfmg_n\cdott d) (\bfmg_1\cdott d)$ has a factor $1/2$ and is confounded with the last term $\tfrac{(\bfmg_1\cdott d)^{2}}{2 !}$. 

 Denoting $\bfmg_n\cdott d$ by $\sfmg_n$, and grouping together the highest-order Lie derivatives, which contain the unknowns (which are some of the components of $(\bfmg_{n+1},\bfmg_n, ...)$), the previous equation becomes 
\begin{align}
	\overline\Ga_n=
	& ~~ \sfmg_{n+1} \Ga_{-1}
\ncr
	& +\sfmg_{n} \big( \sfmg_1 \Ga_{-1}+\Ga_0\big)
\ncr
	& + \sfmg_{n-1}
	\left[  
	\big(  \sfmg_2 +\tfrac{\sfmg_1^2}{2}\big)
	\Ga_{-1}
	+\sfmg_1\Ga_0
	\right]
	\label{OrderForInvertible}
	\\
	& +...
\ncr
	& + \sfmg_1 
	\left[
	\tfrac{(\sfmg_1)^{n}}{n+1 !}\Ga_{-1}
	+\tfrac{(\sfmg_1)^{n-1}}{n !}\Ga_{0}
	\right]
	+ dS_n
	\,.
	\notag
\end{align}
Using (\ref{ReducLagr}) for the lowest-orders reduced Lagrangian $\overline\Ga_k$, that are already known, the previous formula can be rewritten
\begin{align}
	\overline\Ga_n
	&= \sfmg_{n+1} \big(\overline\Ga_{-1}-dS_{-1}\big)
	 + \sfmg_{n} \big(\overline\Ga_0-dS_0\big)
	+ \sfmg_{n-1} \big(\overline\Ga_1-dS_1\big)
\ncr
	& 
	\hspace{12ex}
	+...
	+ \sfmg_1 
	\left[
	\tfrac{(\sfmg_1)^{n}}{n+1 !}\Ga_{-1}
	+\tfrac{(\sfmg_1)^{n-1}}{n !}\Ga_{0}
	\right]
	+ dS_n
\ncr
	& = 
	\bfmg_{n+1}\cdott\overline\om_{-1}
	+\bfmg_{n}\cdott\overline\om_{0}
	+ ...
	+ d S_n
	+ \sfmr_n
	\,,
\label{MatrPivot}
\end{align}
where $\sfmr_n$ indicates all other terms, that are already known, since they do not involve $(\bfmg_{n+1},\bfmg_n, ...)$.

 Equation (\ref{MatrPivot}) shows that the pivotal matrix $\sfmm_n$ is given by the set of matrices $\overline\om_{-1}$, $\overline\om_0$, etc., acting on the set of unknown components of 
$(\bfmg_{n+1},\bfmg_n, ...)$. The matrix $\sfmm_n$ is the same at any (high) order. It is exactly given by the reduced Lagrange matrix at lowest orders. 
 
 As a consequence,the pivotal matrix $\sfmm_\infty$ for all high orders is identified as soon as the set of $(\overline\om_{-1},\overline\om_0, ...)$ is observed to be invertible on the set of unknown components of $(\bfmg_{n+1},\bfmg_n, ...)$, i.e. as soon as $\overline\om_{-1}+\overline\om_0+ ...$ becomes invertible. We will call \label{DefinitNb}$n_b$ the order such that $\overline\om_{-1}+\overline\om_0+ ... + \overline\om_{n_b}$ is invertible, whereas $\overline\om_{-1}+\overline\om_0+ ... + \overline\om_{n_b-1}$ is not. 

 At that point, the basic picture of A) has become a simple picture B), which includes two stages. At low orders it consists in dealing with non invertible pivotal matrices changing at each order, and in choosing $\overline\Ga_n$ such that it both fulfils the requirement and leads to an interesting invertible matrix $\overline\om_{-1}+\overline\om_0+ ... + \overline\om_{n_b}$. Then at high orders, the induction becomes just a matrix inversion $\sfmm_\infty\mun$, as in the initial basic picture A).\\

 C) 
 The simple picture B) has to be refined. Between these two stages, an intermediate stage takes place, since usually $n_c > n_b+1$. For $n\in\{n_b+1,n_b+2, ... , n_c-1\}$, the higher-orders pivotal matrix is already known but not yet efficient. 
 
 The reason is that formulae (\ref{RedLagrEquaGeneric})-(\ref{MatrPivot}) hold only for orders that are high enough, because of the coefficients generated by expanding the exponentials. If the factors involving $(\bfmg_{n+1},\bfmg_n, ... )$ have some coefficients non unity, then formula (\ref{MatrPivot}) does not hold, which spoils the conclusion. 
 
 But this concerns only low orders. Formula (\ref{OrderForInvertible}) shows that the coefficients will be unity as soon as $n > 2n_b+1$, which means that $n_c\leqslant 2n_b+2$. \\

 As an example, consider $\overline\om_n$ for $n=0$. When computing the next order $\overline\Ga_{n+1}=\overline\Ga_1$, then $\overline\Ga_0$ is already known, but it is not yet efficient: in formula (\ref{MatrPivot}) 
\begin{align}
	\overline\Ga_1
	&=
	 \sfmg_{2} \Ga_{-1}
	+ \sfmg_{1} \big( \tfrac{1}{2}\sfmg_1 \Ga_{-1}+\Ga_0\big)
	+dS_1	
	\ncr
	&\neq
	\bfmg_{2} \cdott \om_{-1}
	+ \bfmg_{1} \cdott \overline\om_0 
	+dS_1
	\,,
	\notag
\end{align}
the operator to be inverted is not just the set of $(\overline\om_{-1},\overline\om_0)$ because of the factor $1/2$ in the first line, which comes because the generator $\bfmg_1$ outside the parenthesis has the same order as the generator $\bfmg_1$ inside the parenthesis. For all higher orders, this will not happen, as is illustrated by the next order 
\begin{align}
	\overline\Ga_2
	& =
	 \sfmg_{3} \Ga_{-1}
	+ \sfmg_{2} \big( \sfmg_1 \Ga_{-1}+\Ga_0\big)
	+ o.t. 	
	\ncr
	& =
	\bfmg_{2} \cdott \om_{-1}
	+ \bfmg_{1} \cdott \overline\om_0 
	+ o.t. 
	\,,
	\notag
\end{align}
where $o.t.$ is used for "other terms", in order to avoid writing uninteresting terms. \\

 D) The order mixing can also slightly complicate the picture of C), by spoiling the linear algebraic framework, mainly at order $n=1$. Indeed, the first equation to be solved for $\bfmg_2$ is (\ref{EqForX2}). However, if some of the components of $\bfmg_1$ are still not determined at that point (this is fairly general as $\om_{-1}$ is usually not invertible), then they can be involved in a differential equation. Indeed, $\overline\Ga_0$ can be undetermined at that point, and equation (\ref{EqForX2}) must be let under its initial form (\ref{ReducLagr})
$$
	\overline \Ga_{1}
	=
	\left(\bfmg_2+\tfrac{\bfmg_1\cdott d }{2}\bfmg_1\right)
	\cdott \om_{-1}
	+\bfmg_1\cdott\om_0
	 + dS_1
	 \,,
$$
which is now a differential equation for $\bfmg_1$, and may even be non-linear in the unknown components of $\bfmg_1$. This can make the scheme much more complicated: even solvability conditions may be difficult to identify. \\

 E) 
 Finally, one last point has to be taken into account as well and still makes the scheme more elaborated than the picture D) above. The pivotal matrix $\sfmm_n$ determines the unknown components of the generator $(\bfmg_{n+1}, \bfmg_n, ...)$, but this can generate non-zero time-component $\bfmg_n^t$ for the generator. 
 
 For a time-independent transformation, the requirement $\bfmg_n^t=0$, reduces the dimension of the effective generator $\bfmg_n$. Then the pivotal matrix can be inverted only if some integrability conditions are fulfilled. Another way of saying it is that $\overline\Ga_n$ has seven components (seven requirements) whereas $\bfmg_n$ has only six freedoms. The additional freedom comes from the gauge function $S_n$. 
 
 Actually, the presence of this integrability condition for the pivotal matrix $\sfmm_\infty$ is completely general and comes because $\sfmm_\infty$ is antisymmetric. It is not invertible on the $7$-dimensional space $(\bfq,\bfp,t)$, and can be invertible only on a sub-space, e.g. on the $6$-dimensional phase-space $(\bfq,\bfp)$. For a symplectic Hamiltonian system, the high-orders pivotal matrix $\sfmm_\infty$ is indeed invertible when restricted to the phase-space, since the Lagrange $2$-form $\om_s$ is invertible, and so is $\overline\om_s$. 
 
 So, the gauge function $S_n$ is not determined by the algebraic matrix inversion, but by the solvability condition for the matrix inversion. Furthermore, it appears in a differential equation. Existence of solution for this differential equation can involve other integrability conditions. For instance, in an equation such as 
$$
	\p_\th S_n = f_n
	\,,
$$
inverting $\p_\th S$ implies the function $f_n$ to have no gyro-average. 

 As a result, both the algebraic and the differential integrability conditions must be played with so as to make $\overline\Ga_n=0$. If it is not possible, one has to choose a non-zero reduced Lagrangian $\overline\Ga_n\neq 0$. This means playing with the requirements also, and releasing them slightly, so as to make the integrability conditions fulfilled and at the same time to keep $\overline\Ga$ as strongly reduced as possible.\\

 All these features do not spoil the algorithmic character of the reduction for high orders, because the differential scheme is very simple (the operators are just $\p_{z^k}$) and in addition, at all $n\geqslant n_c$, the algebraic scheme for $(\bfmg_{n+1},\bfmg_n,...)$ is fixed, which makes it possible to conclude about the differential scheme for $S_n$ so as to make the resulting reduction maximal. 
 
 At the end, the induction relations can be written in matrix form provided some coefficients of the matrix are differential operators. By such a redefinition of the matrix $\sfmm_\infty$, the induction relation for high orders $n\geqslant n_c$ just relies on a matrix inverse $\sfmm_\infty^{-1}$. Then, the basic picture of formula (\ref{BasicIdea}) becomes efficient: equations (\ref{ReducLagr}) are solved at arbitrary order through a formula completely analogous to (\ref{BasicIdea}), even if the framework is much more elaborated. We want to stress this fact because the order mixing and the presence of integral operators may hide the triviality of the induction mechanism.

\subsection{Resulting procedure in three stages}

 The previous subsection shows that the reduction is performed in three stages. The first stage corresponds to the first few orders $n\leqslant n_b$. The work consists in verifying that the freedoms can be used both to make the solvability conditions satisfied and to get an interesting invertible matrix $\sfmm_\infty$. At the end $n=n_b$, the invertible high-order pivotal matrix $\sfmm_\infty$ becomes identified, and the first stage is ended. 
 
 The second stage corresponds to a transition stage. The pivotal matrix for high order is identified, but it is still not efficient at that order. The goal is only to check that the solvability conditions can be satisfied at these intermediate orders.
 
 The third stage begins at order $n=n_c$, i.e. as soon as the matrix to be inverted becomes $\sfmm_\infty$. From that order on, it is sure that the reduction can be performed to any order in the Larmor radius. As the matrix is now invertible, the solution exists and is unique to each order, and the process becomes fully algorithmic. \\

 In order to get a formula analogous to (\ref{BasicIdea}), the pivotal matrix must be extended to include the gauge function $S_n$, and some coefficients of the inverted matrix $\sfmm_\infty\mun$ are then integral operators. In addition, in order to deal with the order mixing, some intermediate quantities must be introduced to isolate the components of $(\bfmg_{n+1},\bfmg_n, \bfmg_{n-1})$ that are already known from the ones that are not identified yet. \\

 For example, if the pivotal matrix $\sfmm_\infty$ involves only $\overline\om_{-1}$ and $\overline\om_{0}$. Then, the equation (\ref{ReducLagr}) or (\ref{MatrPivot}) writes
\bq
	\overline\Ga_n=
	\bfmg_{n+1}\cdott\overline\om_{-1}
	+\bfmg_{n}\cdott\overline\om_{0}
	+ d S_n
	+ \sfmr_n
	\,,
\label{EquHighOrd}
\eq
where $\sfmr_n$ indicates all terms of (\ref{ReducLagr}), that do not depend on the unknowns, which are the gauge function $S_n$ and some components of $(\bfmg_{n+1},\bfmg_n)$. These last quantities can be grouped into one single vector
\bq
	\bfg_n:=\big( \bfmg_{n+1},\bfmg_n, S_n \big)
	\,.
\label{DefVectBfgApp}
\eq
The pivotal matrix $\sfmm_n$ is then extended to act on all $\bfg_n$ (including the gauge function) in (\ref{EquHighOrd}) and is defined by 
$$
	\sfmm_n\cdott\bfg_n
	:=
	\bfmg_{n+1}\cdott\overline\om_{-1}
	+\bfmg_{n}\cdott\overline\om_{0}
	+ d S_n
	\,.
$$
As announced, some of its coefficients (the ones acting on the component $S_n$) are differential operators. With these conventions, equation (\ref{EquHighOrd}) writes 
\bq
	\overline\Ga_n=
	\sfmm_n\cdott\bfg_n
	+ \sfmr_n
	\,.
	\notag
\eq

 Now, some of the components of $(\bfmg_{n+1},\bfmg_n)$ are already identified at that order. Let us denote them by the index $a$, and the remaining components of $\bfg$, which are not identified are denoted by the index $\infty$: 
$$
	\bfg=\big( (\bfg_n)_a; (\bfg_n)_\infty \big)
	\,,
$$
with $(\bfg_n)_a$ fully identified and all terms of $(\bfg_n)_\infty$ fully unknown. The Lagrangian writes 
\bq
	\overline\Ga_n=
	(\sfmm_n)_a\cdott(\bfg_n)_a
	+	(\sfmm_n)_\infty\cdott(\bfg_n)_\infty
	+ \sfmr_n
	\,,
\label{DecomposBfgAppen}
\eq
with obvious definitions for the linear operators $(\sfmm_n)_a$ and $(\sfmm_n)_\infty$. By assumption, the quantities $\sfmr_n$ and $(\sfmm_n)_a(\bfg_n)_a$ are known; in addition, $(\sfmm_n)_\infty=\sfmm_n=\sfmm_\infty$ is known and invertible. As a consequence, the induction relation writes
\bq
	(\bfg_n)_\infty
	=
	\sfmm_\infty^{-1}\cdott
	\left[
	\overline\Ga_n
	- (\sfmm_n)_a\cdott(\bfg_n)_a
	- \sfmr_n
	\right]
	\,.
\label{SolutExpl}
\eq
It is explicit and makes the basic picture (\ref{BasicIdea}) apply to all orders $n\geqslant n_c$. Some coefficients of $\sfmm_\infty\mun$ are integral operators, since in the inverse matrix $\sfmm_\infty$ some coefficients are differential operators. 

 A few comments are in place. First, some components of $\bfmg_{n+1}$ remain non-identified after the order $n$; they must be excluded from $(\bfg_n)_\infty$ to get an invertible matrix, because they are elements of the kernel of $\om_{-1}$ and will be determined at the next order; this is well illustrated by (\ref{DefVectBfgText}) and (\ref{DefVectBfgInftyText})-(\ref{DefVectBfgAlpha}). Second, the components $(\bfg_n)_a$ can be extracted from $\bfg_n$ and its term $(\sfmm_n)_a(\bfg_n)_a$ can be grouped with $\sfmr_n$ (see formula (\ref{DefSfmrPrime})), which plays the same role. Last, the reduced Lagrangian $\overline\Ga_n$ is in principle taken to be zero, but it was kept free because integrability conditions for $S_n$ can make it necessary to choose some of its components non-zero; then, it can be included in the vector $\bfg_n$, as is done in (\ref{DefVectBfgText}) and (\ref{DefVectBfgInftyText})-(\ref{DefVectBfgAlpha}). \\

 The final algorithm to be iterated for the $n$-th-order term is trivial: in formula (\ref{SolutExpl}), replace the lowest orders terms by their expression, already known, then compute the Lie derivatives involved in the term $\sfmr_n$, and last apply the matrix product with $\sfmm_\infty^{-1}$. The mechanism involves just two kinds of operations, derivatives and a matrix product, which can be easily implemented to arbitrary order using computer-assisted formal calculus. \\

 The basic idea shown in (\ref{BasicIdea}) and (\ref{SolutExpl}) explains why Lie transforming the Lagrangian $1$-form has the advantage of algebraic equations, which makes it easy to reduce also the averaged part of the reduced motion, and thus to get non-minimal guiding-center reductions.  Indeed, computations for the non-minimal requirements are treated the same way as for the minimal ones, the only difference concerns the priority: if all requirements cannot be satisfied, then the order of priority may impose the requirements to be preferred and the ones to be released. This is an essential advantage of Lie transforming the Lagrangian. 

 But the overall process is much more involved than the method relying on a Lie transform of the equations of motion. This last has the essential advantage of relying on just a gyro-integral, which makes it much more efficient to work on the fluctuating part of the reduced dynamics and to perform the minimal guiding-center reduction, as is clear in \cite{GuilGCmin}.

 In both cases, the reduction relies on explicit induction relations, but when working with the Lagrangian, the algorithmic stage (third stage introduced above) is efficient only for higher orders. For lowest orders, the reduction is not systematic at all, the choices are crucial to make the reduction work or not, but they must be guessed rather than derived. In addition, many solvability conditions appear in the process, and there is no a-priori guarantee that they can be satisfied. 
\\

 In the case of the guiding-center reduction, good choices appear rather naturally, solvability conditions come as easily satisfied, and the reduction can be considered as rather straightforward, but two specificities must be taken into account. 
 
 Indeed, as expected, at each order, the fluctuating part of $\bfmg_n$ is imposed by the minimal requirement (\ref{RequirPrior1}), which means to put to zero the gyro-fluctuating part of the Lagrangian; and the averaged part is imposed by the other requirements (\ref{RequirPrior3}) and (\ref{RequirPrior2}), which mean to put to zero the averaged part of the Lagrangian as well (except that $\overline\Ga_1^\th=\overline\mu$). 
 
 However, one of the components of $(\bfmg_{n+1},\bfmg_n,...)$ that remains not identified is already present in equation (\ref{EquHighOrd}): $\avg(\Ga_n^\mu)$ remains as a parameter in the right-hand side of (\ref{SolutExpl}). 
 
 Furthermore, the integrability conditions on $S_n$ cannot be fully satisfied, one of the optional requirements (\ref{RequirPrior3}) must be dropped; so, the average component $\overline\Ga^\sfb$ (or alternatively $\overline\Ga^t$) is not zero but used to make the integrability condition satisfied. Accordingly, one of the freedoms (the average gauge function $\avg (S_n)$) remains undetermined for the maximal reduction. To determine it, a prescription must be added. For the simplest maximal reduction, it is put to zero.
 
 All these features will suggest to define and decompose the vector $\bfg_n$ in a different way as in (\ref{DefVectBfgApp}) and (\ref{DecomposBfgAppen}), by including in this vector only the unknowns that are involved at that order (see formula (\ref{DefVectBfgText})), and by distinguishing between the unknowns that will be identified and the ones that will remain parameters (see formulae (\ref{DefVectBfgInftyText})-(\ref{DefVectBfgAlpha})).

\end{document}